\documentclass{article} 
\usepackage{amsmath}
\usepackage{paralist}
\usepackage[misc]{ifsym}
\usepackage{epsfig} 
\usepackage{epstopdf} 
\usepackage[colorlinks=true]{hyperref}
\usepackage[a4paper, total={150mm,225mm}, top=32mm]{geometry}
\hypersetup{urlcolor=blue, citecolor=red}
\allowdisplaybreaks 

\usepackage[utf8]{inputenc}
\usepackage{amssymb,bm}
\usepackage{mathrsfs}
\usepackage{xcolor}
\usepackage{scalerel}
\usepackage{theorem}
\usepackage{cite}
\usepackage{url}
\usepackage{array,multirow}
\usepackage{pgfplots}
\usepackage{tikz}
\usetikzlibrary{plotmarks, patterns, tikzmark, positioning}
\usetikzlibrary{matrix,decorations.pathreplacing,calc}
\usepackage{hf-tikz}
\usepackage[ruled,vlined,titlenumbered,linesnumbered]{algorithm2e}
\usepackage[nolist]{acronym}

\definecolor{darkgreen}{rgb}{0,0.5,0}

\def\ve#1{{\mathchoice{\mbox{\boldmath$\displaystyle #1$}}%
		{\mbox{\boldmath$\textstyle #1$}}%
		{\mbox{\boldmath$\scriptstyle #1$}}%
		{\mbox{\boldmath$\scriptscriptstyle #1$}}}}

\pgfplotsset{
compat=1.17,
mystyle/.style={
    scale only axis,
    width=0.75\columnwidth,
    height=0.55\columnwidth,
    label style={inner sep=0, font=\normalsize}, 
    tick label style={font=\scriptsize},
    legend style={font=\scriptsize},
    mark size=3,
    major grid style={dashed},
    line width=0.8pt,
    axis line style = thin}
}

\newcolumntype{M}[1]{>{\centering\arraybackslash}m{#1}}

\newtheorem{theorem}{Theorem}
\newtheorem{definition}{Definition}
\newtheorem{lemma}{Lemma}
\newtheorem{corollary}{Corollary}
\newtheorem{problem}{Problem}
\newtheorem{remark}{Remark}
\newtheorem{proposition}{Proposition}
\newtheorem{example}{Example}

\newtheorem{proof}{Proof}

\newcommand{\Fqm}{\ensuremath{\mathbb F_{q^m}}}
\newcommand{\Fqms}{\ensuremath{\mathbb F_{q^{ms}}}}

\newcommand{\Fq}{\ensuremath{\mathbb F_{q}}}

\newcommand{\F}{\ensuremath{\mathbb F}}
\newcommand{\ZZ}{\ensuremath{\mathbb{Z}}}
\newcommand{\set}[1]{\ensuremath{\mathcal{#1}}}

\newcommand{\intervallincl}[2]{\ensuremath{[#1,#2]}}

\newcommand{\aut}{\ensuremath{\sigma}}
\newcommand{\autMat}[3]{\ensuremath{\aut^{#3}({#1}_{#2})}}
\newcommand{\autVec}[3]{\ensuremath{\aut^{#3}({#1}_{#2})}}
\newcommand{\der}{\ensuremath{\delta}}
\newcommand{\SkewPolyring}{\ensuremath{\Fqm[x;\aut,\der]}}
\newcommand{\SkewPolyringZeroDer}{\ensuremath{\Fqm[x;\aut]}}

\newcommand{\MultSkewPolyringZeroDer}{\ensuremath{\Fqm[x,\allowbreak y_1,\allowbreak \dots,y_\intOrder;\aut]}}
\newcommand{\MultSkewPolyringZeroDerAny}[1]{\ensuremath{\Fqm[x,y_1,\dots,y_{#1},\aut]}}

\newcommand{\opev}[3]{\ensuremath{{#1}(#2)_{#3}}}

\newcommand{\op}[2]{\ensuremath{\mathcal{D}_{\aut}\left(#2\right)_{#1}}}
\newcommand{\opexp}[3]{\ensuremath{\mathcal{D}^{#3}_{\aut}(#2)_{#1}}}

\newcommand{\IPop}[1]{\mathcal{I}_{#1}^{\mathrm{op}}}

\newcommand{\minpolyOp}[2]{\ensuremath{M_{#1}(x)_{#2}}}
\newcommand{\minpolyOpNoX}[1]{\ensuremath{M_{#1}}}

\newcommand{\MABin}{\A}

\newcommand{\MABout}{\B}

\newcommand{\MABoutentry}{B}

\newcommand{\algoname}[1]{{\normalfont\textsc{#1}}} 

\newcommand{\MABnameFull}[3]{$#2$-ordered weak-Popov approximant basis of $#1$ of order $#3$}
\newcommand{\MABnameFullStandard}{\MABnameFull{\MABin}{\s}{d}}
\newcommand{\RMABnameShort}[3]{\mathsf{owPopovApprox}_{\mathsf{R}}(#1,#2,#3)}
\newcommand{\RMABnameShortStandard}{\RMABnameShort{\MABin}{\s}{d}}
\newcommand{\LMABnameShort}[3]{\mathsf{owPopovApprox}_{\mathsf{L}}(#1,#2,#3)}
\newcommand{\LMABnameShortStandard}{\LMABnameShort{\MABin}{\s}{d}}

\SetKwComment{Comment}{}{}
\newcommand{\myAlgoComment}[1]{\Comment{\normalfont // #1}}

\newcommand{\defeq}{:=}

\renewcommand{\bar}{\overline}

\newcommand{\modr}{\; \mathrm{mod}_\mathrm{r} \;}

\DeclareMathOperator{\wt}{wt}
\DeclareMathOperator{\rk}{rk}

\DeclareMathOperator{\diag}{diag}

\newcommand{\LP}[1]{\ensuremath{\textrm{LP}_{\!\wOrder\!}(#1)}}
\newcommand{\mystack}[2]{\ensuremath{\genfrac{}{}{0pt}{}{#1}{#2}}}
\newcommand{\rker}{\ensuremath{\ker_r}}
\newcommand{\lker}{\ensuremath{\ker_l}}

\renewcommand{\vec}[1]{\ve{#1}} 
\newcommand{\mat}[1]{\ensuremath{\bm{#1}}}
\newcommand{\Mat}[1]{\ensuremath{\bm{#1}}}

\newcommand{\genMoore}{{\ensuremath{\aut}-Generalized Moore}}

\newcommand{\opVandermonde}[3]{\ensuremath{\mat{V}_{#1}(#2)_{#3}}}
\newcommand{\genNorm}[2]{\ensuremath{\mathcal{N}^{#1}_{\aut}(#2)}}

\newcommand{\lclm}{\ensuremath{\mathrm{lclm}}}

\renewcommand{\a}{\vec{a}}
\renewcommand{\b}{\vec{b}}
\renewcommand{\c}{\vec{c}}
\newcommand{\e}{\vec{e}}
\newcommand{\f}{\vec{f}}

\renewcommand{\r}{\vec{r}}
\newcommand{\s}{\vec{s}}
\renewcommand{\t}{\vec{t}}

\renewcommand{\v}{\vec{v}}
\newcommand{\w}{\vec{w}}
\newcommand{\x}{\vec{x}}

\newcommand{\A}{\mat{A}}
\newcommand{\B}{\mat{B}}
\newcommand{\C}{\mat{C}}
\newcommand{\D}{\mat{D}}
\newcommand{\E}{\mat{E}}
\newcommand{\G}{\mat{G}}

\newcommand{\Q}{\mat{Q}}
\newcommand{\R}{\mat{R}}

\newcommand{\U}{\mat{U}}
\newcommand{\W}{\mat{W}}
\newcommand{\X}{\mat{X}}

\newcommand{\0}{\ensuremath{\mathbf 0}}
\newcommand{\vecbeta}{\ensuremath{\boldsymbol{\beta}}}

\newcommand{\veczeta}{\ensuremath{\boldsymbol{\zeta}}}
\newcommand{\vecepsilon}{\ensuremath{\boldsymbol{\epsilon}}}

\newcommand{\nVec}{\ensuremath{\vec{n}}}

\newcommand{\intSkewRS}[1]{\ensuremath{\mathrm{I}\mathrm{SRS}[#1]}}

\newcommand{\linRS}[1]{\ensuremath{\mathrm{LRS}[#1]}}
\newcommand{\intLinRS}[1]{\ensuremath{\mathrm{I}\mathrm{LRS}[#1]}}

\newcommand{\SumRankWeight}{\ensuremath{\wt_{\Sigma R}}}
\newcommand{\SkewWeight}{\ensuremath{\wt_{skew}}}

\newcommand{\SumRankDist}{d_{\ensuremath{\Sigma}R}}

\newcommand{\myspace}[1]{\mathcal{#1}}

\newcommand{\Rowspace}[1]{\ensuremath{{\left\langle #1 \right\rangle}_{q}}}
\newcommand{\RowspaceFqm}[1]{\ensuremath{{\left\langle #1 \right\rangle}_{q^m}}}

\newcommand{\List}{\ensuremath{\mathcal{L}}}

\newcommand{\quadbinom}[2]{\left[\genfrac{}{}{0pt}{}{#1\vphantom{N_N}}{#2\vphantom{N}}\right]_{q}}

\newcommand{\quadbinomqm}[2]{\left[\genfrac{}{}{0pt}{}{#1\vphantom{N_N}}{#2\vphantom{N}}\right]_{q^m}}

\newcommand{\gammaq}{\ensuremath{\kappa_q}}

\newcommand{\oh}[1]{\bnd{O}{#1}}
\newcommand{\softoh}[1]{\bnd{\widetilde{O}}{#1}}
\newcommand{\softO}{\tilde{O}} 
\newcommand{\bnd}[2]{\ensuremath{#1\mathopen{}\left(#2\right)\mathclose{}}}

\newcommand{\OMul}[1]{\mathcal{M}(#1)}

\newcommand{\nTransmit}{\ensuremath{n_t}}

\newcommand{\deletions}{\ensuremath{\delta}}

\newcommand{\shot}[2]{\ensuremath{{#1}^{(#2)}}}

\newcommand{\nReceiveShot}[1]{\ensuremath{n_r^{(#1)}}}

\newcommand{\pe}{\ensuremath{\alpha}}

\newcommand{\degConstraint}{\ensuremath{D}}
\newcommand{\degConstraintUnique}{\ensuremath{D_u}}
\newcommand{\intOrder}{\ensuremath{s}}

\newcommand{\shots}{\ensuremath{\ell}}

\DeclareMathOperator{\cdeg}{cdeg}
\DeclareMathOperator{\rdeg}{rdeg}
\newcommand{\IntParam}{\intOrder'}

\newcommand{\LODecMat}{{\vec L}}
\newcommand{\RFmat}{\ensuremath{\Q_R}}
\newcommand{\RFvec}{\ensuremath{\f_R}}
\newcommand{\intMat}{\ensuremath{\R_I}}
\newcommand{\h}{\vec{h}}
\newcommand{\T}{\vec{T}}

\newcommand{\Qspace}{\mathcal{Q}}

\newcommand{\myalpha}{\xi}

\newcommand{\M}{\ve{M}}
\newcommand{\m}{\ve{m}}
\newcommand{\hvec}{\ensuremath{\vec{b}}}

\newcommand{\Tset}{\mathcal{T}}

\newcommand{\sample}{\overset{\$}{\gets}}
\newcommand{\Mset}{\mathcal{M}}
\newcommand{\Eset}{\mathcal{E}}
\newcommand{\tmax}{t_\mathsf{max}}

\newcommand{\q}{\vec{q}}
\newcommand{\Qbar}{\bar{\Q}}

\renewcommand{\b}{\vec{b}}
\newcommand{\that}{\hat{t}}
\newcommand{\ttilde}{\tilde{t}}
\newcommand{\V}{\vec{V}}
\newcommand{\n}{\vec{n}}

\newcommand{\Rtilde}{\tilde{\R}}

\newcommand{\module}[1]{\ensuremath{\myspace{#1}}}
\newcommand{\Groeb}{\ensuremath{\set{B}}}
\newcommand{\GroebStar}{\ensuremath{\Groeb_{<\degConstraint}}}
\newcommand{\cardGroebStar}{\ensuremath{\intOrder'}}
\newcommand{\wOrder}{\ensuremath{\prec_{\vec{w}}}}

\newcommand{\vecbetatilde}{\tilde{\vecbeta}}

\begin{acronym}
 \acro{BCH}{Bose--Chaudhuri--Hocquenghem}
 \acro{BMD}{bounded minimum distance}
 \acro{SRS}{skew Reed--Solomon}
 \acro{ISRS}{interleaved skew Reed--Solomon}
 \acro{lclm}{least common left multiple}
 \acro{LRS}{linearized Reed--Solomon}
 \acro{LLRS}{lifted linearized Reed--Solomon}
 \acro{ILRS}{interleaved linearized Reed--Solomon}
 \acro{LILRS}{lifted interleaved linearized Reed--Solomon}
 \acro{MDS}{maximum distance separable}
 \acro{MRD}{maximum rank distance}
 \acro{MSRD}{maximum sum-rank distance}
 \acro{MSD}{maximum skew distance}
 \acro{RS}{Reed--Solomon}
 \acro{LOlike}[LO-like]{Loidreau--Overbeck-like}
\end{acronym}

\title{Fast Decoding of Interleaved Linearized Reed--Solomon Codes and Variants}
\author{Hannes Bartz and Sven Puchinger}
\date{}

\begin{document}
\maketitle 

\begin{abstract}
  We construct $\intOrder$-\acf{ILRS} codes and variants and propose efficient decoding schemes that can correct errors beyond the unique decoding radius in the sum-rank metric.
  The proposed interpolation-based scheme for \ac{ILRS} codes can be used as a list decoder or as a probabilistic unique decoder that corrects errors of sum-rank up to $t\leq\frac{\intOrder}{\intOrder+1}(n-k)$, where $\intOrder$ is the interleaving order, $n$ the length and $k$ the dimension of the code.
  Upper bounds on the list size and the decoding failure probability are given, where the latter is based on a novel \ac{LOlike} decoder for \ac{ILRS} codes.
  We show how the proposed decoding schemes can be used to decode errors beyond the unique decoding radius in the skew metric by using an isometry between the sum-rank metric and the skew metric.

  We generalize fast minimal approximant basis interpolation techniques to obtain efficient decoding schemes for \ac{ILRS} codes (and variants) with subquadratic complexity in the code length.

  Up to our knowledge, the presented decoding schemes are the first being able to correct errors beyond the unique decoding region in the sum-rank and skew metric.
  The performance of the proposed decoding schemes and the tightness of the upper bound on the decoding failure probability are validated via Monte Carlo simulations.
\end{abstract}

\section{Introduction}\label{sec:introduction}

The sum-rank metric is a mix between the Hamming metric and the rank metric and was first considered in~\cite[Sec.~III]{lu2005unified} for constructing space-time codes. 
Later, the sum-rank metric was discovered as a suitable metric for error control in coherent multishot network coding~\cite{nobrega2010multishot}, i.e. a scenario where the network topology and the in-network combinations are known at the receiver.
Recently, the sum-rank metric has also been considered for applications in code-based quantum-resistant cryptography~\cite{puchinger2020generic,sum_rank_overbeck_cbc}.

In the sum-rank metric vectors are considered in a \emph{block-wise} manner. 
Consider a vector $\x=(\shot{\x}{1} \mid \shot{\x}{2} \mid \dots \mid \shot{\x}{\shots})$ with elements from $\Fqm$ that consists of the $\shots$ blocks $\shot{\x}{1},\dots,\shot{\x}{\shots}$.
The sum-rank weight of $\x$ is then defined as 
\begin{equation*}
  \SumRankWeight(\x) \defeq \sum_{i=1}^{\shots} \rk_q(\shot{\x}{i})
\end{equation*}
where $\rk_q(\shot{\x}{i})$ denotes the $\Fq$-rank of $\shot{\x}{i}$, i.e. the maximum number for $\Fq$-linearly independent elements in $\shot{\x}{i}$.
If the size of each block equals one (i.e. $\shots=n$), the sum-rank metric coincides with the Hamming metric. For a single block ($\shots=1$) the sum-rank metric coincides with the rank-metric.
There is an isometry between the sum-rank metric and the so-called \emph{skew} metric~\cite{martinez2018skew}. 
An overview on fundamentals and applications of codes in the sum-rank metric is given in~\cite{CIT-120}.

Linearized Reed--Solomon (\acs{LRS}) codes~\cite{martinez2018skew,caruso2019residues} are a class of evaluation codes that fulfill the Singleton-like bound in the sum-rank metric with equality.
Hence, \ac{LRS} codes are \ac{MSRD} codes.
Similar to original \ac{RS} codes in the Hamming metric~\cite{reed1960polynomial} and Gabidulin codes~\cite{Gabidulin_TheoryOfCodes_1985} in the rank metric, \ac{LRS} codes are constructed by evaluating degree-restricted polynomials at a set of evaluation points, also called \emph{code locators}.
Other than \ac{RS} codes, that are constructed from ordinary polynomials, and Gabidulin codes, that are constructed from \emph{linearized} polynomials~\cite{ore1933special}, \ac{LRS} codes are constructed from \emph{skew} polynomials~\cite{ore1933theory}, a class of non-commutative polynomials that includes (for particular choices of the automorphism) ordinary and linearized polynomials as special cases (see e.g.~\cite{gluesing2021introduction}). 
\ac{LRS} codes receive their name from the considered skew polynomial evaluation, which is \emph{linear} under certain conditions (i.e. per block).
There exist efficient \ac{BMD} decoders for \ac{LRS} codes that can correct errors of sum-rank up to half the minimum distance of the code~\cite{martinez2019reliable,boucher2014linear,caruso2019residues}.

\emph{Interleaved} codes of interleaving order $\intOrder$ are obtained by stacking $\intOrder$ codewords of a code (e.g. over $\Fqm$) into a matrix.
Interleaving is a common tool in coding theory to design codes and decoders that have an improved \emph{burst} error-correction capability.
In the Hamming metric one gets an improved error-correction capability for errors that occur in a column-wise manner since such error patterns corrupt the same locations in the component codewords.
In the rank metric one obtains an improved performance for errors that share the same $\Fq$-row space.

Decoders for interleaved codes are known in the Hamming metric for Reed--Solomon \cite{krachkovsky1997decoding,bleichenbacher2003decoding,coppersmith2003reconstructing,parvaresh2004multivariate,brown2004probabilistic,parvaresh2007algebraic,schmidt2007enhancing,schmidt2009collaborative,cohn2013approximate,nielsen2013generalised,wachterzeh2014decoding,puchinger2017irs,yu2018simultaneous} and in general algebraic geometry codes \cite{brown2005improved,kampf2014bounds,puchinger2019improved}, and in the rank metric for Gabidulin codes \cite{loidreau2006decoding,sidorenko2010decoding,sidorenko2011skew,sidorenko2014fast,wachter2014list,puchinger2017row,puchinger2017alekhnovich,bartz2020fast}.
All of these decoders have in common that they are either list decoders with exponential worst-case and small average-case list size, or probabilistic unique decoders that fail with a very small probability.

\subsection{Related Results}
\ac{LRS} codes have recently shown to provide reliable and secure coding schemes for multi-shot network coding \cite{martinez2019reliable}.
Furthermore, there is a construction \cite{martinez2019universal} of locally repairable codes with maximal recoverability (also known as partial MDS codes) based on~\ac{LRS} codes, which attains the smallest known field size among all existing code constructions for a wide range of code parameters.
The construction of long \ac{LRS} codes over small field sizes was considered~\cite{martinez2020general} and Cyclic-Skew-Cyclic and sum-rank \ac{BCH} codes were presented in~\cite{martinez2021bch}.
Further constructions include double-extended \ac{LRS} codes~\cite{neri2023geometry}, doubly and triply extended \ac{MSRD} codes~\cite{martinez2023doubly}, twisted \ac{LRS} codes~\cite{neri2022twisted} and codes based on subspace designs~\cite{santonastaso2023subspace,santonastaso2025msrd}.
An efficient syndrome-based error-erasure decoder for horizontally and vertically \ac{ILRS} codes was proposed in~\cite{hormann2024syndrome}.

In~\cite{puchinger2021bounds} the authors generalized the bounds on list decoding of Gabidulin codes in the rank metric~\cite{Wachterzeh_BoundsListDecodingRankMetric_IEEE-IT_2013} to list decoding of \ac{LRS} codes in the sum-rank metric. The results show that polynomial-time list decoding of \ac{LRS} codes beyond the Johnson radius is in general not possible. In particular, some \ac{LRS} codes have an exponential list size directly above the unique decoding radius.

Recently, a Gao-like decoder for \emph{horizontally} interleaved \ac{LRS} codes was proposed in~\cite{hilrs_gao_cbc}. A Metzner--Kapturowski-like decoder that allows to decode \emph{any} $\intOrder$-interleaved sum-rank-metric code with high interleaving order $\intOrder$ was presented in~\cite{sum_rank_metz_kap_cbc}.
It was also shown, that \emph{folded} variants of \ac{LRS} codes can be decoded beyond the unique decoding radius efficiently~\cite{hoermann2021folded,hoermann2022folded}.

Apart from \ac{LRS} codes, there exist several good (but not necessarily \ac{MSRD}) sum-rank-metric codes, such as partial unit memory codes constructed from rank-metric codes \cite{wachter2011partial,wachter2012rank,wachter2015convolutional}, convolutional codes \cite{napp2017mrd,napp2018faster} and constructions with a variable block size~\cite{byrne2021fundamental}.

\subsection{Our Techniques \& Contributions}
We generalize the sum-rank metric to (interleaved) matrices and define a corresponding (burst) sum-rank channel that generalizes the corresponding (burst) channel models in the Hamming metric and the rank metric.
In this channel model the component errors of sum-rank weight at most $t$ share the same row support which, if being stacked into a matrix, yields an error matrix over $\Fqm$ that has a small sum-rank weight $t$.

In Section~\ref{sec:ILRS} we show how to construct \ac{ILRS} codes and propose a \ac{LOlike} decoder and an interpolation-based decoding scheme that both allow for decoding errors beyond the unique decoding radius in the sum-rank metric efficiently.

The \ac{LOlike} decoder for \ac{ILRS} codes generalizes the first decoder for interleaved Gabidulin codes by Loidreau and Overbeck~\cite{loidreau2006decoding,overbeck2007public} and can correct errors of sum-rank weight $t \leq \frac{\intOrder}{\intOrder + 1}(n-k)$ with high probability (Theorem~\ref{thm:LO_decoder_ILRS}), where $\intOrder$ is the interleaving order, $n$ the length and $k$ the dimension of the code.

The proposed efficient interpolation-based decoding scheme for \ac{ILRS} codes is inspired by the Wachter-Zeh--Zeh decoder for interleaved Gabidulin codes~\cite{wachter2014list}.
Similar as the Wachter-Zeh--Zeh decoder for interleaved Gabidulin codes~\cite{wachter2014list}, the proposed interpolation-based decoding scheme can be interpreted as a list decoder (with not necessarily polynomial-time worst-case list size) and a probabilistic unique decoder, which either returns a unique solution or a decoding failure.
The list decoder is capable of correcting errors of sum-rank weight up to $t\leq\frac{\intOrder}{\intOrder+1}(n-k+1)$ (see Theorem~\ref{thm:list_dec_ILRS}), whereas the probabilistic unique decoder can correct up to $t\leq\frac{\intOrder}{\intOrder+1}(n-k)$ (see Theorem~\ref{thm:unique_dec_ILRS}).
The interpolation-based decoder requires at most $\softoh{\intOrder^\omega\OMul{n}}$ operations in $\Fqm$, where $\OMul{n}$ is the cost (in operations in $\Fqm$) of multiplying two skew-polynomials of degree at most $n$, which is \emph{subquadratic} in the code length $n$~\cite{puchinger2017fast}.
The resulting performance is achieved by a fast generalized operator evaluation interpolation algorithm (Algorithm~\ref{alg:fast_interpolation}) that is derived in Section~\ref{sec:fast_interpolation}, which relies on fast minimal approximant bases computations~\cite{bartz2020fast}.

It is shown how the proposed decoding schemes can be used for decoding \ac{ISRS} codes from errors of skew weight up to $t<\frac{\intOrder}{\intOrder+1}(n-k+1)$.
 
For the presented decoding schemes, upper bounds on the worst-case list size and the decoding failure probability are given.
The tightness of the upper bounds on the decoding failure probability are validated by Monte Carlo simulations.
 
Up to our knowledge, the proposed decoding schemes are first ones having an error-correction capability beyond the unique decoding radius in the sum-rank and the skew metric by allowing an exponential worst-case and small average-case list size or a small decoding failure probability.
Therefore, the proposed decoding schemes for \ac{ILRS} codes achieve the best decoding regions compared to all explicit sum-rank-metric code constructions and decoders that are known so far.

The generalization of the results for interleaved Reed--Solomon and interleaved Gabidulin codes to \ac{ILRS} codes in the sum-rank metric is not straightforward, as e.g. the properties of the generalized operator evaluation and the concept of conjugacy have to be taken into account carefully. 
In particular, the $\Fq$-linearity known from the rank metric only holds in a block-wise manner, which in turn requires more sophisticated proof techniques (see e.g. the proof of Lemma~\ref{lem:properties_LODecMat_ILRS}).

The main results of this paper, in particular the improvements upon the existing noninterleaved variants, are illustrated in Table~\ref{tab:overview_decoders}.

\begin{table}[ht!]
  \caption{%
    Overview of new decoding regions. Parameters: code length $n$, interleaving order $\intOrder$ (usually $\intOrder \ll n$), error weight (in resp. metric) $t$ and $\tmax\defeq\frac{\intOrder}{\intOrder+1}(n-k)$.
    $\OMul{n}$ is the cost (in operations in $\Fqm$) of multiplying two skew-polynomials of degree at most $n$ and $\omega$ is the matrix multiplication exponent.
    For the complexity always the lowest of the referenced decoding algorithms is given.
}\label{tab:overview_decoders}
\begin{minipage}{\textwidth}
\small
\begin{center}
\newcommand{\specialcell}[2][c]{{\def\arraystretch{1.2}\begin{tabular}[#1]{@{}l@{}}#2\end{tabular}}}
\newcommand{\specialcellsmalldistleftalign}[2][c]{{\def\arraystretch{1}\begin{tabular}[#1]{@{}l@{}}#2\end{tabular}}}
\newcommand{\specialcellcenter}[2][c]{{\def\arraystretch{1}\begin{tabular}[#1]{@{}c@{}}#2\end{tabular}}}
\def\arraystretch{2.0}
\begin{tabular}{l|c|l|c|l}
Code/Decoder & Metric & Decoding Region & Complexity & Reference(s) \\
\hline \hline
\specialcellsmalldistleftalign{\ac{LRS} Codes\\unique decoder} & sum-rank & $t<\frac{1}{2}(n\!-\!k\!+\!1)$ & $\softoh{\OMul{n}}$ & \cite{boucher2014linear,martinez2019reliable, caruso2019residues}
\\ \hline
\specialcellsmalldistleftalign{\ac{ILRS} Codes \\ list decoder} & sum-rank & $t<\frac{\intOrder}{\intOrder+1}(n\!-\!k\!+\!1)$ & $\softoh{\intOrder^\omega\OMul{n}}$ & \specialcellsmalldistleftalign{Thm.~\ref{thm:list_dec_ILRS} \\ Sec.~\ref{subsubsec:ILRS_list}}
\\ \hline
\specialcellsmalldistleftalign{\ac{ILRS} Codes \\ prob. unique} & sum-rank & $t\leq\frac{\intOrder}{\intOrder+1}(n\!-\!k)$ & $\softoh{\intOrder^\omega\OMul{n}}$ & \specialcellsmalldistleftalign{Thm.~\ref{thm:LO_decoder_ILRS} \&~\ref{thm:unique_dec_ILRS} \\ Sec.~\ref{subsec:LO_dec_ILRS} \&~\ref{subsubsec:ILRS_unique}}
\\ \hline\hline
\specialcellsmalldistleftalign{\acs{SRS} Codes\\unique decoder} & skew & $t<\frac{1}{2}(n\!-\!k\!+\!1)$ & \oh{n^2} & \cite{martinez2019reliable,boucher2019algorithm}
\\ \hline
\specialcellsmalldistleftalign{\ac{ISRS} Codes \\ list decoder} & skew & $t<\frac{\intOrder}{\intOrder+1}(n\!-\!k\!+\!1)$ & \softoh{\intOrder^\omega\OMul{n}} & \specialcellsmalldistleftalign{Thm.~\ref{thm:list_dec_ILRS}, \\ Prop.~\ref{prop:isometry}, Sec.~\ref{subsec:ilrs_skew}}
\\ \hline
\specialcellsmalldistleftalign{\ac{ISRS} Codes \\ prob. unique} & skew & $t\leq\frac{\intOrder}{\intOrder+1}(n\!-\!k)$ & \softoh{\intOrder^\omega\OMul{n}} & \specialcellsmalldistleftalign{Thm.~\ref{thm:LO_decoder_ILRS} \& \ref{thm:unique_dec_ILRS}, \\ Prop.~\ref{prop:isometry}, Sec.~\ref{subsec:ilrs_skew}}
\\ 
\end{tabular}
\end{center}
\end{minipage}
\end{table}

\section{Preliminaries}\label{sec:preliminaries}

\subsection{Notation}
The cardinality of a set $\set{S}=\{s_1,s_2,\dots,s_{r}\}$ is denoted by $|\set{S}|$.
By $\intervallincl{i}{j}$ with $i<j$ we denote the set of integers $\{i,i+1,\dots,j\}$.

Let $\Fq$ be a finite field of order $q$ and denote by $\Fqm$ the extension field of $\Fq$ of degree $m$ with primitive element $\pe$.
The multiplicative group $\Fqm\setminus\{0\}$ of $\Fqm$ is denoted by $\Fqm^*$.
Matrices and vectors are denoted by bold uppercase and lowercase letters like $\mat{A}$ and $\vec{a}$, respectively, and indexed starting from one.
Under a fixed basis of $\Fqm$ over $\Fq$ any element $a\in\Fqm$ can be represented by a corresponding column vector $\vec{a}\in\Fq^{m\times 1}$. 
For a matrix $\mat{A}\in\Fqm^{M\times N}$ we denote by $\rk_q(\mat{A})$ the rank of the matrix $\mat{A}_q\in\Fq^{Mm\times N}$ obtained by column-wise expanding the elements in $\mat{A}$ over~$\Fq$.
Let $\aut:\Fqm\to\Fqm$ be a finite field automorphism given by $\aut(a)=a^{q^r}$ for all $a\in\Fqm$, where we assume that $1 \leq r \leq m$ and $\gcd(r,m)=1$.
For a matrix $\A$ and a vector $\a$ we use the notation $\aut(\A)$ and $\aut(\a)$ to denote the element-wise application of the automorphism $\aut$, respectively.
For $\mat{A}\in\Fqm^{M\times N}$ we denote by $\Rowspace{\mat{A}}$ the $\Fq$-linear rowspace of the matrix $\mat{A}_q\in\Fq^{M\times Nm}$ obtained by \emph{row-wise} expanding the elements in $\mat{A}$ over $\Fq$.
The \emph{left} and \emph{right} kernel of a matrix $\mat{A}\in\Fqm^{M\times N}$ is denoted by $\lker(\A)$ and $\rker(\A)$, respectively.

For a set $\set{I}\subset\ZZ_{>0}$ we denote by $[\A]_{\set{I}}$ (respectively $[\a]_{\set{I}}$) the matrix (vector) consisting of the columns (entries) of the matrix $\A$ (vector $\a$) indexed by $\set{I}$.

Vector spaces are denoted by calligraphic letters such as e.g. $\myspace{V}$.
For non-negative integers $a$ and $b$, the number of $b$-dimensional subspaces of $\Fq^a$ is given by the Gaussian binomial $\quadbinom{a}{b}$ which is defined as
\begin{equation*}
\quadbinom{a}{b} = \prod_{i=1}^{b} \frac{q^{a-b+i}-1}{q^i-1}.
\end{equation*}
The Gaussian binomial satisfies~\cite{koetter2008coding}
\begin{equation}
q^{(a-b)b} \leq \textstyle\quadbinom{a}{b} \leq \gammaq q^{(a-b)b}, \label{eq:bounds_gaussian_binomial}
\end{equation}
where
\begin{equation}
\gammaq := \prod_{i=1}^{\infty} (1-q^{-i})^{-1}. \label{eq:gamma_q}
\end{equation}
Note that $\gammaq$ is monotonically decreasing in $q$ with a limit of $1$, and e.g.~$\kappa_2 \approx 3.463$, $\kappa_3 \approx 1.785$, and $\kappa_4 \approx 1.452$.

The notion of conjugacy is an integral part for the definition of \ac{LRS} codes.

\begin{definition}[Conjugacy~\cite{lam1985general}]\label{def:conjugates}
 For any two elements $a\in\Fqm$ and $c\in\Fqm^*$ define
 \begin{equation}
   a^c\defeq \aut(c)ac^{-1}.
 \end{equation}
\begin{itemize}
 \item Two elements $a,b\in\Fqm$ are called $\aut$-conjugates, if there exists an element $c\in\Fqm^*$ such that $b=a^c$.
 \item Two elements that are not $\aut$-conjugates are called $\aut$-distinct.
\end{itemize}
\end{definition}
The notion of $\aut$-conjugacy defines an equivalence relation on $\Fqm$ and thus a partition of $\Fqm$ into conjugacy classes~\cite{lam1988vandermonde}.
The set 
\begin{equation}
  \set{C}(a)\defeq\left\{a^c:c\in\Fqm^*\right\}
\end{equation}
is called \emph{conjugacy class} of $a$.
A finite field $\Fqm$ has at most $q-1$ distinct nontrivial conjugacy classes.
The elements $1,\pe,\pe^2,\dots,\pe^{q-2}$ are representatives of all nontrivial disjoint conjugacy classes of $\Fqm$.

\subsection{Sum-Rank Metric}

The sum-rank metric was defined in~\cite{martinez2018skew} and generalized the Hamming metric and the rank metric.
For the sum-rank metric, we consider vectors $\vec{x}=\left(\vec{x}^{(1)} \mid \vec{x}^{(2)} \mid \dots \mid \vec{x}^{(\shots)}\right)\in\Fqm^n$ that consist of $\shots$ blocks $\vec{x}^{(1)},\vec{x}^{(2)},\dots,\vec{x}^{(\shots)}$ of lengths $n_1, n_2, \dots, n_\shots$, respectively. 
The vector $\n=(n_1,n_2,\dots,n_\shots)\in\ZZ_{\geq 0}^\shots$ containing the block-lengths such that $\sum_{i=1}^{\shots}n_i=n$ is called the \emph{length partition} of $\x$.

\begin{definition}[Sum-Rank Weight~\cite{nobrega2010multishot}]
   Let $\shots\in\ZZ_{\geq 0}$, let $\nVec=(n_1,n_2,\dots,n_\shots)$ be the length partition with $n_i\in\ZZ_{\geq 0}$ for all $i=1,\dots,\shots$ and let $n\defeq\sum_{i=1}^{\shots} n_i$.
   Let $\vec{x}=\left(\vec{x}^{(1)} \mid \vec{x}^{(2)} \mid \dots \mid \vec{x}^{(\shots)}\right)\in\Fqm^n$ where $\vec{x}^{(i)}\in\Fqm^{n_i}$ for all $i=1,\dots,\shots$.
   The sum-rank weight of $\vec{x}$ is defined as
   \begin{equation}
    \SumRankWeight(\vec{x})\defeq\sum_{i=1}^{\shots}\rk_q\left(\vec{x}^{(i)}\right).
   \end{equation}
   The vector 
   \begin{equation}
    \r\defeq\left(\rk_q(\vec{x}^{(1)}),\rk_q(\vec{x}^{(2)}),\dots,\rk_q(\vec{x}^{(\shots)})\right)\in\ZZ_{\geq0}^\shots
   \end{equation}
   is called the \emph{rank partition} of $\x$.
\end{definition}

Note, that for $\shots=n$ we have that the sum-rank metric coincides with the Hamming metric, whereas for $\shots=1$ we obtain the rank metric.
For any vector $\x\in\Fqm^n$ we have that $\SumRankWeight(\x)$ is always less than or equal to its Hamming weight.
By~\cite{martinez2016similarities,martinez2018skew} there always exists a basis of $\Fqm$ over $\Fq$ such that equality holds.
The sum-rank distance between two vectors $\vec{x},\vec{y}\in\Fqm^n$ is defined as
\begin{equation}
  \SumRankDist(\vec{x},\vec{y})\defeq\SumRankWeight(\vec{x}-\vec{y})=\sum_{i=1}^{\shots}\rk_q\left(\vec{x}^{(i)}-\vec{y}^{(i)}\right).
\end{equation}
We define the (burst) sum-rank weight of a matrix $\mat{X}=(\mat{X}^{(1)} \mid \mat{X}^{(2)} \mid \dots \mid \mat{X}^{(\shots)})\in\Fqm^{\intOrder\times n}$ as 
\begin{equation}
  \SumRankWeight(\mat{X})\defeq\sum_{i=1}^{\shots}\rk_q\left(\mat{X}^{(i)}\right),
\end{equation}
where $\mat{X}^{(i)}\in\Fqm^{\intOrder\times n_i}$ for all $i=1,\dots,\shots$.
The sum-rank distance between two matrices $\mat{X},\mat{Y}\in\Fqm^{\intOrder\times n}$ is then defined as
\begin{equation}
  \SumRankDist(\mat{X},\mat{Y}) \defeq \SumRankWeight(\mat{X}-\mat{Y})=\sum_{i=1}^{\shots}\rk_q\left(\mat{X}^{(i)}-\mat{Y}^{(i)}\right).
\end{equation}

\begin{remark}
 We want to emphasize that the sum-rank weight and the sum-rank distance depend on the length partition $\n$ of the considered vector $\x$.
 To simplify the notation, we implicitly assume that the sum-rank weight and distance is computed with respect to the length partition of $\x$, since this will be clear from the context. 
\end{remark}

\subsection{Skew Polynomials}\label{subsec:skew_polys}

\emph{Skew polynomials} are a special class of non-commutative polynomials that were introduced by Ore~\cite{ore1933theory}.
A \emph{skew polynomial} is a polynomial of the form 
\begin{equation}
 \textstyle f(x)=\sum_{i}f_i x^i
\end{equation}
with a finite number of coefficients $f_i\in\Fqm$ being nonzero.
The degree $\deg(f)$ of a skew polynomial $f$ is defined as $\max\{i:f_i\neq 0\}$ if $f\neq0$ and $-\infty$ otherwise.

The set of skew polynomials with coefficients in $\Fqm$ together with ordinary polynomial addition and the multiplication rule
\begin{equation}\label{eq:skew_poly_mult_rule}
  xa=\aut(a)x,\qquad a\in\Fqm
\end{equation}
forms a non-commutative ring denoted by $\SkewPolyringZeroDer$.

The set of skew polynomials in $\SkewPolyringZeroDer$ of degree less than $k$ is denoted by $\SkewPolyringZeroDer_{<k}$.
For any $a,b\in\Fqm$ we define the operator
\begin{equation}\label{eq:def_op}
  \op{a}{b} \defeq \aut(b)a.
\end{equation}
For an integer $i\geq0$, we define (see~\cite[Proposition~32]{martinez2018skew})
\begin{equation}\label{eq:def_op_exp}
  \opexp{a}{b}{i+1}
  =\op{a}{\opexp{a}{b}{i}}
  =\aut^{i+1}(b)\genNorm{i+1}{a}
\end{equation}
where $\opexp{a}{b}{0}=b$ and $\genNorm{i}{a}=\aut^{i-1}(a)\aut^{i-2}(a)\dots\aut(a)a$ is the generalized power function (see~\cite{lam1988vandermonde}).
For an integer $i<0$ we define
\begin{equation}
  \opexp{a}{b}{-i}=\aut^{-i}(b)/\aut^{-i}(\genNorm{i}{a}).
\end{equation}
Observe, that for any integers $i,j$ we have that
\begin{equation}\label{eq:conc_prop_op}
  \opexp{a}{\opexp{a}{b}{i}}{j}=\opexp{a}{b}{i+j}.
\end{equation}

The \emph{generalized operator evaluation} of a skew polynomial $f\in\SkewPolyringZeroDer$ at an element $b$ w.r.t. $a$, where $a,b\in\Fqm$, is defined as (see~\cite{leroy1995pseudolinear,martinez2018skew})
\begin{equation}\label{eq:def_gen_op_eval}
  \opev{f}{b}{a}=\sum_{i}f_i\opexp{a}{b}{i}.
\end{equation}
The generalized operator evaluation forms an $\Fq$-linear map, i.e. for any $f\in\SkewPolyringZeroDer$, $\beta,\gamma\in\Fq$ and $a,b,c\in\Fqm$ we have that
\begin{equation}
  \opev{f}{\beta b+\gamma c}{a}=\beta\opev{f}{b}{a}+\gamma\opev{f}{c}{a}.
\end{equation}
For an element $a\in\Fqm$, a vector $\b\in\Fqm^n$ and a skew polynomial $f\in\SkewPolyringZeroDer$ we define
\begin{equation}
  \opev{f}{\b}{a}\defeq(\opev{f}{b_1}{a},\opev{f}{b_2}{a},\dots,\opev{f}{b_n}{a}).
\end{equation}

\begin{proposition}[Number of Roots~\cite{caruso2019residues}]\label{prop:number_of_roots}
Let $a_1,\dots,a_\shots$ be representatives from distinct nontrivial conjugacy classes of $\Fqm$ and let $\b^{(i)}=\big(b_1^{(i)},\dots,b_{n_i}^{(i)}\big)$ contain elements from $\Fqm$ for all $i=1,\dots,\shots$.
Then for any nonzero $f\in\SkewPolyringZeroDer$ satisfying
\begin{equation}
  \opev{f}{b_j^{(i)}}{a_i}=0,\forall i=1,\dots,\shots,j=1,\dots,n_i
\end{equation}
we have that $\sum_{i=1}^{\shots}\rk_q\big(\b^{(i)}\big)\leq\deg(f)$ where equality holds if and only if the $b_1^{(i)},\dots,b_{n_i}^{(i)}$ are $\Fq$-linearly independent for each $i=1,\dots,\shots$.
\end{proposition}

For two skew polynomials $f,g\in\SkewPolyringZeroDer$ we denote by $f \modr g$ the right modulo operation, i.e. the remainder of the right division of $f$ by $g$. 

The existence of a (generalized operator evaluation) interpolation polynomial is considered in Lemma~\ref{lem_int_poly_op_ev} (see e.g.~\cite{caruso2019residues}).
\begin{lemma}[Lagrange Interpolation (Generalized Operator Evaluation)]\label{lem_int_poly_op_ev}
 Let $b_1^{(i)},\allowbreak\dots,b_{n_i}^{(i)}$ be $\Fq$-lin\-early independent elements from $\Fqm$ for all $i=1,\dots,\shots$.
 Let $c_1^{(i)},\dots,c_{n_i}^{(i)}$ be elements from $\Fqm$ and let $a_1,\dots,a_\shots$ be representatives for different nontrivial conjugacy classes of $\Fqm$.
 Define the set of tuples $\set{B} \defeq \{(b_j^{(i)}, c_j^{(i)}, a_i):i=1,\dots,\shots, \, j=1,\dots,n_i\}$.
 Then there exists a unique interpolation polynomial $\IPop{\set{B}}\in\SkewPolyringZeroDer$ such that
 \begin{equation}
  \opev{\IPop{\set{B}}}{b_j^{(i)}}{a_i}=c_j^{(i)},\qquad \forall i=1,\dots,\shots, \, \forall j=1,\dots, n_i,
 \end{equation}
 and $\deg(\IPop{\set{B}})<\sum_{i=1}^{\shots}n_i$.
\end{lemma}

\begin{lemma}[Product Rule~\cite{martinez2019private}]\label{lem:opev_product_rule}
For two skew polynomials $f,g\in\SkewPolyringZeroDer$ and elements $a,b\in\Fqm$ the generalized operator evaluation of the product $f\cdot g$ at $b$ w.r.t $a$ is given by
\begin{equation}
  \opev{(f\cdot g)}{b}{a} = \opev{f}{\opev{g}{b}{a}}{a}.
\end{equation}
\end{lemma}
The set of all skew polynomials of the form
\begin{equation}\label{eq:def_mult_var_skew_poly}
    Q(x, y_1,\dots, y_\intOrder)=Q_0(x)+Q_1(x)y_1+\dots+Q_\intOrder(x)y_\intOrder,
\end{equation}
where $Q_j\in\SkewPolyringZeroDer$ for all $j=0,\dots,\shots$ is denoted by $\MultSkewPolyringZeroDer$.

\begin{definition}[$\vec{w}$-weighted Degree]
  Given a vector $\vec{w}\in\ZZ_{\geq0}^{\intOrder+1}$, the $\vec{w}$-weighted degree of a multivariate skew polynomial from $Q\in\MultSkewPolyringZeroDer$ is defined as 
  \begin{equation}
    \deg_{\vec{w}}(Q)=\max_j\{\deg(Q_j)+w_j\}.
  \end{equation}
\end{definition}

Given a vector $\vec{w}=(w_0,w_1,\dots,w_{\intOrder})$ the $\vec{w}$-weighted total order $\wOrder$ on monomials in $\MultSkewPolyringZeroDer$ is defined for all $j,j'\in\intervallincl{0}{\intOrder}$ and some $l,l'\geq0$ as 
\begin{align*}
  x^ly_j\wOrder x^{l'}y_{j'}
  \quad\Longleftrightarrow\quad
  \left\{
  \begin{array}{l}
     l+w_{j}< l' + w_{j'} \text{ or }  
     \\
     l+w_{j}= l' + w_{j'} \text{ and }j<j'.
  \end{array}
  \right.
\end{align*}
The $\vec{w}$-weighted monomial ordering is also called $\vec{w}$-weighted \emph{term over position} ordering~\cite{adams1994introduction} since first the $\vec{w}$-weighted degree of the term is considered and the position~$j$ is considered only if two monomials have the same $\vec{w}$-weighted degree.

We identify the \emph{leading position} of a multivariate polynomial $Q\in\MultSkewPolyringZeroDer$ as the as index $j$ of the maximum monomial $x^ly_j$ under $\wOrder$ and denote it by $\LP{Q}$.
For a set $\set{S}\subseteq\MultSkewPolyringZeroDer$ we denote the set of all leading positions of the elements in $\set{S}$ by $\LP{\set{S}}\defeq\{\LP{Q} : Q\in\set{S}\}$.

For an element $a\in\Fqm$ and a vector $\vec{b}=(b_1,b_2,\dots,b_n)\in\Fqm^n$ we define the vector
\begin{equation*}
  \opexp{a}{\b}{j} \defeq \left(\opexp{a}{b_1}{j}, \opexp{a}{b_2}{j}, \dots, \opexp{a}{b_n}{j}\right)
\end{equation*}
and the matrix
\begin{equation}
  \opVandermonde{d}{\vec{b}}{a}
  \defeq
  \begin{pmatrix}
    \b 
    \\ 
    \opexp{a}{\b}{1}
    \\ 
    \opexp{a}{\b}{2}
    \\ 
    \vdots 
    \\
    \opexp{a}{\b}{d-1}
  \end{pmatrix}
  =
  \begin{pmatrix}
   b_1 & b_2 & \dots & b_n
   \\
   \opexp{a}{b_1}{1} & \opexp{a}{b_2}{1} & \dots & \opexp{a}{b_n}{1} 
   \\
   \opexp{a}{b_1}{2} & \opexp{a}{b_2}{2} & \dots & \opexp{a}{b_n}{2} 
   \\
   \vdots & \vdots & \ddots & \vdots
   \\
   \opexp{a}{b_1}{d-1} & \opexp{a}{b_2}{d-1} & \dots & \opexp{a}{b_n}{d-1} 
  \end{pmatrix}
  \in\Fqm^{d\times n}.
\end{equation}

For a vector $\vec{x}=\left(\vec{x}^{(1)} \mid \vec{x}^{(2)} \mid \dots \mid \vec{x}^{(\shots)}\right)\in\Fqm^n$ with $\vec{x}^{(i)}\in\Fqm^{n_i}$ for all $i=1,\dots,\shots$, a length partition $\n=(n_1,n_2,\dots,n_\shots)\in\ZZ_{\geq0}^\shots$ such that $\sum_{i=1}^{\shots}n_i=n$ and a vector $\a = (a_1,a_2,\dots,a_\shots)\in\Fqm^\shots$ we define the vector\footnote{To simplify the notation we omit the length partition $\n$ from the vector operator $\opexp{\a}{\x}{i}$ since it will be always clear from the context (i.e. as the length partition of the vector $\x$).}
\begin{align*}
\opexp{\a}{\x}{i}
:= \begin{pmatrix}
& \opexp{a_1}{\x^{(1)}}{i} & \big\lvert & \opexp{a_2}{\x^{(2)}}{i}  & \big\lvert & \dots & \big\lvert &  \opexp{a_\ell}{\x^{(\ell)}}{i} &
\end{pmatrix} \in \Fqm^n.
\end{align*}

By the properties of the operator $\opexp{a}{\cdot}{i}$ (see~\eqref{eq:conc_prop_op}), we have that
\begin{equation}\label{eq:rho_i+j_property}
\opexp{\a}{\x}{i+j} = \opexp{\a}{\opexp{\a}{\x}{i}}{j}
\end{equation}
and
\begin{equation}\label{eq:rho_linearity_property}
\opexp{\a}{\myalpha\x}{i} = \aut^i(\myalpha) \opexp{\a}{\x}{i} \quad \forall \, \myalpha \in \Fqm. 
\end{equation}

For a matrix $\X\in\Fqm^{d\times n}$ with rows $\x_1,\x_2,\dots,\x_d$, an integer $j$ and a vector $\a = (a_1,a_2,\dots,a_\shots)\in\Fqm^\shots$ we define $\opexp{\a}{\cdot}{j}$ applied to $\X$ as
\begin{equation}
  \opexp{\a}{\X}{j}\defeq
  \begin{pmatrix}
   \opexp{\a}{\x_1}{j}
   \\ 
   \opexp{\a}{\x_2}{j}
   \\ 
   \vdots 
   \\ 
   \opexp{\a}{\x_d}{j}
  \end{pmatrix}.
\end{equation}

Lemma~\ref{lem:rank_row_op} relates the rank of a matrix $\X$ with the rank of $\opexp{\a}{\X}{j}$. 
The proof is proceeds similar as for the special case of the element-wise Frobenius automorphism (see e.g.~\cite{wachter2013decoding}) and is therefore omitted.

\begin{lemma}[Rank of Row-Operator Matrix]\label{lem:rank_row_op}
 Let $\n=(n_1,n_2,\dots,n_\shots)\in\ZZ_{\geq0}^\shots$ be a length partition such that $\sum_{i=1}^{\shots}n_i=n$ and let $\X = (\shot{\X}{1} \mid \shot{\X}{2} \mid \dots \mid \shot{\X}{\shots})\in\Fqm^{d \times n}$ with $\shot{\X}{i} \in \Fqm^{d \times n_i}$ for all $i=1,\dots,\shots$. 
 Let the vector $\a = (a_1,a_2,\dots,a_\shots)\in\Fqm^\shots$ contain representatives from different (nontrivial) conjugacy classes of $\Fqm$.
 Then for any integer $j$ we have that
 \begin{equation}
   \rk_{q^m}(\opexp{\a}{\X}{j})=\rk_{q^m}(\X).
 \end{equation}
\end{lemma}

\begin{definition}[{\genMoore} Matrix]
  For an integer $d \in \ZZ_{>0}$, a length partition $\n=(n_1,n_2,\allowbreak\dots,n_\shots)\in\ZZ_{\geq0}^\shots$ such that $\sum_{i=1}^{\shots}n_i=n$ and the vectors $\a=(a_1,a_2,\dots,a_\shots)\in\Fqm^\shots$ and $\vec{x}=\left(\vec{x}^{(1)} \mid \vec{x}^{(2)} \mid \dots \mid \vec{x}^{(\shots)}\right) \allowbreak \in\Fqm^n$ with $\vec{x}^{(i)}\in\Fqm^{n_i}$ for all $i=1,\dots,\shots$, the {\genMoore} matrix is defined as
  \begin{align*}
  \lambda_d(\x)_{\a}
  \defeq
  \begin{pmatrix}
  \opexp{\a}{\x}{0} \\
  \opexp{\a}{\x}{1} \\
  \vdots \\
  \opexp{\a}{\x}{d-1}
  \end{pmatrix}
  =
  \left(
  \begin{array}{c|c|c|c}
  \opVandermonde{d}{\x^{(1)}}{a_1} & \opVandermonde{d}{\x^{(2)}}{a_2} & \cdots & \opVandermonde{d}{\x^{(\shots)}}{a_\shots}
  \end{array}\right) 
  \in \Fqm^{d \times n}.
  \end{align*}
\end{definition}

We denote the {\genMoore} matrix with respect to the inverse automorphism by $\lambda_d^{\aut^{-1}}(\x)_{\a}$.

Similar as for ordinary polynomials and Vandermonde matrices, there is a relation between the generalized operator evaluation and product with a {\genMoore} matrix. 
In particular, for a skew polynomial $f(x)=\sum_{i=0}^{k-1} f_i x^i \in \SkewPolyringZeroDer_{<k}$ and vectors $\a=(a_1,a_2,\dots,a_\shots)\in\Fqm^\shots$ and $\vec{x}=\left(\vec{x}^{(1)} \mid \vec{x}^{(2)} \mid \dots \mid \vec{x}^{(\shots)}\right)\in\Fqm^n$ we have that
\begin{equation}\label{eq:relation_op_ev_moore}
  \opev{f}{\a}{\x} = (f_0, f_1, \dots, f_{k-1}) \cdot \lambda_k(\x)_{\a}.
\end{equation}

Proposition~\ref{prop:rankGenOpVandermonde} provides an important result on the rank of {\genMoore} matrices.

\begin{proposition}[Rank of {\genMoore} Matrix]\label{prop:rankGenOpVandermonde}
 For a vector $\vec{x}=\left(\vec{x}^{(1)} \mid \vec{x}^{(2)} \mid \dots \mid \vec{x}^{(\shots)}\right)\in\Fqm^n$ where $\vec{x}^{(i)}\in\Fqm^{n_i}$ for all $i=1,\dots,\shots$ and a vector $\a = (a_1,a_2,\dots,a_\shots)$, the $\Fqm$-rank of $\lambda_d(\x)_{\a}$ satisfies
 \begin{equation}
   \rk_{q^m}(\lambda_d(\x)_{\a})=\min\{d,n\}
 \end{equation}
 if and only if we have that $\rk_q(\x_i)=n_i$ for all $i=1,\dots,\shots$ and the elements $a_1,\dots,a_\shots$ belong to different conjugacy classes.
\end{proposition}

The statement in Proposition~\ref{prop:rankGenOpVandermonde} follows directly from~\cite[Theorem~4.5]{lam1988vandermonde} and~\cite[Theorem~2]{martinez2018skew}.

\begin{remark}
 To simplify the notation we omit the rank partition $\n$ in $\lambda_j(\cdot)_\a$ since it will be always clear from the context (i.e. the length partition of the considered vector).
\end{remark}

\subsection{Linearized Reed--Solomon Codes}

Linearized Reed--Solomon (\acs{LRS}) codes were first defined by Mart{\'\i}nez-Pe{\~n}as in~\cite{martinez2018skew} and also considered by Caruso in~\cite{caruso2019residues}.
\ac{LRS} codes are a class of sum-rank-metric evaluation codes that generalize \ac{RS} in the Hamming metric as well as Gabidulin codes~\cite{Gabidulin_TheoryOfCodes_1985} in the rank metric. 
\ac{LRS} receive their name from the generalized operator evaluation of skew polynomials that is used for the code construction, which is \emph{$\Fq$-linear} for a fixed evaluation parameter. 

There exists an isometry between the sum-rank metric and the skew metric~\cite{martinez2018skew} that relates \ac{LRS} to \ac{SRS} codes in the skew metric. Hence, \ac{LRS} codes can be seen as generalized~\ac{SRS} codes (see~\cite{liu2015construction}).

\begin{definition}[Linearized Reed--Solomon Code~\cite{martinez2018skew}]\label{def:LRS_codes}
 Let $\a=(a_1,a_2,\dots,a_\shots)\in\Fqm^\shots$ be a vector containing representatives from different conjugacy classes of $\Fqm$.
 Let $\nVec := (n_1,n_2,\dots,n_\shots)\in\ZZ_{\geq0}^\shots$ be a length partition and let $n=\sum_{i=1}^{\shots}n_i$.
 Let the vectors $\vecbeta^{(i)}=(\beta_1^{(i)},\beta_2^{(i)},\dots,\beta_{n_i}^{(i)})\in\Fqm^{n_i}$ contain $\Fq$-linearly independent elements from $\Fqm$ for all $i=1,\dots,\shots$ and define $\vecbeta=\left(\vecbeta^{(1)}\mid\vecbeta^{(2)}\mid\dots\mid\vecbeta^{(\shots)}\right)\in\Fqm^n$.
 A linearized Reed--Solomon (LRS) code $\linRS{\vecbeta,\a,\shots;\nVec,k}$ of length $n$ and dimension $k$ is defined as the set
 \begin{equation}
  \left\{\left(
  \begin{array}{c|c|c|c}
   \opev{f}{\vecbeta^{(1)}}{a_1} & \opev{f}{\vecbeta^{(2)}}{a_2} & \dots & \opev{f}{\vecbeta^{(\shots)}}{a_\shots}
  \end{array}
  \right): f \in \SkewPolyringZeroDer_{<k}\right\} \subseteq \Fqm^{n}.
 \end{equation}
\end{definition}

Each codeword $\c \in \linRS{\vecbeta,\a,\shots;\nVec,k}$ has the form
\begin{equation*}
  \c = \left(\shot{\c}{1} \mid \shot{\c}{2} \mid \dots \mid \shot{\c}{\shots}\right)
\end{equation*}
where $\shot{\c}{i}=\opev{f}{\shot{\vecbeta}{i}}{a_i}$ for all $i=1,\dots,\shots$ and some $f \in \SkewPolyringZeroDer_{<k}$. 
Note, that an \ac{LRS} code $\linRS{\vecbeta,\a,\shots;\nVec,k}$ can be described by a generator matrix $\lambda_k(\vecbeta)_{\a}$.

\ac{LRS} codes achieve the Singleton-like bound in the sum-rank metric (see~\cite[Proposition~34]{martinez2018skew}) with equality, i.e. the minimum sum-rank distance equals $n-k+1$, and thus are \ac{MSRD} codes.

There exist efficient decoding algorithms that allow for \ac{BMD} decoding errors of sum-rank weight up to $t \leq \lfloor \frac{n-k}{2}\rfloor$ (see~\cite{martinez2019reliable, caruso2019residues, boucher2019algorithm}).

In Section~\ref{sec:decodingILRS} we construct vertically $\intOrder$-interleaved \ac{LRS} codes by stacking $\intOrder$ codewords of an \ac{LRS} code and show that the error-correction capability can be increased to $t <\frac{\intOrder}{\intOrder+1}(n-k+1)$ by either admitting a list of candidate codewords or a small decoding failure probability.
Compared to \ac{BMD} decoding this is a gain of almost a factor of two, even for moderately large values of $\intOrder$.

\subsection{Cost Model}\label{ssec:cost_model}

We use the \emph{big-O} notation $\oh{\cdot}$ and the \emph{soft-O} notation $\softO{(\cdot)}$, which neglects logarithmic factors in the input parameter, to state the asymptotic cost of algorithms, which is expressed in terms of arithmetic operations (additions, multiplications and applications of a (specific) automorphism $\aut$) in the field $\Fqm$.
For the complexity of the corresponding arithmetic operations in the subfield $\Fq$, the reader is referred to the work by Couveignes and Lercier~\cite{couveignes2009elliptic}.

By $\omega$ we denote the matrix multiplication exponent, i.e.~the infimum of values $\omega_0 \in [2; 3]$ such that there is an algorithm for multiplying $n \times n$ matrices over $\Fqm$ in $O(n^{\omega_0})$ operations in $\Fqm$.
The best currently known bound is $\omega < 2.37286$ \cite{le_gall_powers_2014}.

We denote by $\OMul{n}$ the cost of multiplying two skew polynomials with coefficients in $\Fqm$ of degree $n$.
The currently best-known cost bound on $\OMul{n}$ is
\begin{equation}\label{eq:comp_OMul_Fqm}
\OMul{n} \in O\!\left( n^{\min\!\left\{\frac{\omega+1}{2},1.635\right\}} \right)
\end{equation}
operations in $\Fqm$ using the algorithm in \cite{puchinger2017fast} (see~\cite{caruso2017fast,caruso2017new} for algorithms with a cost bound over $\Fq$).
Overall, the algorithms in~\cite{caruso2017fast,caruso2017new,puchinger2017fast} are faster than classical multiplication (exponent is reduced from $2$ to $\leq 1.635$ in~\cite{puchinger2017fast}), which has quadratic complexity. 

\section{Decoding of Interleaved Linearized Reed--Solomon Codes}\label{sec:ILRS}

In this section, we consider \ac{ILRS} codes with respect to the sum-rank metric. 
In the Hamming metric, the gain from interleaving comes from the fact, that \emph{burst} errors, i.e. errors that act in a column-wise manner share the same location.
This principle can be extended to the rank metric~\cite{loidreau2006decoding,wachter2013decoding,sidorenko2010decoding} and as we will show in this section, to the sum-rank metric.
By fixing a basis of $\Fqms$ over $\Fqm$, each column of the interleaved matrix can be seen as an element from $\Fqms$.

After defining and analyzing \ac{ILRS} codes, we propose an \ac{LOlike} decoder for \ac{ILRS} codes that is capable of correcting (burst) sum-rank errors beyond the unique decoding radius at a cost of a (very) small decoding failure probability.
We derive an upper bound on the decoding failure probability that accounts for the distribution of the error matrices.
The \ac{LOlike} decoder allows a rigorous analysis of the decoding failure probability and gives insights about the decoding process.

We propose an interpolation-based decoding scheme for \ac{ILRS} codes that can correct sum-rank errors beyond the unique decoding radius, which can be used as a (not necessarily polynomial-time) list decoder or as a probabilistic unique decoder that either returns a unique solution or a decoding failure.
For the list decoder, an upper bound on the worst-case list size is proposed, and for the probabilistic unique decoder an upper bound on the decoding failure probability that is based on the \ac{LOlike} decoder, is derived. 

We generalize the isometry between the sum-rank metric and the skew metric (see~\cite{martinez2018skew}) to interleaved matrices and we define \acf{ISRS} codes, which are considered in~Section~\ref{subsec:ilrs_skew}.

Before defining \ac{ILRS} codes, we start by introducing the (burst) sum-rank error channel.

\subsection{Sum-Rank Error Channel}\label{subsec:sum_rank_channel}

As a channel model we consider the (burst) sum-rank error channel which is defined as follows.
The output $$\R=\left(\mat{R}^{(1)} \mid \mat{R}^{(2)} \mid \dots \mid \mat{R}^{(\shots)}\right)\in\Fqm^{\intOrder\times n}$$ is related to the input $\C=\left(\mat{C}^{(1)} \mid \mat{C}^{(2)} \mid \dots \mid \mat{C}^{(\shots)}\right)\in\Fqm^{\intOrder\times n}$ by
\begin{align}\label{eq:sum_rank_channel}
  \mat{R}=\mat{C}+\mat{E}.
\end{align}
The error matrix $\mat{E}$ has the form
\begin{equation}
 \mat{E}=\left(\mat{E}^{(1)} \mid \mat{E}^{(2)} \mid \dots \mid \mat{E}^{(\shots)}\right)\in\Fqm^{\intOrder\times n}
\end{equation}
where $\mat{E}^{(i)}\in\Fqm^{\intOrder\times n_i}$ has $\rk_q(\mat{E}^{(i)})=t_i$ for all $i=1,\dots,\shots$ and $\SumRankWeight(\mat{E})=\sum_{i=1}^{\shots}t_i=t$.
Alternatively, we may write the sum-rank channel in~\eqref{eq:sum_rank_channel} as
\begin{equation}\label{eq:def_sum_rank_channel}
 \begin{pmatrix}
  \r_1 \\
  \r_2 \\ 
  \vdots \\ 
  \r_\intOrder
 \end{pmatrix}
 =
 \begin{pmatrix}
  \c_1 \\
  \c_2 \\ 
  \vdots \\ 
  \c_\intOrder
 \end{pmatrix}
 +
 \begin{pmatrix}
  \e_1 \\
  \e_2 \\ 
  \vdots \\ 
  \e_\intOrder
 \end{pmatrix}
\end{equation}
where $\r_j=\big(\r_j^{(1)} \mid \r_j^{(2)} \mid \dots \mid \r_j^{(\shots)}\big)\in\Fqm^n$, $\c_j=\big(\c_j^{(1)} \mid \c_j^{(2)} \mid \dots \mid \c_j^{(\shots)}\big)\in\Fqm^n$ and $\e_j=\big(\e_j^{(1)} \mid \e_j^{(2)} \mid \dots \mid \e_j^{(\shots)}\big)\in\Fqm^n$ for all $j=1,\dots,\intOrder$.

\subsection{Interleaved Linearized Reed--Solomon Codes}\label{sec:decodingILRS}

Motivated by the results on interleaved Reed--Solomon codes~\cite{krachkovsky1997decoding,krachkovsky1998decoding} and interleaved Gabidulin codes~\cite{loidreau2006decoding}, we define \ac{ILRS} codes as follows.

\begin{definition}[Interleaved Linearized Reed--Solomon Code]\label{def:ILRS_codes}
 Let $\a=(a_1,a_2,\dots,\allowbreak a_\shots)\in\Fqm^\shots$ be a vector containing representatives from different conjugacy classes of $\Fqm$.
 Let $\nVec := (n_1,n_2,\dots,n_\shots)\in\ZZ_{\geq0}^\shots$ be a length partition and let $n=\sum_{i=1}^{\shots}n_i$.
 Let the vectors $\vecbeta^{(i)}=(\beta_1^{(i)},\beta_2^{(i)},\dots,\beta_{n_i}^{(i)})\in\Fqm^{n_i}$ contain $\Fq$-linearly independent elements from $\Fqm$ for all $i=1,\dots,\shots$ and define $\vecbeta=\left(\vecbeta^{(1)}\mid\vecbeta^{(2)}\mid\dots\mid\vecbeta^{(\shots)}\right)\in\Fqm^n$.
 A (homogeneous) $\intOrder$-interleaved linearized Reed--Solomon (ILRS) code $\intLinRS{\vecbeta,\a,\shots,\intOrder;\nVec,k}$ of length $n$ and dimension $k$ is defined as the set
 \begin{equation}
  \left\{\left(
  \begin{array}{c|c|c|c}
   \opev{f_1}{\vecbeta^{(1)}}{a_1} & \opev{f_1}{\vecbeta^{(2)}}{a_2} & \dots & \opev{f_1}{\vecbeta^{(\shots)}}{a_\shots}
   \\
   \opev{f_2}{\vecbeta^{(1)}}{a_1} & \opev{f_2}{\vecbeta^{(2)}}{a_2} & \dots & \opev{f_2}{\vecbeta^{(\shots)}}{a_\shots}
   \\
   \vdots & \vdots & \ddots & \vdots
   \\
   \opev{f_\intOrder}{\vecbeta^{(1)}}{a_1} & \opev{f_\intOrder}{\vecbeta^{(2)}}{a_2} & \dots & \opev{f_\intOrder}{\vecbeta^{(\shots)}}{a_\shots}
  \end{array}
  \right): \begin{array}{c}f_j\in\SkewPolyringZeroDer_{<k}, \\\forall j\in\intervallincl{1}{\intOrder}\end{array}\right\} \subseteq \Fqm^{\intOrder \times n}.
 \end{equation}
\end{definition}

The \ac{ILRS} codes from Definition~\ref{def:ILRS_codes} include \ac{LRS} codes (see Definition~\ref{def:LRS_codes}) as a special case for $\intOrder=1$.
Besides that, \ac{ILRS} codes generalize several code families in the Hamming, rank and sum-rank metric. 
For $\shots=1$ we obtain interleaved Gabidulin codes~\cite{loidreau2006decoding} with ordinary Gabidulin codes~\cite{Gabidulin_TheoryOfCodes_1985} for $\intOrder=1$.
Interleaved generalized Reed--Solomon codes are obtained by setting $\aut$ to be the identity and $\shots=n$ implying that $n_i=1$ for all $i=1,\dots,\shots$.
The generator matrix of an $\intOrder$-interleaved \ac{LRS} code $\intLinRS{\vecbeta,\a,\shots,\intOrder;\nVec,k}$ is the same as for a non-interleaved \ac{LRS} code $\linRS{\vecbeta,\a,\shots;\nVec,k}$ and is given by $\G=\lambda_k(\vecbeta)_{\a}$.
\begin{remark}
 Let $\pe$ be a primitive element of $\Fqm$.
 Then $\pe^0,\dots,\pe^{q-2}$ are representatives of all disjoint conjugacy classes (except the trivial one).
 Hence, we have that $\shots\leq(q-1)$ and that the length is bounded by $n\leq(q-1)m$.
 Further, we may choose the vector $\a$ in Definition~\ref{def:ILRS_codes} as $\a=(1,\pe,\dots,\pe^{q-2})$.
\end{remark}

Any codeword $\C\in\intLinRS{\vecbeta,\a,\shots,\intOrder;\nVec,k}$ has the form 
\begin{equation}
 \mat{C}\defeq
  (\vec{C}^{(1)} \mid \mat{C}^{(2)} \mid \dots \mid \mat{C}^{(\shots)}) 
\end{equation}
where 
\begin{equation}\label{eq:comp_mat_ILRS}
  \vec{C}^{(i)}\defeq
  \begin{pmatrix}
   \c_1^{(i)}
   \\ 
   \c_2^{(i)}
   \\ 
   \vdots 
   \\ 
   \c_\intOrder^{(i)}
  \end{pmatrix}
  =
  \begin{pmatrix}
   \opev{f_1}{\vecbeta^{(i)}}{a_i}
   \\ 
   \opev{f_2}{\vecbeta^{(i)}}{a_i}
   \\ 
   \vdots 
   \\ 
   \opev{f_\intOrder}{\vecbeta^{(i)}}{a_i}
  \end{pmatrix}
  \in\Fqm^{\intOrder\times n_i}
\end{equation}
for all $i=1,\dots,\shots$. 
To emphasize the interleaving we may write any codeword $\mat{C}\in\intLinRS{\vecbeta,\a,\shots,\intOrder;\nVec,k}$ as
\begin{equation}
\C=
  \begin{pmatrix}
   \vec{c}_1
   \\
   \vec{c}_2
   \\
   \vdots 
   \\
   \vec{c}_\intOrder
  \end{pmatrix}
\end{equation}
where each row $\c_j=(\c_j^{(1)} \mid \c_j^{(2)} \mid \dots \mid \c_j^{(\intOrder)})$ is a codeword of the component code $\linRS{\vecbeta,\a,\shots;\nVec,k}$.
The structure of the codeword matrices of an \ac{ILRS} code is illustrated in Figure~\ref{fig:ilrs_codeword}.
\begin{figure}[ht!]
  \centering
  \pgfkeys{tikz/mymatrixenv/.style={decoration={brace},every left delimiter/.style={xshift=8pt},every right delimiter/.style={xshift=-8pt}}}

\pgfkeys{tikz/mymatrix/.style={matrix of math nodes,nodes in empty cells,left delimiter={(},right delimiter={)},inner sep=1pt,outer sep=1.5pt,column sep=0pt,row sep=8pt,nodes={minimum width=60pt,minimum height=15pt,anchor=center,inner sep=0pt,outer sep=0pt}}}

\pgfkeys{tikz/mymatrixbrace/.style={decorate,thick}}
\tikzset{
  style cyan/.style={
    set fill color=cyan!90!blue!60, draw opacity=0.6,
    set border color=blue!70!cyan!30,fill opacity=0.3,
  },
  style blue/.style={
    set fill color=blue!90!blue!60, draw opacity=0.6,
    set border color=blue!70!blue!30,fill opacity=0.2,
  },
  kwad/.style={
    above left offset={-0.83,0.4},
    below right offset={0.83,-0.3},
    #1
  },
  pion/.style={
    above left offset={-0.72,0.45},
    below right offset={0.72,-0.3},
    #1
  },
  set fill color/.code={\pgfkeysalso{fill=#1}},
  set border color/.style={draw=#1}
}

\[
  \C=
    \begin{tikzpicture}[baseline={-0.5ex},mymatrixenv]
    \matrix [mymatrix,inner sep=5pt] (m) {
      \c_1^{(1)}  & \tikzmarkin[pion=style blue]{mat}  \c_1^{(2)} &  \dots  & \c_1^{(\shots)} 
      \\[-5pt]
      \tikzmarkin[kwad=style cyan]{row}\c_2^{(1)}  & \c_2^{(2)}  &  \dots  & \c_2^{(\shots)}\tikzmarkend{row} 
      \\[-5pt]
      \vdots  & \vdots & \ddots & \vdots   
      \\[-5pt]
      \c_\intOrder^{(1)}  &  \c_\intOrder^{(2)} \tikzmarkend{mat} &  \dots  & \c_\intOrder^{(\shots)}    
      \\    
    };

    \draw[]  (m-1-2.north west) -- (m-4-2.south west);
    \draw[]  (m-1-3.north west) -- (m-4-3.south west);
    \draw[]  (m-1-4.north west) -- (m-4-4.south west);

  \node[above of=m-1-3, xshift=50pt, yshift=0pt] (vec_node) {$\c_{2}$};
    \node[below of=m-4-2, xshift=50pt, yshift=5pt] (mat_node) {$\C^{(2)}$};
    \draw[color=lightgray, thick]  (m-2-3) to[bend left] (vec_node.west);
    \draw[color=lightgray, thick]  (m-4-2) to[bend right] (mat_node.west);

    \draw[<->, thick, color=gray]  ([yshift=-10pt]m-4-1.south west) -- ([yshift=-10pt,xshift=-0.1pt]m-4-1.south east) node[midway,yshift=-8pt] {$n_1$};
    \draw[<->, thick, color=gray]  ([yshift=10pt]m-1-1.north west) -- ([yshift=10pt,xshift=-0.1pt]m-1-4.north east) node[midway,yshift=8pt] {$n$};
    \draw[<->, thick, color=gray]  ([xshift=10pt,yshift=4pt]m-1-4.north east) -- ([xshift=10pt,yshift=-5pt]m-4-4.south east) node[midway,xshift=8pt] {$\intOrder$};

    \end{tikzpicture}
\]
  \caption{Illustration of the structure a codeword matrix from an \ac{ILRS} code.
  }
  \label{fig:ilrs_codeword}
\end{figure}

To indicate the relation between codewords and the corresponding message polynomials we define $\f\defeq(f_1,f_2,\dots,f_\intOrder)\in\SkewPolyringZeroDer^{\intOrder}$ and write 
\begin{equation*}
  \C(\f)=(\C^{(1)}(\f) \mid \C^{(2)}(\f) \mid \dots \mid \C^{(\intOrder)}(\f)).
\end{equation*}

Proposition~\ref{prop:ILRS_msrd} shows that \ac{ILRS} codes fulfill the Singleton-like bound in the sum-rank metric (see~\cite[Proposition~34]{martinez2018skew}) with equality and thus are \ac{MSRD} codes.

\begin{proposition}[Minimum Distance]\label{prop:ILRS_msrd}
 The minimum sum-rank distance of an \ac{ILRS} code $\intLinRS{\vecbeta,\a,\shots,\intOrder;\nVec,k}$ satisfies
 \begin{equation}
  \SumRankDist\left(\intLinRS{\vecbeta,\a,\shots,\intOrder;\nVec,k}\right)=n-k+1.
 \end{equation}
\end{proposition}

\begin{proof}
 The statement follows directly by considering a codeword containing only one nonzero row corresponding to a codeword having minimum sum-rank weight among all codewords of $\linRS{\vecbeta,\a,\shots;\nVec,k}$. By~\cite[Theorem~4]{martinez2018skew} and the $\Fqm$-linearity, the minimum distance is thus $n-k+1$.
\end{proof}

\subsection{Loidreau--Overbeck-like Decoder for ILRS Codes}\label{subsec:LO_dec_ILRS}
 
Based on the decoder by Loidreau and Overbeck for interleaved Gabidulin codes from~\cite{loidreau2006decoding,overbeck2007public,overbeck2008structural}, we now derive a decoding scheme for \ac{ILRS} codes.
This \ac{LOlike} decoding scheme allows to decode errors beyond the \ac{BMD} radius by allowing a small decoding failure probability.
The main result is summarized in Theorem~\ref{thm:LO_decoder_ILRS} and proved in the remainder of this section.

\begin{theorem}[\ac{LOlike} Decoder for~\ac{ILRS} Codes]\label{thm:LO_decoder_ILRS}
Let $\R=\C(\f)+\E\in\Fqm^{\intOrder\times n}$ where $\C(\f)\in\intLinRS{\vecbeta,\a,\shots,\intOrder;\nVec,k}$ and $\E \in \Fqm^{\intOrder \times n}$ is chosen uniformly at random from the set
\begin{align*}
  \left\{\E \in \Fqm^{\intOrder \times n} \, : \, \SumRankWeight(\E) = t\right\},
\end{align*}
where
\begin{align*}
t \leq \tmax := \tfrac{s}{s+1}(n-k).
\end{align*}
Then, Algorithm~\ref{alg:LO_ILRS} with input $\R$ returns the correct message polynomial vector $\f$ with success probability at least
\begin{align}\label{eq:succ_prob_LO}
\Pr(\text{success}) \geq 1 - \gammaq^{\shots+1} q^{-m((s+1)(\tmax-t)+1)}.
\end{align}
Furthermore, the algorithm has complexity $O(\intOrder n^\omega)$ operations in $\Fqm$ plus $O(m n^{\omega-1})$ operations in $\Fq$.
\end{theorem}
Although the \ac{LOlike} decoder has a higher computational complexity than the interpolation-based decoder, which we derive in Section~\ref{subsec:decodingILRS}, it plays a central role in bounding the decoding failure probability by relating the conditions for successful decoding of the two decoding schemes.
A similar approach was used for bounding the decoding failure probability of the interpolation-based decoding scheme for interleaved Gabidulin codes in~\cite{wachter2013decoding}.

Compared to the original Loidreau--Overbeck decoder for interleaved Gabidulin codes the main challenge for deriving an \ac{LOlike} decoder for \ac{ILRS} codes is to obtain the transformation matrices that allow for transforming the received word, such that the rank error and the non-corrupted part are aligned in particular columns, in a block-wise manner.

Suppose we transmit a codeword $\C\in\intLinRS{\vecbeta,\a,\shots,\intOrder;\nVec,k}$ over a sum-rank channel~\eqref{eq:def_sum_rank_channel} and receive
\begin{align*}
\R = \begin{pmatrix}
\r_1 \\
\r_2 \\
\vdots \\
\r_{\intOrder} \\
\end{pmatrix}
= \begin{pmatrix}
\c_1 \\
\c_2 \\
\vdots \\
\c_\intOrder \\
\end{pmatrix}
+\begin{pmatrix}
\e_1 \\
\e_2 \\
\vdots \\
\e_\intOrder \\
\end{pmatrix}
= \C + \E  \in \Fqm^{\intOrder \times n}
\end{align*}
where the error matrix $\E$ has sum-rank weight $t$ with $\Fq$-rank partition $\t=(t_1,t_2,\dots,\allowbreak t_\shots)$.
Now consider the \ac{LOlike} decoding matrix
\begin{equation}\label{eq:LO_matrix_ILRS}
\LODecMat := 
\begin{pmatrix}
\lambda_{n-t-1}(\vecbeta)_{\a} \\
\lambda_{n-t-k}\!\left(\r_1\right)_{\a} \\
\vdots \\
\lambda_{n-t-k}\!\left(\r_\intOrder\right)_{\a} \\
\end{pmatrix}
\in \Fqm^{((\intOrder+1)(n-t)-\intOrder k -1) \times n}.
\end{equation}
and with slightly abusing the notation above the matrix
\begin{align*}
\lambda_{n-t-k}(\E)_{\a} := 
\begin{pmatrix}
\lambda_{n-t-k}\!\left(\e_1\right)_{\a} \\
\vdots \\
\lambda_{n-t-k}\!\left(\e_\intOrder\right)_{\a} \\
\end{pmatrix}
\in \Fqm^{\intOrder(n-t-k) \times n}.
\end{align*}

\begin{lemma}[Properties of Decoding Matrix]\label{lem:properties_LODecMat_ILRS}
  Consider the transmission of a codeword $\C$ from the \ac{ILRS} code $\intLinRS{\vecbeta,\a,\shots,\intOrder;\nVec,k}$ over a sum-rank channel~\eqref{eq:def_sum_rank_channel} where the error matrix $\E$ has sum-rank weight $t$ with $\Fq$-rank partition $\t=(t_1,t_2,\dots,t_\shots)$.
  Suppose that $\lambda_{n-t-k}(\E)_{\a}$ has $\Fqm$-rank $t$.
  Then the decoding matrix $\LODecMat$ in~\eqref{eq:LO_matrix_ILRS} has the following properties:
  \begin{enumerate}
  \item \label{itm:transformed_LO_matrix} The $\Fqm$-linear row space of $\LODecMat$ satisfies
  \begin{align*}
  \RowspaceFqm{\LODecMat} = \RowspaceFqm{\begin{pmatrix}
  \lambda_{n-t-1}(\vecbeta)_{\a} \\
  \lambda_{n-t-k}(\E)_{\a}
  \end{pmatrix}}.
  \end{align*}
  \item \label{itm:LO_lemma_non-zero_colums_of_EU} There are invertible matrices $\W^{(i)} \in \Fq^{n_i \times n_i}$ such that $$\lambda_{n-t-k}(\E)_{\a} \cdot \diag\!\left(\W^{(1)},\allowbreak \dots, \W^{(\ell)}\right)$$ has exactly $t$ non-zero columns. Moreover, these columns are $\Fqm$-linearly independent. 
  \item \label{itm:LO_lemma_rank_L} We have $\rk_{q^m}(\LODecMat) = n-1$.
\end{enumerate}
\end{lemma}

The proof of Lemma~\ref{lem:properties_LODecMat_ILRS} can be found in Appendix~\ref{app:proofs_ILRS}.
Note, that by the rank-nullity theorem statement~\ref{itm:transformed_LO_matrix}) in Lemma~\ref{lem:properties_LODecMat_ILRS} also implies that also the $\Fqm$-linear right kernels of the two matrices are the same. 
We now derive properties of elements in the right $\Fqm$-kernel of the decoding matrix, that lay the foundations for an \ac{LOlike} decoder for \ac{ILRS} codes.

\begin{lemma}[Properties of Right Kernel]\label{lem:LO_correctness_lemma}
  Suppose that $\lambda_{n-t-k}(\E)_{\a}$ has $\Fqm$-rank $t$.
  Let $\h = (\h^{(1)} \mid \h^{(2)} \mid \dots \mid \h^{(\ell)}) \in \Fqm^n$ be a non-zero vector in the right kernel of the decoding matrix $\LODecMat$ in~\eqref{eq:LO_matrix_ILRS}.
  Then:
  \begin{enumerate}
    \item \label{itm:LO_lemma_rank_of_kernel_element} We have $\rk_{q}(\h^{(i)})=n_i-t_i$ for all $i=1,\dots,\ell$, i.e., $\h$ has sum-rank weight $\SumRankWeight(\h) = n-t$. 
  \item \label{itm:LO_lemma_Ti_matrices} There are invertible matrices $\T^{(i)} \in \Fq^{n_i \times n_i}$, for all $i=1,\dots,\ell$, such that the first (leftmost) $t_i$ positions of $\h^{(i)} \T^{(i)}$ are zero.
  \item \label{itm:LO_lemma_EDi_zero} For the matrices $\T^{(i)} \in \Fq^{n_i\times n_i}$ above, define $\D^{(i)} := \left({\T^{(i)}}^{-1}\right)^\top$. Then, the rightmost $n_i-t_i$ columns of $\E^{(i)} \D^{(i)}$ are zero.
  \item \label{itm:LO_lemma_retrieve_error} Write $\vecbetatilde^{(i)} := \vecbeta^{(i)} \D^{(i)}$ and denote by $\tilde{r}^{(i)}_{j,\mu}$ the $\mu$-th entry of $\r^{(i)}_j\D^{(i)}$. Then, independently for any $j=1,\dots,\intOrder$, the $j$-th message polynomial $f_j$ can be uniquely reconstructed from the received word as the interpolation polynomial
  \begin{align*}
  f_j = \IPop{\set{B}_j}
  \end{align*}
  where $\set{B}_j \defeq \{(\tilde{\beta}^{(i)}_\mu,\tilde{r}^{(i)}_{j,\mu}, a_i):i=1,\dots,\shots, \mu=t_i+1,\dots,n_i\}$.
  \end{enumerate}
\end{lemma} 

The proof of Lemma~\ref{lem:LO_correctness_lemma} can be found in Appendix~\ref{app:proofs_ILRS}.

\begin{figure}[ht!]
\centering
\includegraphics[width=\textwidth]{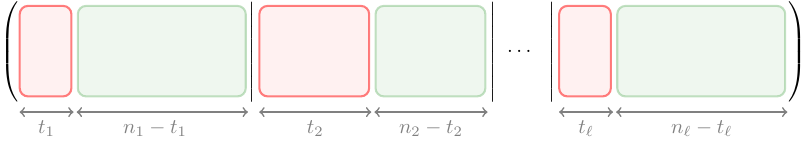}
\caption{Qualitative illustration of the transformed received matrix $\Rtilde$. The red parts correspond to the corrupted columns whereas the green parts correspond to the non-corrupted columns.}
\label{fig:R_hat}
\end{figure}

The structure of the transformed received matrices $\tilde{\R}^{(i)}=\R^{(i)}\D^{(i)}$ is illustrated in Figure~\ref{fig:R_hat_i}.
\begin{figure}[ht!]
\centering
\pgfkeys{tikz/mymatrixenv/.style={decoration={brace},every left delimiter/.style={xshift=8pt},every right delimiter/.style={xshift=-8pt}}}

\pgfkeys{tikz/mymatrix/.style={matrix of math nodes,nodes in empty cells,left delimiter={(},right delimiter={)},inner sep=1pt,outer sep=1.5pt,column sep=8pt,row sep=8pt,nodes={minimum width=20pt,minimum height=10pt,anchor=center,inner sep=0pt,outer sep=0pt}}}

\pgfkeys{tikz/mymatrixbrace/.style={decorate,thick}}

\tikzset{
  style green/.style={
    set fill color=green!50!black!60,draw opacity=0.4,
    set border color=green!50!black!60,fill opacity=0.1,
  }, 
  style red/.style={
    set fill color=red!90!pink!20, draw opacity=0.5,
    set border color=red, fill opacity=0.3,    
  },
  kwad/.style={
    above left offset={-0.1,0.5},
    below right offset={0.15,-0.4},
    #1
  },
  pion/.style={
    above left offset={-0.1,0.5},
    below right offset={0.05,-0.4},
    #1
  },
  set fill color/.code={\pgfkeysalso{fill=#1}},
  set border color/.style={draw=#1}
}
\[
  \Rtilde^{(i)}=\R^{(i)}\D^{(i)}=
    \begin{tikzpicture}[baseline={-0.5ex},mymatrixenv]
    \matrix [mymatrix,inner sep=5pt] (mat) {
      \tikzmarkin[kwad=style red]{err} \tilde{r}_{1,1}^{(i)}  &  \dots &  \tilde{r}_{1,t_i}^{(i)}  &  \tikzmarkin[pion=style green]{code}\tilde{c}_{1,t_i+1}^{(i)} & \dots & \tilde{c}_{1,n_i}^{(i)} 
      \\
      \tilde{r}_{2,1}^{(i)}  &  \dots &  \tilde{r}_{2,t_i}^{(i)}  & \tilde{c}_{2,t_i+1}^{(i)} & \dots & \tilde{c}_{2,n_i}^{(i)}
      \\ 
      \vdots  & \ddots & \vdots & \vdots  & \ddots & \vdots 
      \\
      \tilde{r}_{\intOrder,1}^{(i)}  &  \dots &  \tilde{r}_{\intOrder,t_i}^{(i)}\tikzmarkend{err}  & \tilde{c}_{\intOrder,t_i+1}^{(i)} & \dots & \tilde{c}_{\intOrder,n_i}^{(i)}\tikzmarkend{code}
      \\    
    };

    \draw[<->, thick, color=gray]  ([yshift=-15pt]mat-4-1.south west) -- ([yshift=-15pt,xshift=3.8pt]mat-4-3.south east) node[midway,yshift=-8pt] {$t_i$};
    \draw[<->, thick, color=gray]  ([yshift=-15pt,xshift=-3.5pt]mat-4-4.south west) -- ([yshift=-15pt]mat-4-6.south east) node[midway,yshift=-8pt] {$n_i-t_i$};
    \end{tikzpicture}
\]
\caption{Illustration of the transformed received matrices $\tilde{\R}^{(i)}=\R^{(i)}\D^{(i)}$. The red part is corrupted by an error of rank $t_i$ whereas the green part corresponds to the last (rightmost) $n_i-t_i$ columns of the transformed codeword matrix $\tilde{\C}(\f)^{(i)}=\C^{(i)}(\f)\D^{(i)}$ that is obtained by evaluating $\f$ at the transformed code locators $\vecbeta^{(i)}\D^{(i)}$.}
\label{fig:R_hat_i}
\end{figure}
A qualitative illustration of the transformed received matrix $\Rtilde=(\Rtilde^{(1)}\mid\Rtilde^{(2)}\mid\dots\mid\Rtilde^{(\shots)})$ is illustrated in Figure~\ref{fig:R_hat}.

Lemma~\ref{lem:LO_correctness_lemma} provides an efficient algorithm to retrieve the message polynomial vector from a received word under the condition that the matrix $\lambda_{n-t-k}(\E)_{\a}$ has $\Fqm$-rank $t$. The method is outlined in Algorithm~\ref{alg:LO_ILRS}. 

The complete procedure for the \ac{LOlike} decoder for \ac{ILRS} codes is given in Algorithm~\ref{alg:LO_ILRS}.

\begin{algorithm}[ht!]
  \caption{\algoname{Loidreau--Overbeck Decoder for \ac{ILRS} Codes}}\label{alg:LO_ILRS}
  \SetKwInOut{Input}{Input}\SetKwInOut{Output}{Output}
  
  \Input{A received matrix $\R=\C(\f)+\E\in\Fqm^{\intOrder\times n}$ where $\C(\f)\in\intLinRS{\vecbeta,\a,\shots;\nVec,k}$ and $\SumRankWeight(\E)=t$.}
  
  \Output{Message polynomial vector $\f=(f_1,\dots,f_\intOrder)\in\SkewPolyringZeroDer_{<k}^{\intOrder}$ or ``decoding failure''}
  \BlankLine
  
  Set up the matrix $\mat{L}$ as in~\eqref{eq:LO_matrix_ILRS} 
  \\ 
  Compute right kernel $\myspace{H}=\rker(\LODecMat)$
  \\
  \If{$\dim(\myspace{H})>1$ \label{line:check_if_kernel_dimension>1}}{
    \Return{``decoding failure''}
  }
  \Else{
    Compute an element $\h=\left(\vec{h}^{(1)} \mid \dots \mid \vec{h}^{(\shots)}\right)\in\myspace{H}\setminus\{\vec{0}\}$ \label{line:LO_h}
    \\ 
    \For{$i=1,\dots,\shots$}{
      Compute $n_i-t_i\gets\rk_{q}\left(\vec{h}^{(i)}\right)$ \label{line:LO_rk_hi} \\
      Compute full-rank matrix $\T^{(i)}\in\Fq^{n_i\times n_i}$ such that the first $t_i$ entries of $\vec{h}^{(i)}\T^{(i)}$ are zero \label{line:LO_Ti} \\
      $\vecbetatilde^{(i)} \gets \vecbeta^{(i)} \left({\T^{(i)}}^{-1}\right)^\top$ \label{line:LO_beta_i_computation}
    }
    \For{$j=1,\dots,\intOrder$}{
      $\tilde{\r}_{j} = \begin{pmatrix}
      \tilde{\r}_{j}^{(1)} & \cdots & \tilde{\r}_{j}^{(\ell)}
      \end{pmatrix}
      \gets \r_j \diag\!\left(\left({\T^{(1)}}^{-1}\right)^\top, \dots, \left({\T^{(\shots)}}^{-1}\right)^\top\right)$ \label{line:LO_R_trans} \hfill \tcp{$\r_j$ is the $j$-th row of $\R$}
      $f_j \gets \IPop{\set{B}_j}$ where $\set{B}_j \defeq \{(\tilde{\beta}^{(i)}_\mu,\tilde{r}^{(i)}_{j,\mu}, a_i):i=1,\dots,\shots, \mu=t_i+1,\dots,n_i\}$ \label{line:LO_IP}
    }
    \Return{$\f=(f_j,\dots,f_\intOrder)$}
  }
\end{algorithm}

\begin{remark}
  Since an operation in $\Fqm$ costs at least $\Omega(m)$ operations in $\Fq$ (cf.~Section~\ref{ssec:cost_model}), the cost ``$O(\intOrder n^\omega)$ operations in $\Fqm$'' dominates the complexity of Algorithm~\ref{alg:LO_ILRS}.
\end{remark}

Lemma~\ref{lem:LO_technical_lemma_for_failure_prob} provides a condition on the $\Fqm$-rank of stacked $\aut$-generalized Moore matrices that we will later on use to derive the probability of success of Algorithm~\ref{alg:LO_ILRS}.

\begin{lemma}\label{lem:LO_technical_lemma_for_failure_prob}
  Let $\M \in \Fqm^{\intOrder \times t}$ with $\SumRankWeight(\M) = t$, where $\t = (t_1,\dots,t_\ell)$ with $t_i \geq 0$ and $\sum_{i=1}^{\ell} t_i = t$. Then, we have
  \begin{equation}
  \rk_{q^m}\left(\lambda_{n-t-k}(\M)_{\a}\right)<t \label{eq:LO_lemma_Fqm_rank<t}
  \end{equation}
  if and only if
  \begin{equation}
  \exists \, \hvec \in \Fqm^t \, : \quad  \SumRankWeight(\hvec)>n-t-k \quad \text{and} \quad \lambda_{n-t-k}^{\aut^{-1}}(\myalpha \hvec)_{\aut^{-1}(\a)} \M^\top = \0 \quad  \forall \, \myalpha \in \Fqm^\ast. \label{eq:LO_lemma_existence_h}
  \end{equation} 
\end{lemma}

The proof of Lemma~\ref{lem:LO_technical_lemma_for_failure_prob} proceeds similarly as the proof of~\cite[Lemma~3.14]{overbeck2007public} and can be found in Appendix~\ref{app:proofs_ILRS}.

Lemma~\ref{lem:LO_probability_full_Fqm_rank} provides an upper bound on the probability that the matrix $\lambda_{n-t-k}(\E)_{\a}$ is rank deficient if the error matrix $\E$ is chosen uniformly at random from the set of all matrices from $\Fqm^{\intOrder \times n}$ having sum-rank weight $t$. 

\begin{lemma}\label{lem:LO_probability_full_Fqm_rank}
Let $t\leq \tmax := \tfrac{s}{s+1}(n-k)$.
Let $\n=(n_1,n_2,\dots,n_\shots)\in\ZZ_{\geq0}^\shots$ be a length partition such that $\sum_{i=1}^{\shots}n_i=n$ and let $\E \in \Fqm^{\intOrder \times n}$ be chosen uniformly at random from the set
\begin{align*}
\left\{\E \in \Fqm^{\intOrder \times n} \, : \, \SumRankWeight(\E) = t\right\}.
\end{align*}
Then, we have
\begin{align*}
\Pr\!\left(\rk_{q^m}(\lambda_{n-t-k}(\E)_{\a})<t\right) \leq \gammaq^{\ell+1} q^{-m\left((s+1)(\tmax-t)+1\right)}.
\end{align*}
\end{lemma}

The proof of Lemma~\ref{lem:LO_probability_full_Fqm_rank} can be found in Appendix~\ref{app:proofs_ILRS}.

Finally, we are now equipped with all results that are needed to proof Theorem~\ref{thm:LO_decoder_ILRS} stated at the beginning or this section.

\begin{proof}
Due to Lemma~\ref{lem:LO_correctness_lemma}, the algorithm returns the correct message polynomial vector if the $\Fqm$-rank of $\lambda_{n-t-k}(\E)_\a$ is at least $t$. Hence, the success probability is lower bounded by the probability that $\rk_{q^m}(\lambda_{n-t-k}(\E)_{\a})\allowbreak=t$, which is given in Lemma~\ref{lem:LO_probability_full_Fqm_rank}.

The lines of the algorithm have the following complexities:
\begin{itemize}
  \item Lines~\ref{line:check_if_kernel_dimension>1} and~\ref{line:LO_h}: 
  This can be done by solving the linear system of equations $\LODecMat \h^\top = \0$. Since $\LODecMat \in \Fqm^{((\intOrder+1)(n-t)-\intOrder k -1) \times n}$, it costs $\oh{\intOrder n^\omega}$ operations in $\Fqm$.
  \item Line~\ref{line:LO_rk_hi} can be implemented by transforming the matrix representation of $\h^{(i)}$, which is an $m\times n_i$ matrix over $\Fq$, into column echelon form. For each $i$, this costs $\oh{m n_i^{\omega-1}}$ operations in $\Fq$. In total, all $\shots$ calls of this line cost $\oh{\shots m \sum_i n_i^{\omega-1}} \subseteq \oh{m n^{\omega-1}}$ operations in $\Fq$.
  \item Line~\ref{line:LO_Ti} can be implemented by transforming the matrix representation of $\h^{(i)}$ into column echelon form, which was already accomplished in Line~\ref{line:LO_rk_hi}.
  \item Line~\ref{line:LO_beta_i_computation} is a matrix-matrix multiplication over $\Fq$, which costs $\oh{m n_i^{\omega-1}}$ operations in $\Fq$ for each $i$. All $\shots$ iterations of this line cost together $\oh{m n^{\omega-1}}$ operations in $\Fq$.
  \item Line~\ref{line:LO_R_trans} requires $\oh{\sum_in_i^2}\subseteq\oh{n^2}$ multiplications over $\Fqm$ and thus $\oh{\intOrder n^2}$ operations in $\Fqm$ in total.
  \item Line~\ref{line:LO_IP} computes $\intOrder$ interpolation polynomials of degree less than $k\leq n$ point tuples. This costs in total $\softoh{\intOrder \OMul{n}}$ operations in $\Fqm$.
\end{itemize}
This proves the complexity statement.
\end{proof}

Note, that the decoding radius $\tmax$ defined above does not necessarily need to be an integer.
The lower bound on the probability of successful decoding in~\eqref{eq:succ_prob_LO} corresponds to an upper bound on the decoding failure probability, i.e. we have that
\begin{equation}
  \Pr(\text{failure}) \leq \gammaq^{\shots+1} q^{-m((s+1)(\tmax-t)+1)}.
\end{equation}

An execution of the \ac{LOlike} decoder is illustrated in Example~\ref{ex:LO_ILRS}.
\begin{example}[\ac{LOlike} Decoder]\label{ex:LO_ILRS}
  Consider the finite field $\F_{3^3}$ with primitive element $\pe$ defined by the primitive polynomial $x^3 + 2x + 1$ and let $\aut$ be the Frobenius automorphism.
  Consider the interleaved \ac{LRS} code $\intLinRS{\vecbeta,\a,\shots,\intOrder;\nVec,k}$ over $\F_{3^3}$ with code locators $\vecbeta = ((1,\pe,\pe^2) \mid (1,\pe,\pe^2))$, evaluation parameters $\a=(1,\pe)$, $\intOrder=2$, length partition $\n = (3,3)$ and dimension $k=3$.
  Suppose we transmit the codeword
  \begin{equation}\label{eq:ilrs_ex_codeword}
    \C(\f) =
    \left(
    \begin{array}{ccc|ccc}
      2\pe^2 & 2\pe+1 & 2\pe^2+\pe & 2\pe^2 & 2\pe+1 & 2\pe^2+\pe 
      \\
      \pe+1 & 2\pe^2+1 & \pe^2+1 & \pe+1 & \pe^2+\pe+2 & 0
    \end{array}
    \right) 
  \end{equation}
  from $\intLinRS{\vecbeta,\a,\shots,\intOrder;\nVec,k}$ that corresponds to the message polynomials $\f=(f_1, f_2) = (2\pe^2, x^2 + (2\pe^2+\pe)x+\pe^2)$ over a sum-rank channel that adds an error 
  \begin{equation*}
    \E =
    \left(
    \begin{array}{ccc|ccc}
       0 & 2\pe^2+1 & 2\pe^2+1 & 0 & 0 & 2
       \\
       0 & \pe^2+\pe+1 & \pe^2+\pe+1 & 0 & 0 & 2\pe^2+2
    \end{array}
    \right) 
  \end{equation*}
  of sum-rank weight $t=2$ and we receive
  \begin{equation}\label{eq:ilrs_ex_recword}
    \R =
    \left(
    \begin{array}{ccc|ccc}
      2\pe^2 & 2\pe^2+2\pe+2 & \pe^2+\pe+1 & 2\pe^2 & 2\pe+1 & 2\pe^2+\pe+2
      \\ 
      \pe+1 & \pe+2 & 2\pe^2+\pe+2 & \pe+1 & \pe^2+\pe+2 & 2\pe^2+2
    \end{array}
    \right).
  \end{equation}
  Note, that a \ac{BMD} decoder could only correct errors up to sum-rank weight $\lfloor\frac{n-k}{2}\rfloor=1$.
  According to~\eqref{eq:LO_matrix_ILRS} the \ac{LOlike} decoding matrix is
  \begin{equation*}
    \LODecMat =
    \left(
    \begin{array}{ccc|ccc}
      1 & \pe & \pe^2 & 1 & \pe & \pe^2
      \\ 
      1 & \pe+2 & \pe^2+\pe+1 & \pe & \pe^2+2\pe & \pe^2+2\pe+2
      \\ 
      1 & \pe+1 & \pe^2+2\pe+1 & \pe^2+2\pe & 2 & 2\pe+2
      \\ 
      2\pe^2 & 2\pe^2+2\pe+2 & \pe^2+\pe+1 & 2\pe^2 & 2\pe+1 & 2\pe^2+\pe+2
      \\ 
      \pe+1 & \pe+2 & 2\pe^2+\pe+2 & \pe+1 & \pe^2+\pe+2 & 2\pe^2+2
    \end{array}
    \right).
  \end{equation*}
  The decoding matrix $\LODecMat$ has $\Fqm$-rank $n-1=5$ implying the right $\F_{3^3}$-kernel of $\LODecMat$ has dimension one.
  We pick $$\h = (\h^{(1)} \mid \h^{(2)}) = ( (\pe, 2\pe^2+2\pe+1, \pe^2+\pe+2) \mid (\pe+1, \pe^2+\pe, 0)) \in \rker(\LODecMat)$$ as non-zero element from the right kernel of $\LODecMat$ and recover the rank partition of the error as $$\t = (n_1 - \rk_3(\h^{(1)}) \mid n_2 - \rk_3(\h^{(2)}))=(1 \mid 1).$$
  Next, we compute the transformation matrices 
  \begin{equation*}
    \T^{(1)}=
    \begin{pmatrix}
      0 & 1 & 1
      \\ 
      1 & 0 & 1
      \\ 
      1 & 0 & 0 
    \end{pmatrix}
    \quad \text{and} \quad
    \T^{(2)}=
    \begin{pmatrix}
      0 & 1 & 1
      \\ 
      0 & 1 & 2
      \\ 
      1 & 2 & 2 
    \end{pmatrix}
  \end{equation*}
  such that the first entry of $\h^{(1)}\T^{(1)}$ and $\h^{(2)}\T^{(2)}$ are zero, i.e. we have
  \begin{equation*}
    \h^{(1)}\T^{(1)} = (0, \pe, 2\pe^2+1)
    \quad\text{and}\quad
    \h^{(2)}\T^{(2)} = (0, \pe^2+2\pe+1, 2\pe^2+1).
  \end{equation*}
  Defining the block diagonal matrix $\D=\diag\left(\left({\T^{(1)}}^{-1}\right)^\top, \left({\T^{(2)}}^{-1}\right)^\top\right)$ we can compute the invertible transformed code locators $\tilde{\vecbeta}$ and the transformed received word $\tilde{\R}$ as
  \begin{align*}
    \tilde{\vecbeta}&=\vecbeta \cdot \D = ((\pe^2, \pe^2+2\pe+1, 2\pe^2+\pe) \mid (\pe^2+1,2\pe+2,\pe+2))
    \\ 
    \tilde{\R}&=\R \cdot \D 
    \\ 
    &=
    \small
    \left(
    \begin{array}{ccc|ccc}
      \pe^2+\pe+1 & \pe^2+2\pe+2 & \pe^2+\pe+1 & \pe^2+\pe+2 & \pe^2+\pe+2 & \pe^2+2\pe+1
      \\ 
      2\pe^2+\pe+2 & 2\pe^2+\pe+1 & \pe^2 & 2\pe^2+\pe & 2\pe^2+\pe & \pe^2+1
    \end{array}
    \right).
  \end{align*}
  Observe, that $\tilde{\R}$ is the transformed codeword $\tilde{\C}(\f)=\C(\f)\D \in \intLinRS{\tilde{\vecbeta},\a,\shots,\intOrder;\nVec,k}$ corresponding to the message polynomials in $\f = (f_1, f_2)$ that is corrupted by the transformed error
  \begin{equation*}
    \tilde{\E}=\E\D=
    \left(
    \begin{array}{ccc|ccc}
      2\pe^2+1 & 0 & 0 & 2 & 0 & 0
      \\ 
      \pe^2+\pe+1 & 0 & 0 & 2\pe^2+1 & 0 & 0
    \end{array}
    \right)
  \end{equation*}
  of sum-rank weight $t$ whose $n_i-t_i=2$ rightmost columns in each block are zero.
  Hence, the two rightmost columns in each block of $\tilde{\R}$ are equal to the two rightmost columns in each block of $\tilde{\C}(\f)$ which allows for recovering the message polynomials $f_1(x)$ and $f_2(x)$ via Lagrange interpolation. 
\end{example}   

\subsection{An Interpolation-Based Decoding Approach for ILRS Codes}\label{subsec:decodingILRS}
In the previous subsection we derived an \ac{LOlike} probabilistic unique decoder for \ac{ILRS} codes that requires at most $\oh{\intOrder n^\omega}$ operations in $\Fqm$(see Theorem~\ref{thm:LO_decoder_ILRS}).
We now derive a fast Wachter-Zeh--Zeh-like~\cite{wachter2014list} interpolation-based decoding scheme for~\ac{ILRS} codes which can either be used as a list decoder (with not necessarily polynomial-time list size) or as probabilistic unique decoder, which either returns a unique solution (if it exists) or a decoding failure.
In the course of this section we will derive the main result which is stated in Theorem~\ref{thm:list_dec_ILRS}.

\begin{theorem}[List Decoding of \ac{ILRS} Codes]\label{thm:list_dec_ILRS}
 Consider a received word $\R=\C+\E\in\Fqm^{\intOrder\times n}$ where $\C\in\intLinRS{\vecbeta,\a,\shots,\intOrder;\nVec,k}$ is a codeword of an $\intOrder$-interleaved \ac{ILRS} code. 
 If $t=\SumRankWeight(\E)$ satisfies
 \begin{equation}
  t<\frac{\intOrder}{\intOrder+1}(n-k+1)
 \end{equation}
 then a list $\List$ of size 
 \begin{equation}
  |\List|\leq q^{m(k(\intOrder-1))}
 \end{equation}
 containing all message polynomial vectors $\f\in\SkewPolyringZeroDer_{<k}^{\intOrder}$ that correspond to codewords $\C(\f)$ in the \ac{ILRS} code $\intLinRS{\vecbeta,\a,\shots,\intOrder;\nVec,k}$ satisfying $\SumRankDist(\C(\f),\R)<\frac{\intOrder}{\intOrder+1}(n-k+1)$ can be found requiring at most $\softoh{\intOrder^\omega\OMul{n}}$ operations in $\Fqm$.
\end{theorem}

Since $\OMul{n}\in\oh{n^{1.635}}$ (see~\eqref{eq:comp_OMul_Fqm}) and for most applications we have that $\intOrder\ll n$, the proposed interpolation-based decoder is \emph{subquadratic} in the code length $n$ and thus faster compare to the \ac{LOlike} decoder from Section~\ref{subsec:LO_dec_ILRS}.

Suppose we transmit a codeword $\mat{C}(\f)\in\intLinRS{\vecbeta,\a,\shots,\intOrder;\nVec,k}$ over a sum-rank channel~\eqref{eq:sum_rank_channel} and receive a matrix $\R\in\Fqm^{\intOrder\times n}$ that is corrupted by an error matrix $\E\in\Fqm^{\intOrder\times n}$ of sum-rank weight $t$.

\subsubsection{Interpolation Step}

For a multivariate skew polynomial of the form
\begin{equation}\label{eq:mult_var_skew_poly_ILRS}
  Q(x, y_1,\dots, y_\intOrder)=Q_0(x)+Q_1(x)y_1+\dots+Q_\intOrder(x)y_\intOrder
\end{equation}
where $Q_l(x)\in\SkewPolyringZeroDer$ for all $l\in\intervallincl{0}{\intOrder}$ define the $n$ generalized operator evaluation maps 
\begin{align}
\MultSkewPolyringZeroDer \times \Fqm^{\intOrder+1} &\to \Fqm \notag \\
  \left(Q, (\beta_j^{(i)},r_{1,j}^{(i)},\dots,r_{\intOrder,j}^{(i)})\right) &\mapsto \mathscr{E}_j^{(i)}(Q)\defeq \opev{Q_0}{\beta_j^{(i)}}{a_i}+\sum_{l=1}^{\intOrder}\opev{Q_l}{r_{l,j}^{(i)}}{a_i} \label{eq:defDiGenOpILRS}
\end{align}
for all $i=1,\dots,\shots$ and $j=1,\dots,n_i$.

Consider the following interpolation problem in the skew polynomial ring $\SkewPolyringZeroDer$. 

\begin{problem}[ILRS Interpolation Problem]\label{prob:skewIntProblemGenOpILRS}
 Given the integers $\degConstraint,\intOrder,\shots\in\ZZ_{\geq0}$, a set  
 \begin{equation}
  \set{E}=\left\{\mathscr{E}_j^{(i)}:i=1,\dots,\shots, j=1,\dots,n_i\right\}
 \end{equation}
 containing the generalized operator evaluation maps defined in~\eqref{eq:defDiGenOpILRS} and a vector $\vec{w}=(0,k-1,\dots,k-1)\in\ZZ_{\geq0}^{\intOrder+1}$, find a nonzero polynomial of the form 
 \begin{equation}
  Q(x, y_1,\dots, y_\intOrder)=Q_0(x)+Q_1(x)y_1+\dots+Q_\intOrder(x)y_\intOrder
 \end{equation}
 with $Q_l(x)\in\SkewPolyringZeroDer$ for all $l\in\intervallincl{0}{\intOrder}$ that satisfies:
 \begin{enumerate}
  \item $\mathscr{E}_j^{(i)}(Q)=0, \qquad\forall i=1,\dots,\shots$, $j=1,\dots,n_i$, \label{item:int_cond_1_ILRS}
  \item $\deg_{\vec{w}}(Q)<\degConstraint$.
 \end{enumerate}
\end{problem}

Defining the skew polynomials 
\begin{equation}
  Q_0(x)=\sum_{i=0}^{\degConstraint-1}q_{0,i}x^i
  \qquad\text{and}\qquad
  Q_j(x)=\sum_{i=0}^{\degConstraint-k}q_{j,i}x^i,
\end{equation}
a solution of Problem~\ref{prob:skewIntProblemGenOpILRS} can be found by solving the $\Fqm$-linear system
\begin{equation}\label{eq:intSystemILRS}
  \intMat\vec{q}=\vec{0}
\end{equation}
for
\begin{equation}\label{eq:defIntVec_ILRS}
  \vec{q} = \left(q_{0,0},q_{0,1},\dots,q_{0,\degConstraint-1}\mid q_{1,0},q_{1,1},\dots,q_{1,\degConstraint-k}\mid \dots \mid  q_{\intOrder,0},q_{\intOrder,1},\dots,q_{\intOrder,\degConstraint-k}\right)
\end{equation}
where the interpolation matrix $\intMat\in\Fqm^{n\times \degConstraint(\intOrder+1)-\intOrder(k-1)}$ is given by
\begin{equation}\label{eq:intMatrixILRS}
  \intMat=
  \begin{pmatrix}
   \lambda_{\degConstraint}(\vecbeta)_\a^\top & \lambda_{\degConstraint-k+1}(\r_1)_\a^\top & \dots & \lambda_{\degConstraint-k+1}(\r_\intOrder)_\a^\top
  \end{pmatrix}.
\end{equation}
Problem~\ref{prob:skewIntProblemGenOpILRS} can be solved using the Kötter interpolation over skew polynomial rings~\cite{liu2014kotter} in $\oh{\intOrder^2 n^2}$ operations in $\Fqm$.
A solution of Problem~\ref{prob:skewIntProblemGenOpILRS} can be found efficiently requiring only $\softoh{\intOrder^\omega \OMul{n}}$ operations in $\Fqm$ using a variant of the minimal approximant bases approach from~\cite{bartz2020fast}, which we derive in Section~\ref{subsec:fast_interpolation} (cf. Corollary~\ref{cor:complexity_interpolation}).
Another approach yielding the same computational complexity of $\softoh{\intOrder^\omega \OMul{n}}$ operations in $\Fqm$ is given by the fast divide-and-conquer Kötter interpolation from~\cite{bartz2021fastSkewKNH}.
\begin{lemma}[Existence of Solution]\label{lem:existence_ILRS}
 A nonzero solution of Problem~\ref{prob:skewIntProblemGenOpILRS} exists if
\begin{equation}\label{eq:degConstraintILRS}
 \degConstraint=\left\lceil\tfrac{n+\intOrder(k-1)+1}{\intOrder+1}\right\rceil.
\end{equation}
\end{lemma}

\begin{proof}
 Problem~\ref{prob:skewIntProblemGenOpILRS} corresponds to a system of $n$ $\Fqm$-linear equations in $\degConstraint(\intOrder+1)-\intOrder(k-1)$ unknowns (see~\eqref{eq:intSystemILRS}) which has a nonzero solution if the number of equations is less than the number of unknowns, i.e. if
 \begin{equation}
  n<\degConstraint(\intOrder+1)-\intOrder(k-1)\label{eq:existenceCondILRS}
  \quad\Longleftrightarrow\quad
  \degConstraint\geq\tfrac{n+\intOrder(k-1)+1}{\intOrder+1}.
 \end{equation} 
\end{proof}

The $\Fqm$-linear solution space $\Qspace$ of Problem~\ref{prob:skewIntProblemGenOpILRS} is defined as
\begin{equation}\label{eq:def_sol_space_ILRS}
 \Qspace\defeq\{Q\in\MultSkewPolyringZeroDer:\vec{q}(Q)\in\rker(\intMat)\}  
\end{equation}
where $\vec{q}(Q)\in\Fqm^{\degConstraint(\intOrder+1)-\intOrder(k-1)}$ is the coefficient vector of $Q$ as defined in~\eqref{eq:defIntVec_ILRS}.
The dimension of the $\Fqm$-linear solution space $\Qspace$ of Problem~\ref{prob:skewIntProblemGenOpILRS} (i.e. the dimension of the right kernel of $\intMat$ in~\eqref{eq:intMatrixILRS}) is denoted by 
\begin{equation}
  d_I\defeq\dim(\Qspace)=\dim(\rker(\intMat)).
\end{equation}

All polynomials of the form~\eqref{eq:mult_var_skew_poly_ILRS} that satisfy Condition~\ref{item:int_cond_1_ILRS} of Problem~\ref{prob:skewIntProblemGenOpILRS} form a (free) left $\SkewPolyringZeroDer$-module
\begin{equation}
  \module{K}=\{Q\in\MultSkewPolyringZeroDer: \mathscr{E}_j^{(i)}(Q)=0, \forall i=1,\dots,\shots, j=1,\dots,n_i\},
\end{equation}
which we further call the \emph{interpolation module}.
Note, that $\module{K}$ contains also polynomials that have degree larger or equal to $\degConstraint$.
By restricting the degree of the elements in $\module{K}$ to at most $\degConstraint-1$, we have that $\Qspace$ coincides with $\module{K}\cap\MultSkewPolyringZeroDer_{<\degConstraint}$.

\subsubsection{Root-Finding Step}\label{subsubsec:RF_ILRS}

The goal of the root-finding step is to recover the message polynomials $f_1,\dots,f_\intOrder \in\SkewPolyringZeroDer_{<k}$ from the multivariate polynomial constructed in the interpolation step. 
Therefore, we need the following results.

\begin{lemma}[Roots of Polynomial]\label{lem:decConditionILRS}
 Let
 \begin{equation}\label{eq:def_Px}
  P(x)\defeq Q_0(x)+Q_1(x)f_1(x)+\dots+Q_\intOrder(x)f_\intOrder(x).
 \end{equation}
 Then there exist elements $\zeta_1^{(i)},\dots,\zeta_{n_i-t_i}^{(i)}$ in $\Fqm$ that are $\Fq$-linearly independent for each $i=1,\dots,\shots$ such that
 \begin{equation}
  \opev{P}{\zeta_j^{(i)}}{a_i}=0
 \end{equation}
 for all $i=1,\dots,\shots$ and $j=1,\dots,n_i-t_i$.
\end{lemma}

\begin{proof}
 The proof exploits the $\Fq$-linearity of the generalized operator evaluation (per block) which allows to transform the $\Fq$-rank errors (per block) into corrupted and non-corrupted columns.
 By definition, the sum-rank weight of $\mat{E}$ equals $t=\sum_{i=1}^{\shots}t_i$, where $t_i=\rk_q(\shot{\mat{E}}{i})$. 
 Hence, there exist nonsingular matrices $\mat{T}^{(i)}\in\Fq^{n_i\times n_i}$ such that the $\mat{E}^{(i)}\mat{T}^{(i)}$ has only $t_i$ nonzero columns for all $i=1,\dots,\shots$.
 Now assume w.l.o.g. that the matrices $\mat{T}^{(i)}$ are chosen such that only the last $t_i$ columns of $\mat{E}^{(i)}\mat{T}^{(i)}$ are nonzero for all $i=1,\dots,\shots$.
 Define $\veczeta^{(i)}=\vecbeta^{(i)}\mat{T}^{(i)}$. Since we have that $\rk_q(\vecbeta^{(i)})=n_i$ and $\mat{T}^{(i)}$ is invertible, we have that $\rk_q(\veczeta^{(i)})=n_i$ for all $i=1,\dots,\shots$.
 Then we have that the first $n_i-t_i$ columns of $\mat{R}^{(i)}\mat{T}^{(i)}$ are non-corrupted and given by
 \begin{equation}
  \begin{pmatrix}
   \opev{f_1}{\zeta^{(i)}_1}{a_i} & \dots & \opev{f_1}{\zeta^{(i)}_{n_i-t_i}}{a_i}
   \\ 
   \vdots & \ddots & \vdots
   \\ 
   \opev{f_\intOrder}{\zeta^{(i)}_1}{a_i} & \dots & \opev{f_\intOrder}{\zeta^{(i)}_{n_i-t_i}}{a_i}
  \end{pmatrix}
  \in\Fqm^{\intOrder\times (n_i-t_i)},
\end{equation}
 for all $i=1,\dots,\shots$.
 By Lemma~\ref{lem:opev_product_rule} and the interpolation conditions in Problem~\ref{prob:skewIntProblemGenOpILRS} we have
 \begin{equation}
  \opev{P}{\zeta_j^{(i)}}{a_i}=\opev{Q_0}{\zeta_j^{(i)}}{a_i}+\opev{Q_1}{\opev{f_1}{\zeta_j^{(i)}}{a_i}}{a_i}+\dots+\opev{Q_\intOrder}{\opev{f_\intOrder}{\zeta_j^{(i)}}{a_i}}{a_i}=0
 \end{equation}
 for all $i=1,\dots,\shots$ and $j=1,\dots,n_i-t_i$ and the statement follows.
\end{proof} 

\begin{theorem}[Decoding Radius]\label{thm:decRadiusILRS}
 Let $\mat{C}(\f)$ be a codeword from $\intLinRS{\vecbeta,\a,\shots,\intOrder;\nVec,k}$ and let $\mat{R}=\mat{C}(\vec{f})+\mat{E}$ be the received word. Further, let $Q(x,y_1,\dots,y_\intOrder)\neq0$ fulfill the constraints in Problem~\ref{prob:skewIntProblemGenOpILRS}. 
 If $t=\SumRankWeight(\mat{E})$ satisfies
 \begin{equation}\label{eq:listDecRegionILRS}
  t<\frac{\intOrder}{\intOrder+1}(n-k+1)
 \end{equation}
 then 
 \begin{equation}\label{eq:rootFindingEquationILRS}
  P(x)=Q_0(x)+Q_1(x)f_1(x)+\!\dots\!+Q_\intOrder(x)f_\intOrder(x)=0.
 \end{equation}
\end{theorem}

\begin{proof}
 By Lemma~\ref{lem:decConditionILRS} there exist elements $\zeta_1^{(i)},\dots,\zeta_{n_i-t_i}^{(i)}$ in $\Fqm$ that are $\Fq$-linearly independent for each $i=1,\dots,\shots$ such that
 \begin{equation}
  \opev{P}{\zeta_j^{(i)}}{a_i}=0
 \end{equation}
 for all $i=1,\dots,\shots$ and $j=1,\dots,n_i-t_i$.
 By choosing 
 \begin{equation}\label{eq:decDegreeConstraintILRS}
  \degConstraint\leq n-t
 \end{equation}
 the degree of $P(x)$ exceeds the degree bound from Proposition~\ref{prop:number_of_roots} which is possible only if $P(x)=0$.
 Combining~\eqref{eq:existenceCondILRS} and~\eqref{eq:decDegreeConstraintILRS} we get
 \begin{align*}
  n+\intOrder(k-1)&<\degConstraint(\intOrder+1)\leq(\intOrder+1)(n-t)
  \\\Longleftrightarrow\qquad
  t&<\frac{\intOrder}{\intOrder+1}(n-k+1).
 \end{align*}
\end{proof}

Theorem~\ref{thm:decRadiusILRS} shows, that the message polynomials $f_1,\dots,f_\intOrder \in \SkewPolyringZeroDer_{<k}$ satisfy~\eqref{eq:listDecRegionILRS} if the sum-rank weight of the error lies within the decoding radius in~\eqref{eq:listDecRegionILRS}. 
The decoding region in~\eqref{eq:listDecRegionILRS} shows a significantly improved (burst) error-correction performance due to interleaving.

In the root-finding step, all polynomials $f_1,\dots,f_{\intOrder}\in\SkewPolyringZeroDer_{<k}$ that satisfy~\eqref{eq:rootFindingEquationILRS} need to be found. 
In order to minimize the number of solutions of the root-finding problem one may use a basis of the $\Fqm$-linear solution space $\Qspace$ of Problem~\ref{prob:skewIntProblemGenOpILRS} instead of only considering only a single solution (see~\cite{wachter2014list}).

In~\cite{bartz2017algebraic} it was shown that using a degree-restricted subset of a Gröbner basis (of cardinality at most $\intOrder$) for the interpolation module $\module{K}$ w.r.t. $\wOrder$ is sufficient to achieve the minimal number of solutions of the root-finding problem. 
Although the results in~\cite{bartz2017algebraic} we derived for linearized polynomial modules, they carry over to skew polynomial modules since the structure of the corresponding problems (including the noncommutativity) is the same.
For details about solving the root-finding problem using Gröbner bases the reader is referred to~\cite{bartz2017algebraic}.

We now use this approach to obtain the minimal number of solutions of the root-finding problem.
Let $\Groeb\subset\MultSkewPolyringZeroDer$ be a basis for the interpolation module $\module{K}$ such that the left $\SkewPolyringZeroDer$-span of the polynomials in the degree-restricted subset 
\begin{equation}\label{eq:degree_restr_subset}
  \GroebStar\defeq\{Q\in\Groeb:\deg_\w(Q)<\degConstraint\}\subseteq\Qspace
\end{equation}
contains $\Qspace$.
Examples of bases $\Groeb$ where the degree-restricted subset $\GroebStar\subseteq\Qspace$ spans $\Qspace$ are minimal Gröbner bases w.r.t. $\wOrder$ and $\w$-ordered weak-Popov approximant bases for the interpolation module $\module{K}$.
Mininal Gröbner bases for $\module{K}$ w.r.t. $\wOrder$ can be computed efficiently using the multivariate Kötter interpolation over skew polynomial rings from~\cite{liu2014kotter,bartz2021fastSkewKNH}.
An efficient method to construct $\w$-ordered weak-Popov approximant bases for $\module{K}$ is given in Section~\ref{subsec:fast_interpolation}.

Let the polynomials $Q^{(1)},\dots,Q^{(\cardGroebStar)}\in\GroebStar$ be given by
\begin{equation}
  Q^{(r)}=Q_0^{(r)}+Q_1^{(r)}y_1+\dots+Q_\intOrder^{(r)}y_\intOrder
\end{equation}
with
\begin{equation}
  Q_0^{(r)}(x)=\textstyle\sum_{i=0}^{\degConstraint-1}q_{0,i}^{(r)}x^i \qquad\text{and}\qquad
  Q_j^{(r)}(x)=\textstyle\sum_{i=0}^{\degConstraint-k}q_{j,i}^{(r)}x^i,\qquad\forall j=1,\dots,\intOrder
\end{equation}
for all $r=1,\dots,\cardGroebStar$.
By using the same arguments as in~\cite[Lemma~5.4]{bartz2017algebraic} one can show that $\cardGroebStar\defeq|\GroebStar|$ satisfies $1\leq\cardGroebStar\leq\intOrder$.

Define the matrices 
 \begin{equation}\label{eq:def_RF_blocks}
  \autMat{\Q}{j}{i}\defeq
  \begin{pmatrix}
   \aut^i\left(q_{1,j}^{(1)}\right) & \aut^i\left(q_{2,j}^{(1)}\right) & \dots & \aut^i\left(q_{\intOrder,j}^{(1)}\right)
   \\
   \vdots & \vdots & \ddots & \vdots
   \\ 
   \aut^i\left(q_{1,j}^{(\cardGroebStar)}\right) & \aut^i\left(q_{2,j}^{(\cardGroebStar)}\right) & \dots & \aut^i\left(q_{\intOrder,j}^{(\cardGroebStar)}\right)
  \end{pmatrix}\in\Fqm^{\cardGroebStar\times s}
 \end{equation}
 and the vectors
 \begin{equation}
  \autVec{\f}{j}{i}
  \defeq\left(\aut^i\left(f_{j,1}\right),\dots,\aut^i\left(f_{j, \intOrder}\right)\right)\in\Fqm^{\intOrder}
 \end{equation}
 and
 \begin{equation}\label{eq:def_RF_vecs}
  \autVec{\q}{0,j}{i}
  \defeq\left(\aut^i\left(q_{0,j}^{(1)}\right),\dots,\aut^i\left(q_{0,j}^{(\cardGroebStar)}\right)\right)\in\Fqm^{\cardGroebStar}.
 \end{equation}
Defining the root-finding matrix
 \begin{equation}\label{eq:rootFindingMatrixILRS}
 \RFmat\defeq
 \begin{pmatrix}
 \phantom{\aut^{-}(}\Q_0        &     &  &              \\
 \autMat{\Q}{1}{-1}         & \autMat{\Q}{0}{-1}  &        &              \\[-3pt]
 \vdots             & \autMat{\Q}{1}{-2}  & \ddots &              \\[-3pt]
 \autMat{\Q}{\degConstraint-k}{-(\degConstraint-k)} & \vdots    & \ddots & \autMat{\Q}{0}{-(k-1)}       \\[-3pt]
              & \autMat{\Q}{\degConstraint-k}{-(\degConstraint-k-1)}    & \ddots & \autMat{\Q}{1}{-k\phantom{(-1)}}           \\[-5pt]
                                          &     & \ddots & \vdots          \\
              &           & & \autMat{\Q}{\degConstraint-k}{-(\degConstraint-1)}
 \end{pmatrix}\in\Fqm^{\degConstraint \cardGroebStar\times \intOrder k}
\end{equation}
and the vectors
\begin{align}\label{eq:RFvectorsILRS}
 \RFvec\!&\defeq\!\left(\vec{f}_0,\autVec{\f}{1}{-1},\dots,\autVec{\f}{k-1}{-(k-1)}\right)^\top\!\!\!\in\Fqm^{\intOrder k} \quad\text{and}\quad \notag
 \\ 
 \vec{q}_{0} \!&\defeq\! \left(\vec{q}_{0,0},\autVec{\vec{q}_{0,1}}{}{-1},\dots, \autVec{\vec{q}_{0,\degConstraint-1}}{}{-(\degConstraint-1)}\right)^\top\!\!\!\in\Fqm^{\degConstraint \cardGroebStar}
\end{align}
we can write the root-finding system~\eqref{eq:rootFindingEquationILRS} as
\begin{equation}\label{eq:rootFindingSystemFqmILRS}
 \RFmat\cdot\RFvec=-\vec{q}_0.
\end{equation}
The root-finding step can be solved efficiently by the minimal approximant bases algorithm in~\cite{bartz2019fast,bartz2020fast} with at most $\softoh{\intOrder^\omega \OMul{n}}$ operations in $\Fqm$.

Proposition~\ref{prop:RFproperties} summarizes some results from~\cite{bartz2017algebraic} on solving the root-finding problem using $\GroebStar$.

\begin{proposition}[Root-Finding with $\GroebStar$]\label{prop:RFproperties}
 Let the sum-rank weight of the error matrix $\E$ satisfy $t<\frac{\intOrder}{\intOrder+1}(n-k+1)$.
 Let $\GroebStar\subseteq\Qspace\setminus \{0\}$ be a set of $\SkewPolyringZeroDer$-linearly independent polynomials with distinct leading positions whose left $\SkewPolyringZeroDer$-linear span contains the $\Fqm$-linear solution space $\Qspace$ of Problem~\ref{prob:skewIntProblemGenOpILRS}.
 Then:
 \begin{enumerate}
  \item \label{itm:cardinalityGroebStar} We have that $\cardGroebStar\defeq|\GroebStar|$ satisfies
    \begin{equation*}
      1\leq\cardGroebStar\leq\intOrder.
    \end{equation*}
  \item \label{itm:rankRFmat} The rank of the root-finding matrix $\RFmat$ in~\eqref{eq:rootFindingMatrixILRS} satisfies
 \begin{equation*}
   \rk_{q^m}(\RFmat)\geq\cardGroebStar k.
 \end{equation*}
 \item \label{itm:numSolRFsystem} The root-finding system in~\eqref{eq:rootFindingEquationILRS} has at most $q^{m(k(\intOrder-\intOrder'))}$ solutions $f_1,\dots,f_\intOrder \in \SkewPolyringZeroDer_{<k}$.
 \item \label{itm:unqiueSolRFsystem} The root-finding system in~\eqref{eq:rootFindingEquationILRS} has a unique solution if and only if $\cardGroebStar=\intOrder$.
 \end{enumerate}
\end{proposition}

A proof of Proposition~\ref{prop:RFproperties} can be found in Appendix~\ref{app:proofs_ILRS}.

Observe, that Proposition~\ref{prop:RFproperties} allows to derive the actual number of solutions of the root-finding problem right after the interpolation step by considering $\cardGroebStar$.

\subsubsection{List Decoding}\label{subsubsec:ILRS_list}

We now interpret the proposed interpolation-based decoding scheme for \ac{ILRS} as a list decoder.
In general, the root-finding matrix $\RFmat$ in~\eqref{eq:rootFindingMatrixILRS} can be rank deficient.
In this case we obtain a \emph{list} $\List$ of potential message polynomials $f_1,\dots,f_\intOrder$.
By Proposition~\ref{prop:RFproperties} the list size $|\List|$, i.e. the maximum number of solutions of \eqref{eq:rootFindingSystemFqmILRS}, is upper bounded by $q^{m(k(\intOrder-1))}$.
Note, that Proposition~\ref{prop:RFproperties} provides an upper bound on the actual list size by considering the cardinality of $\GroebStar$ right after the interpolation step.

In general, we have that $k\leq n$, where $n\leq\shots m$.
Hence, for $m\approx n/\shots$ we get a worst-case list size of $q^{\frac{n}{\shots}(k(\intOrder-1))}$.
Although Proposition~\ref{prop:RFproperties} shows, that the worst-case list size is exponential in $n$, we will later see that the average list size is close to one for most parameters of interest.

Algorithm~\ref{alg:list_dec_ILRS} and Theorem~\ref{thm:list_dec_ILRS} summarize the interpolation-based list decoder for~\ac{ILRS} codes.

\begin{algorithm}[ht]
  \caption{\algoname{List Decoding of \ac{ILRS} Codes}}\label{alg:list_dec_ILRS}
  \SetKwInOut{Input}{Input}\SetKwInOut{Output}{Output}

  \Input{Channel output $\R=\C+\E\in\Fqm^{\intOrder\times n}$ where $\C\in\intLinRS{\vecbeta,\a,\shots,\intOrder;\nVec,k}$ and $\SumRankWeight(\E)=t<\frac{\intOrder}{\intOrder+1}(n-k+1)$.}

  \Output{A list $\List$ containing message polynomial vectors $\f=(f_1,\dots,f_\intOrder) \in \SkewPolyringZeroDer_{<k}^\intOrder$ that satisfy~\eqref{eq:rootFindingEquationILRS}.}
  \BlankLine
  Find left $\SkewPolyringZeroDer$-linearly independent $Q^{(1)},\dots,Q^{(\IntParam)} \in \Qspace \setminus \{0\}$ whose left $\SkewPolyringZeroDer$-span contains the $\Fqm$-linear solution space $\Qspace$ of Problem~\ref{prob:skewIntProblemGenOpILRS}.
  \\ 
  Using $Q^{(1)},\dots,Q^{(\IntParam)}$, find the list $\List\subseteq\SkewPolyringZeroDer^{\intOrder}_{<k}$ of all $\f=(f_1,\dots,f_\intOrder)\in\SkewPolyringZeroDer_{<k}^\intOrder$ that satisfy~\eqref{eq:rootFindingEquationILRS}.
  \\
  \Return{$\List$}
\end{algorithm}

\subsubsection{Probabilistic Unique Decoding}\label{subsubsec:ILRS_unique}

We now consider the proposed interpolation-based decoder for \ac{ILRS} codes as a probabilistic unique decoder which either returns a unique solution (if the list size is equal to one) or a decoding failure.

Using similar arguments as in~\cite[Lemma~3]{wachter2014list} we can lower bound the dimension $d_I$ of the $\Fqm$-linear solution space $\Qspace$ of Problem~\ref{prob:skewIntProblemGenOpILRS}.

\begin{lemma}[Dimension of Solution Space]\label{lem:dim_Qspace_ILRS}
 Let $t$ satisfy~\eqref{eq:listDecRegionILRS}.
 Then the dimension $d_I=\dim(\Qspace)$ of the $\Fqm$-linear solution space $\Qspace$ of Problem~\ref{prob:skewIntProblemGenOpILRS} satisfies
 \begin{equation}
  d_I\geq \intOrder(\degConstraint+1)-\intOrder k-t.
 \end{equation}
\end{lemma}

\begin{proof}
 By $\Fq$-linear row operations and permutations and $\Fqm$-linear column operations we can bring the interpolation matrix $\intMat$ in~\eqref{eq:intMatrixILRS} into a matrix of the form
 \begin{equation}
  \widetilde{\R}_{I}=
   \begin{pmatrix}
    \lambda_{\degConstraint}(\veczeta)_\a^\top & \0 & \dots & \0
    \\
    \lambda_{\degConstraint}(\vecepsilon)_\a^\top & \lambda_{\degConstraint-k+1}(\tilde{\e}_1)_\a^\top & \dots & \lambda_{\degConstraint-k+1}(\tilde{\e}_\intOrder)_\a^\top
   \end{pmatrix}
   \in\Fqm^{n\times \degConstraint(\intOrder+1)-\intOrder(k-1)}
 \end{equation}
 where $\veczeta=(\veczeta^{(1)} \mid \veczeta^{(2)} \mid \dots \mid \veczeta^{(\shots)})\in\Fqm^{n-t}$ with $\veczeta^{(i)}=(\zeta_1^{(i)},\dots,\zeta_{n_i-t_i}^{(i)})\in\Fqm^{n_i-t_i}$ and $\rk_q(\veczeta^{(i)})=n_i-t_i$ for all $i=1,\dots,\shots$ and $\vecepsilon,\tilde{\e}_j\in\Fqm^t$ for all $j=1,\dots,\intOrder$.
 By Proposition~\ref{prop:rankGenOpVandermonde} and the fact that $\degConstraint\leq n-t$ (see~\eqref{eq:decDegreeConstraintILRS}) the matrix $\lambda_{\degConstraint}(\veczeta)_\a^\top\in\Fqm^{(n-t)\times\degConstraint}$ has $\Fqm$-rank $\degConstraint$ since the entries in $\veczeta$ are block-wise $\Fq$-linearly independent and the entries in $\a$ are representatives from different conjugacy classes of $\Fqm$.
 The last $t$ rows of $\widetilde{\R}_{I}$ can increase the $\Fqm$-rank by at most $t$ and thus $\rk_{q^m}(\intMat)=\rk_{q^m}(\widetilde{\R}_{I})\leq\degConstraint+t$.
 Hence, the dimension $d_I$ of the $\Fqm$-linear solution space $\Qspace$ of Problem~\ref{prob:skewIntProblemGenOpILRS} satisfies
 \begin{align*}
  d_I=\dim(\rker(\intMat))&=\degConstraint(\intOrder+1)-\intOrder(k-1)-\rk_{q^m}(\intMat)
  \\ 
  &\geq \intOrder(\degConstraint+1)-\intOrder k -t.
 \end{align*}
\end{proof}

The rank of the root-finding matrix $\RFmat$ can be full only if the dimension of the solution space of the interpolation problem $d_I=\dim(\Qspace)$ is at least $\intOrder$, i.e. if
\begin{align}
 d_I\geq\intOrder\qquad\Longleftrightarrow\qquad
 t&\leq\intOrder\degConstraint-\intOrder k\nonumber
 \\\Longleftrightarrow\qquad
 t&\leq\frac{\intOrder}{\intOrder+1}(n-k).\label{eq:decRegionILRSprob}
\end{align}
The probabilistic unique decoding region in~\eqref{eq:decRegionILRSprob} is only sightly smaller than the list decoding region in~\eqref{eq:listDecRegionILRS}.
Combining the decoding condition $\degConstraint \leq \nTransmit - \deletions$ and~\eqref{eq:decRegionILRSprob} we get the degree constraint for the probabilistic unique decoder (see also~\cite{bartz2017algebraic})
\begin{equation}
  \degConstraintUnique=\left\lceil\frac{n+\intOrder k}{\intOrder+1}\right\rceil.
\end{equation}
In order to get an estimate probability of successful decoding, we may use similar assumptions as in~\cite{wachter2014list} to derive a heuristic upper bound on the decoding failure probability.
Under the assumption that the coefficients $q_{i,j}^{(r)}$ are uniformly distributed over $\Fqm$ (see~\cite[Lemma~9]{wachter2014list}) one can derive a heuristic upper bound on the decoding failure probability $P_f$ as
\begin{equation}\label{eq:heur_upper_bound_Psucc_ILRS}
  P_f
  \leq \gammaq q^{-m(d_I-\intOrder+1)}
  \leq \gammaq q^{-m\left(\intOrder\left(\left\lceil\frac{n+\intOrder k}{\intOrder+1}\right\rceil-k\right)-t+1\right)}.
\end{equation} 

By reducing the conditions of successful decoding of the interpolation-based decoder to the conditions of the \ac{LOlike} decoder from Section~\ref{subsec:LO_dec_ILRS} we obtain an upper bound on the decoding failure probability that takes into account the distribution of the error matrix $\E$.
The results of the interpolation-based probabilistic unique decoder are summarized in Algorithm~\ref{alg:unique_dec_ILRS} and Theorem~\ref{thm:unique_dec_ILRS}.

\begin{algorithm}[ht]
  \caption{\algoname{Probabilistic Unique Decoding of \ac{ILRS} Codes}}\label{alg:unique_dec_ILRS}
  \SetKwInOut{Input}{Input}\SetKwInOut{Output}{Output}

  \Input{Channel output $\R=\C+\E\in\Fqm^{\intOrder\times n}$ where $\C\in\intLinRS{\vecbeta,\a,\shots,\intOrder;\nVec,k}$ and $\SumRankWeight(\E)=t$.}

  \Output{Message polynomial vector $\f=(f_1,\dots,f_\intOrder)\in\SkewPolyringZeroDer_{<k}^\intOrder$ or ``decoding failure''}
  \BlankLine
  Find left $\SkewPolyringZeroDer$-linearly independent $Q^{(1)},\dots,Q^{(\IntParam)} \in \Qspace \setminus \{0\}$ whose left $\SkewPolyringZeroDer$-span contains the $\Fqm$-linear solution space $\Qspace$ of Problem~\ref{prob:skewIntProblemGenOpILRS}.
  \\ 
  \If{$\IntParam=\intOrder$}
  {
  Use $Q^{(1)},\dots,Q^{(\intOrder)}$ to find the unique vector $\f=(f_1,\dots,f_\intOrder)\in\SkewPolyringZeroDer_{<k}^\intOrder$ that satisfies~\eqref{eq:rootFindingEquationILRS}
  \\ 
    \Return{Message polynomial vector $\f=(f_1,\dots,f_\intOrder)\in\SkewPolyringZeroDer_{<k}^\intOrder$}
  }
  \Else{
    \Return{``decoding failure''}
  }
\end{algorithm}

\begin{theorem}[Probabilistic Unique Decoding of \ac{ILRS} Codes]\label{thm:unique_dec_ILRS}
 Consider a received word $\R=\C+\E\in\Fqm^{\intOrder\times n}$ where $\C\in\intLinRS{\vecbeta,\a,\shots,\intOrder;\nVec,k}$ is a codeword of an $\intOrder$-interleaved \ac{ILRS} code and $\E$ is chosen uniformly at random from all matrices from $\Fqm^{\intOrder\times n}$ of sum-rank weight $t$. 
 If $t=\SumRankWeight(\E)$ satisfies
 \begin{equation}
  t\leq\tmax\defeq\frac{\intOrder}{\intOrder+1}(n-k)
 \end{equation}
 then the unique message polynomial vector $\f\in\SkewPolyringZeroDer_{<k}^{\intOrder}$ corresponding to the codeword $\C(\f)$ in the code $\intLinRS{\vecbeta,\a,\shots,\intOrder;\nVec,k}$ 
 can be found with probability at least
 \begin{equation}\label{eq:lower_bound_Psucc_ILRS}
  1 - \gammaq^{\ell} q^{-m((s+1)(\tmax-t)+1)}
 \end{equation}
 requiring at most $\softoh{\intOrder^\omega\OMul{n}}$ operations in $\Fqm$.
\end{theorem}

\begin{proof}
For the purpose of the proof (but not algorithmically), we consider the root-finding problem set up with an $\Fqm$-basis $Q^{(1)},\dots,Q^{(d_I)}$ of $\Qspace$.    
The unique decoder fails if there are at least two distinct roots $\f$ and $\f'$.
In this case, the $\Fqm$-linear system $\RFmat\cdot\RFvec=-\vec{q}_0$ in~\eqref{eq:rootFindingSystemFqmILRS} set up with the $\Fqm$-basis $\widetilde{Q}^{(r)}\in\Qspace$ for $r=1,\dots,d_I$ has at least two solutions.
This means that $\RFmat \in \Fq^{\degConstraint d_I\times \intOrder k}$ must have rank $<\intOrder k$.

The matrix $\RFmat$ contains a lower block triangular matrix with matrices $$\Q_0,\aut^{-1}(\Q_0),\dots,\aut^{-(k-1)}(\Q_0)$$ on the upper diagonal, which have all $\Fqm$-rank $\rk_{q^m}(\Q_0)$ (see Lemma~\ref{lem:rank_row_op}).
Thus, if $\rk_{q^m}(\Q_0)=\intOrder$ the matrix $\RFmat$ has full $\Fqm$-rank $\intOrder k$. Therefore, $\rk_{q^m}(\RFmat)<\intOrder k$ implies that $\Q_0$ has rank $<s$.

Since the root-finding system~\eqref{eq:rootFindingSystemFqmILRS} has at least one solution $\RFvec$, there is a vector $\f_0 \in \Fqm^\intOrder$ such that
\begin{equation*}
\Q_0 \f_0 = - \q_{0,0}^\top.
\end{equation*}
Thus, the matrix
\begin{equation*}
\Qbar_0 := 
\begin{pmatrix}
\Q_0 &  \q_{0,0}^\top
\end{pmatrix} \in \Fqm^{d_I\times (s+1)}
\end{equation*}
has rank $\rk_{q^m}(\Qbar_0) = \rk_{q^m}(\Q_0) < s$.
Hence, there are at least $d_I-s+1$ $\Fqm$-linearly independent polynomials $\widetilde{Q}^{(1)},\dots,\allowbreak\widetilde{Q}^{(d_I-s+1)}\in\Qspace$ such that their zeroth coefficients $\widetilde{q}_{l,0}^{\,(1)},\dots,\widetilde{q}_{l,0}^{\,(d_I-\intOrder+1)}$ are zero for all $l=0,\dots,\intOrder$ (obtained by suitable) $\Fqm$-linear combinations of the original basis polynomials $Q^{(1)},\dots,Q^{(d_I)}$, such that the corresponding $\Fqm$-linear row operations on $\Qbar_0$ give a $(d_I-\intOrder+1)\times(\intOrder+1)$ zero matrix (recall that $\Qbar_0$ has $d_I$ rows, but rank at most $s-1$).

The $d_I-s+1$ $\Fqm$-linearly independent coefficient vectors of $\widetilde{Q}^{(1)},\dots,\widetilde{Q}^{(d_I-s+1)}$ of the form~\eqref{eq:defIntVec_ILRS} are in the left kernel of the matrix
\begin{equation*}
\intMat^\top=
\begin{bmatrix}
  \lambda_{\degConstraint}(\vecbeta)_\a \\
  \lambda_{\degConstraint-k+1}(\r_1)_\a \\
  \vdots \\
  \lambda_{\degConstraint-k+1}(\r_\intOrder)_\a
\end{bmatrix}\in\Fqm^{\degConstraint(\intOrder+1)-\intOrder(k-1)\times n}.
\end{equation*}
Since the zeroth components $\widetilde{q}_{l,0}^{\,(r)}$ of all $\widetilde{Q}^{(r)}$ are zero for all $l=0,\dots,\intOrder$ and $r=1,\dots,d_I-\intOrder+1$, this means that the left kernel of the matrix
\begin{equation*}
  \widetilde{\R}_I^\top=
  \op{\a}{
  \begin{bmatrix}
    \lambda_{\degConstraint-1}(\vecbeta)_\a \\
    \lambda_{\degConstraint-k}(\r_1)_\a \\
    \vdots \\
    \lambda_{\degConstraint-k}(\r_\intOrder)_\a
  \end{bmatrix}  
  }
  \in \Fqm^{\degConstraint(\intOrder+1)-\intOrder k -1 \times n}
\end{equation*}
has dimension at least $d_I-s+1$. 
The maximum decoding radius $\tmax$ corresponds to the degree constraint $\degConstraint=n-\tmax$ (see~\eqref{eq:decDegreeConstraintILRS}) and thus 
\begin{align*}
 \dim(\lker(\widetilde{\R}_I^\top))&\geq d_I-s+1
 \\ 
 &\geq\intOrder(n-\tmax+1)-\intOrder k-\tmax-\intOrder+1
 \\ 
 &\geq1.
\end{align*}
Therefore, we have that
\begin{align*}
 \rk_{q^m}(\widetilde{\R}_I^\top)
 &\leq\degConstraint(\intOrder+1)-\intOrder k -1 -\dim(\lker(\widetilde{\R}_I^\top))
 \\ 
 &<\degConstraint(\intOrder+1)-\intOrder k -1
 \\ 
 &=n-1. 
\end{align*}
Observe, that for $\degConstraint=n-\tmax$ we have that
\begin{equation*}
  \widetilde{\R}_I^\top = \op{\a}{\LODecMat}
\end{equation*}
where $\LODecMat$ is the \ac{LOlike} decoding matrix from~\eqref{eq:LO_matrix_ILRS}.
By Lemma~\ref{lem:rank_row_op} the $\Fqm$-rank of $\LODecMat$ and $\op{\a}{\LODecMat}$ is the same and thus we have that
\begin{align*}
  \rk_{q^m}(\LODecMat)=\rk_{q^m}(\widetilde{\R}_I)< n-1
\end{align*}
which shows that in this case the \ac{LOlike} decoder fails as well. 
Therefore, we conclude that 
\begin{equation}
  \Pr(\rk_{q^m}(\RFmat)<\intOrder k)\leq\Pr(\rk_{q^m}(\Q_0)<\intOrder)\leq\Pr(\rk_{q^m}(\LODecMat)<n-1)
\end{equation}
and thus the lower bound on the probability of successful decoding follows from Theorem~\ref{thm:LO_decoder_ILRS}.
The complexity statement follows from Corollary~\ref{cor:complexity_interpolation} and the efficient root-finding method in~\cite{bartz2019fast,bartz2020fast}.
\end{proof}

The lower bound on the probability of successful decoding in Theorem~\ref{thm:unique_dec_ILRS} yields also an upper bound on the decoding failure probability $P_f$, i.e. we have that
\begin{equation}\label{eq:upper_bound_Pfail_ILRS}
  P_f\leq\gammaq^{\ell+1} q^{-m((s+1)(\tmax-t)+1)}.
\end{equation}

The normalized decoding radius $\tau\defeq t/n$ for \ac{ILRS} codes is $\tau\approx\frac{\intOrder}{\intOrder+1}\left(1-R\right)$.
The improvement of the normalized decoding radius upon the normalized \ac{BMD} radius is illustrated in Figure~\ref{fig:decRadiusILRS}.
\begin{figure}[ht!]
\centering
\begin{tikzpicture}

\tikzstyle{BMD} 	   = [color=red!80!black, dashed]
\tikzstyle{SB}         = [color=black, solid]
\tikzstyle{ILRSs2}     = [color=blue!80!black, dash pattern=on 2pt off 4pt on 6pt off 4pt,mark=o, mark options={solid}]
\tikzstyle{ILRSs5}     = [color=green!80!black, dash pattern=on 2pt off 4pt on 6pt off 4pt,mark=triangle, mark options={solid}]
\tikzstyle{ILRSs10}    = [color=orange!80!black, dash pattern=on 2pt off 4pt on 6pt off 4pt,mark=square, mark options={solid}]

\begin{axis}[%
scale only axis,
xmin=0,
xmax=1,
xlabel={Code Rate R},
xmajorgrids,
ymin=0,
ymax=1,
ylabel={Normalized decoding radius $\tau$},
ymajorgrids,
legend style={legend cell align=left,align=left,draw=white!15!black},
mystyle
]
\addplot [style=SB]
  table[row sep=crcr]{%
0	1\\
0.1	0.9\\
0.2	0.8\\
0.3	0.7\\
0.4	0.6\\
0.5	0.5\\
0.6	0.4\\
0.7	0.3\\
0.8	0.2\\
0.9	0.1\\
1	0\\
};
\addlegendentry{Singleton Bound / \\List Decoding Capacity};

\addplot [style=BMD]
  table[row sep=crcr]{%
0	0.5\\
0.1	0.45\\
0.2	0.4\\
0.3	0.35\\
0.4	0.3\\
0.5	0.25\\
0.6	0.2\\
0.7	0.15\\
0.8	0.1\\
0.9	0.05\\
1	0\\
};
\addlegendentry{BMD radius};

\addplot [style=ILRSs2]
  table[row sep=crcr]{%
0.000000	0.666667\\
0.100000	0.600000\\
0.200000	0.533333\\
0.300000	0.466667\\
0.400000	0.400000\\
0.500000	0.333333\\
0.600000	0.266667\\
0.700000	0.200000\\
0.800000	0.133333\\
0.900000	0.066667\\
1.000000	0.000000\\
};
\addlegendentry{ILRS, \intOrder=2};

\addplot [style=ILRSs5]
  table[row sep=crcr]{%
0.000000	0.833333\\
0.100000	0.750000\\
0.200000	0.666667\\
0.300000	0.583333\\
0.400000	0.500000\\
0.500000	0.416667\\
0.600000	0.333333\\
0.700000	0.250000\\
0.800000	0.166667\\
0.900000	0.083333\\
1.000000	0.000000\\
};
\addlegendentry{ILRS, \intOrder=5};

\addplot [style=ILRSs10]
  table[row sep=crcr]{%
0.000000	0.909091\\
0.100000	0.818182\\
0.200000	0.727273\\
0.300000	0.636364\\
0.400000	0.545455\\
0.500000	0.454545\\
0.600000	0.363636\\
0.700000	0.272727\\
0.800000	0.181818\\
0.900000	0.090909\\
1.000000	0.000000\\
};
\addlegendentry{ILRS, \intOrder=10};

\end{axis}
\end{tikzpicture}%
\caption{Normalized decoding radius $\tau$ of \ac{ILRS} codes over the code rate $R$ for  interleaving orders $\intOrder\in\{2,5,10\}$.} 
\label{fig:decRadiusILRS}
\end{figure}
The simulations results in Section~\ref{subsec:sim_results_ILRS} show that the upper bound on the decoding failure probability in~\eqref{eq:upper_bound_Pfail_ILRS} gives a good estimate on the expected success probability of the probabilistic unique decoder.

The interpolation-based probabilistic unique decoding scheme for \ac{ILRS} codes is illustrated in Example~\ref{ex:int_based_ILRS}.

\begin{example}[Interpolation-Based Decoding]\label{ex:int_based_ILRS}
  Consider the code $\intLinRS{\vecbeta,\a,\shots,\intOrder;\nVec,k}$, the codeword $\C(\f)$ from~\eqref{eq:ilrs_ex_codeword} and received word $\R$ from~\eqref{eq:ilrs_ex_recword} considered in Example~\ref{ex:LO_ILRS}.
  The interpolation points corresponding to the $n=6$ evaluation maps $\mathscr{E}_1^{(1)},\dots,\mathscr{E}_3^{(2)}$ are the columns of the matrix
  \begin{equation*}
    \begin{pmatrix}
      \vecbeta 
      \\
      \R
    \end{pmatrix}
    =
    \left(
    \begin{array}{ccc|ccc}
      1 & \pe & \pe^2 & 1 & \pe & \pe^2
      \\ 
      2\pe^2 & 2\pe^2+2\pe+2 & \pe^2+\pe+1 & 2\pe^2 & 2\pe+1 & 2\pe^2+\pe+2
      \\ 
      \pe+1 & \pe+2 & 2\pe^2+\pe+2 & \pe+1 & \pe^2+\pe+2 & 2\pe^2+2
    \end{array}
    \right).
  \end{equation*}
  First, we compute $\SkewPolyringZeroDer$-linearly independent polynomials of minimal $\w=(0,2,2)$-weighted degree that span the solution space of Problem~\ref{prob:skewIntProblemGenOpILRS} as
  \begin{align*}
    Q^{(1)} 
    \!&=\! \left((\pe^2 + 2\pe)x^2 + (2\pe^2 + \pe + 2)x + \pe^2 + \pe + 2\right) 
    + \left(x + 2\pe^2 + 2\pe + 1\right)y_1
    + \left(2\pe^2 + \pe\right)y_2
    \\
    Q^{(2)} 
    \!&=\! \left(2x^3 + (2\pe^2 + 2\pe + 1)x^2 + \pe^2x + 2\right) 
    + \left((2\pe^2 + 2)x + 2\pe\right)y_1
    + \left(x + 2\pe^2 + \pe + 1\right)y_2
  \end{align*}
  using e.g. the skew Kötter interpolation from~\cite{liu2014kotter}.
  Since the $\w$-weighted degree of the $\intOrder=2$ polynomials $Q^{(1)}$ and $Q^{(2)}$ is less than $\degConstraint=4$, our decoding problem has a unique solution (cf. Proposition~\ref{prop:RFproperties}).

  Next, using the coefficients of $Q^{(1)}$ and $Q^{(2)}$ we set up the root-finding matrix $\RFmat$ as (see~\eqref{eq:rootFindingMatrixILRS}) 
  \begin{equation}
  \RFmat = 
    \small
    \left(\begin{array}{rrrrrr}
    2 \pe^{2} + 2 \pe + 1 & 2 \pe^{2} + \pe & 0 & 0 & 0 & 0 \\
    2 \pe & 2 \pe^{2} + \pe + 1 & 0 & 0 & 0 & 0 \\
    1 & 0 & 2 \pe^{2} + 2 & 2 \pe^{2} + 2 \pe & 0 & 0 \\
    2 \pe^{2} + \pe + 1 & 1 & 2 \pe + 2 & 2 \pe^{2} + 2 \pe + 1 & 0 & 0 \\
    0 & 0 & 1 & 0 & 2 \pe^{2} + \pe + 1 & 2 \pe^{2} + 1 \\
    0 & 0 & 2 \pe^{2} + 2 \pe + 1 & 1 & 2 \pe + 1 & 2 \pe^{2} + 2 \\
    0 & 0 & 0 & 0 & 1 & 0 \\
    0 & 0 & 0 & 0 & 2 \pe^{2} + 2 & 1
    \end{array}\right)
  \end{equation}
  and the vector $\q_0$ as (see~\eqref{eq:RFvectorsILRS})
  \begin{equation}
    \q_0 = \left(\pe^{2} + \pe + 2,\,2,\,2 \pe^{2} + 2 \pe + 2,\,\pe^{2} + 2 \pe + 1,\,\pe^{2} + 2,\,2 \pe^{2} + \pe + 1,\,0,\,2\right)^\top.
  \end{equation}
  The unique solution of the $\Fqm$-linear root-finding system $\RFmat\cdot\RFvec=-\vec{q}_0$ in~\eqref{eq:rootFindingSystemFqmILRS} is
  \begin{align*}
    \RFvec 
    \defeq& \left(f_0^{(1)}, f_0^{(2)}, \aut^{-1}(f_1^{(1)}), \aut^{-1}(f_1^{(2)}), \aut^{-2}(f_2^{(1)}), \aut^{-2}(f_2^{(2)})\right) 
    \\
    =& \left(2 \pe^{2},\,\pe^{2},\,0,\,2 \pe^{2} + 2 \pe,\,0,\,1\right)^\top.
  \end{align*}
  Considering the structure of $\RFvec$ (cf.~\eqref{eq:RFvectorsILRS}) we can recover the message polynomials
  \begin{alignat}{2}
    f_1(x) &= 2\pe^2 + \aut(0)x + \aut^2(0)x^2 &&= 2\pe^2
    \\ 
    f_2(x) &= \pe^2 + \aut(2\pe^2+2\pe)x + \aut^2(1)x^2 &&= \pe^2 + (2\pe^2+\pe)x + x^2
  \end{alignat}
  which correspond to the transmitted codeword $\C(\f)$.
\end{example}

\subsection{Comparison to Previous Work and Simulation Results}\label{subsec:sim_results_ILRS}

In order to evaluate the upper bound on the decoding failure probability in~\eqref{eq:upper_bound_Pfail_ILRS} we performed a Monte Carlo simulation ($100$ errors) of a code $\intLinRS{\vecbeta,\a,\shots=2,\intOrder=4;\nVec=(4,4),k=3}$ over $\mathbb{F}_{3^4}$ over a sum-rank channel~\eqref{eq:sum_rank_channel}, where the error matrices were chosen uniformly at random from the set of all error matrices of sum-rank $t\in\{3,4\}$.

The channel realization is chosen uniformly at random from all possible realizations of the sum-rank channel with exactly this number of weight $t$ errors. For the implementation of the channel model the procedure from~\cite[Appendix~A]{puchinger2020generic} was used, which is a variant of enumerative coding~\cite{cover1973enumerative}.

The results in Figure~\ref{fig:simILRS} show, that the upper bound in~\eqref{eq:upper_bound_Pfail_ILRS} gives a good estimate of the actual decoding failure probability of the decoder.
\begin{figure}[ht!]
\centering
\definecolor{mycolor1}{rgb}{1.00000,0.00000,1.00000}%
\definecolor{mycolor2}{rgb}{0.00000,1.00000,1.00000}%

\newcommand{\setthr}[2]{\draw[thin,black] (axis cs:#1,1e-12) -- (axis cs:#1,3e-12);\node at (axis cs: #1,5e-12) {\tiny \textcolor{black}{#2}};}

\begin{tikzpicture}

\begin{axis}[%
xmin=1.5,
xmax=4.5,
xtick={1,2,3,4,5},
xlabel={Sum-rank error weight $t$},
compat=newest,
xmajorgrids,
ymode=log,
ymin=1e-12,
ymax=1,
yminorticks=false,
label style={anchor=near ticklabel, font=\footnotesize},
label style={inner sep=0}, 
ylabel={Decoding failure probability $P_{f}$},
ymajorgrids,
yminorgrids,
tick label style={font=\scriptsize},
legend style={at={(0.01,0.989)},anchor=north west,legend cell align=left,align=left,draw=white!15!black, font=\footnotesize},
mystyle]

\addplot [color=red!80!black,
solid,
mark=x,
mark options={solid}
]
table[row sep=crcr]{%
4 7.021496e-02\\
3 2.013745e-11\\  
};
\addlegendentry{Upper bound on $P_f$ \eqref{eq:upper_bound_Pfail_ILRS}};

\addplot [color=cyan!80!black,dotted,
mark=o,
mark options={solid}
]
table[row sep=crcr]{%
4	2.203704e-02\\ 
3	6.320160e-12\\ 
};
\addlegendentry{Heuristic upper bound on $P_f$ \eqref{eq:heur_upper_bound_Psucc_ILRS}};


\addplot [color=blue,
dashed,
mark=asterisk,
mark options={solid}
]
table[row sep=crcr]{%
4 1.438021e-02\\
};
\addlegendentry{Simulation};

\setthr{2}{BMD}
\setthr{4}{$\tmax$}

\end{axis}
\end{tikzpicture}%
\caption{Result of a Monte Carlo simulation of the code $\intLinRS{\vecbeta,\a,\shots=2,\intOrder=4;n=8,k=3}$ over $\F_{3^4}$ transmitted over a sum-rank error channel with overall $t\in\{3,4\}$.}
\label{fig:simILRS}
\end{figure}  
For the same parameters a (non-interleaved) linearized Reed--Solomon code~\cite{martinez2019reliable} (i.e. $\intOrder=1$) can only correct errors of sum-rank weight up to $t=2$ (\ac{BMD} radius).

\subsection{Applications to Decoding Errors in the Skew Metric}\label{subsec:ilrs_skew}

The sum-rank metric is closely related to the skew metric, also defined in~\cite{martinez2018skew}.
In particular, there exists an isometry between the sum-rank metric and the skew metric~\cite{martinez2018skew}.
We now show how the isometry from~\cite[Theorem~9]{martinez2019reliable} can be modified in order to use \ac{ILRS} codes for decoding errors beyond the unique decoding radius in the skew metric.

Consider an \ac{ILRS} code $\intLinRS{\vecbeta,\a,\shots,\intOrder;\nVec,k}$ and define the vectors
\begin{equation}
    \vecbeta^{-1}\defeq\left((\beta_1^{(1)})^{-1},(\beta_2^{(1)})^{-1},\allowbreak\dots,\allowbreak(\beta_{n_\shots}^{(\shots)})^{-1}\right)
\end{equation} 
and 
\begin{equation}
 \vec{b}\defeq\left(\op{a_1}{\vecbeta^{(1)}}\mid\op{a_2}{\vecbeta^{(2)}}\mid\dots\mid\op{a_\shots}{\vecbeta^{(\shots)}} \right)\cdot\diag(\vecbeta^{-1}).
\end{equation}

Then the \emph{skew weight} of a vector $\vec{\x}=\left(\shot{\x}{1} \mid \shot{\x}{2} \mid \dots \mid \shot{\x}{\shots}\right) \in \Fqm^n$ with $\shot{\x}{i} \in \Fqm^{n_i}$ for all $i=1,\dots,\shots$ is defined as (see~\cite{boucher2019algorithm})
\begin{equation}\label{eq:def_skew_weight}
 \SkewWeight(\x) \defeq \deg\left(\lclm\left(x-\op{b_j^{(i)}}{x_j^{(i)}}\left(x_j^{(i)}\right)^{-1}\right)_{x_j^{(i)} \neq 0}\right).
\end{equation}

By fixing a basis of $\Fqms$ over $\Fqm$ we can consider a matrix $\X\in\Fqm^{\intOrder\times n}$ as a vector $\vec{x}=(x_1,x_2,\dots,x_n)\in\Fqms^n$. 
The skew weight of a matrix $\X\in\Fqm^{\intOrder\times n}$ with respect to $\set{B}$ is then as the skew weight of the vector $\x\in\Fqms^n$, i.e. as (see~\eqref{eq:def_skew_weight})
\begin{equation}\label{eq:def_skew_weight_mat}
 \SkewWeight(\X) \defeq \deg\left(\lclm\left(x-\op{b_j^{(i)}}{x_j^{(i)}}\left(x_j^{(i)}\right)^{-1}\right)_{x_j^{(i)} \neq 0}\right)
\end{equation}
where the polynomial on the right-hand side is now from $\Fqms[\aut;x]$ since we have that $x_i\in\Fqms$ for all $i=1,\dots,n$.

Proposition~\ref{prop:isometry} summarizes the isometry between the sum-rank metric and the skew metric for interleaved matrices, which directly follows from~\cite[Theorem~9]{martinez2019reliable} and~\eqref{eq:def_skew_weight_mat}.

\begin{proposition}[Isometry Between the Sum-Rank and the Skew Metric]\label{prop:isometry}
 Let $$\vecbeta=\left(\shot{\vecbeta}{1} \mid \shot{\vecbeta}{2} \mid \dots \mid \shot{\vecbeta}{\shots}\right) \in \Fqm^n$$ be a vector with $\SumRankWeight(\vecbeta) = n$ and define the vectors
 \begin{equation}
    \vecbeta^{-1}\defeq\left((\beta_1^{(1)})^{-1},(\beta_2^{(1)})^{-1},\allowbreak\dots,\allowbreak(\beta_{n_\shots}^{(\shots)})^{-1}\right).
 \end{equation} 
 Then for a matrix $\X=\left(\shot{\X}{1} \mid \shot{\X}{2} \mid \dots \mid \shot{\X}{\shots}\right)$ we have that 
 \begin{equation}
  \SkewWeight(\X \cdot \diag\left(\vecbeta^{-1})\right) = \SumRankWeight(\X).
 \end{equation}
\end{proposition}

Using the results from above we can define $\intOrder$-interleaved skew Reed--Solomon \acs{ISRS} codes as
\begin{equation}
  \intSkewRS{\vec{b},\shots,\intOrder;n,k} \defeq \left\{\mat{C}\cdot\diag(\vecbeta^{-1}):\mat{C}\in\intLinRS{\vecbeta,\a,\shots,\intOrder;n,k}\right\}.
\end{equation}

Now suppose we transmit a codeword $\C \in \intSkewRS{\vec{b},\intOrder;n,k}$ over a skew error channel
\begin{equation}
  \R = \C+\E
\end{equation}
where $\E\in\Fqm^{\intOrder\times n}$ has skew weight $t=\SkewWeight(\E)$.
Then the decoding schemes from Section~\ref{sec:decodingILRS} can be used to decode \ac{ISRS} codes as follows:
\begin{enumerate}
 \item Compute $\R'=\R \cdot \diag\left(\vecbeta^{-1}\right)$. This step requires $\oh{\intOrder n}$ operations in $\Fqm$.
 \item Use the list or probabilistic unique decoders from Section~\ref{sec:ILRS} to decode errors of skew weight up to $t \leq \frac{\intOrder}{\intOrder+1}(n-k)$ requiring at most $\softoh{\intOrder^\omega \OMul{n}}$ operations in $\Fqm$.
\end{enumerate}

\section{Fast Interpolation via Minimal Approximant Bases}\label{sec:fast_interpolation}

In this section we show how to speed up the above described decoding schemes for \ac{ILRS} codes by reducing the core computation, namely the \emph{interpolation} aproblem, to a \emph{minimal approximant basis} computation of matrices over the relevant skew polynomial ring.
This work continues the speed-ups obtained for several code families in the rank, sum-rank and subspace metric in~\cite{bartz2020fast}.
In particular, we generalize the vector operator interpolation problem~\cite[Problem~13]{bartz2020fast} to the \emph{generalized} operator evaluation and use the ideas of~\cite[Algorithm~6]{bartz2020fast} to derive a new interpolation algorithm to solve it efficiently via fast minimal approximant bases computations.
By using the relation between the remainder evaluation and the generalized operator evaluation (see~\cite{martinez2018skew,leroy1995pseudolinear}), the proposed algorithm can be used to solve the multi-dimensional generalization of the two-dimensional vector remainder interpolation problem~\cite[Problem~27]{bartz2020fast}.

Since the derivation of the fast minimal approximant bases interpolation is rather technical and not at the core of the paper, we only provide the final complexity result in Corollary~\ref{cor:complexity_interpolation}. 
For details on the derivation and the actual algorithm (Algorithm~\ref{alg:fast_interpolation}) the reader is referred to Appendix~\ref{app:proofs_fast_interpolation}.

\begin{corollary}[Complexity of Interpolation Problems]\label{cor:complexity_interpolation}
 Algorithm~\ref{alg:fast_interpolation} can solve the interpolation problem Problem~\ref{prob:skewIntProblemGenOpILRS} in at most $\softoh{\intOrder^\omega \OMul{n}}$ operations in $\Fq$.
\end{corollary}

\section{Conclusion}\label{sec:conclusion}

\subsection{Summary}\label{subsec:generality_remarks}
 We considered $\intOrder$-\acf{ILRS} codes and showed, that they are capable of correcting errors beyond the unique decoding radius in the sum-rank metric.
 We proposed an efficient interpolation-based decoding scheme for \ac{ILRS} codes, which can be used as a list decoder or as a probabilistic unique decoder and can correct errors of sum-rank up to $t\leq\frac{\intOrder}{\intOrder+1}(n-k+1)$ and $t\leq\frac{\intOrder}{\intOrder+1}(n-k)$, respectively, where $\intOrder$ is the interleaving order, $n$ the length and $k$ the dimension of the code.
 We derived an \ac{LOlike} decoder for \ac{ILRS} codes, which provides arguments to upper bound on the decoding failure probability for the interpolation-based probabilistic unique decoder.

 By using the isometry between the sum-rank an the skew metric we defined \ac{ISRS} codes and showed how to use the proposed decoding schemes for correcting errors in the skew metric.

 Up to our knowledge, the proposed decoding schemes are the first being able to correct errors beyond the unique decoding region in the sum-rank and the skew metric efficiently.
 
 We presented an efficient minimal approximant bases interpolation algorithm, that allows to implement the interpolation-based decoding scheme for \ac{ILRS} and \ac{ISRS} codes requiring at most $\softoh{\intOrder^\omega\OMul{n}}$ operations in $\Fqm$, where $\OMul{n}$ is the cost (in operations in $\Fqm$) of multiplying two skew-polynomials of degree at most $n$ and $\omega$ is the matrix multiplication exponent.
 As a result, we obtained the currently fastest known decoding algorithms in the sum-rank and the skew metric.

\subsection{Remarks on Generality}

For the sake of simplicity, we considered codes constructed by skew polynomials with zero derivations, i.e. polynomials from~$\SkewPolyringZeroDer$, only.
All considered decoding algorithms as well as the isometry between the (burst) sum-rank and skew metric work as well over $\SkewPolyring$.

In order to not further complicate the quite involved notation we considered decoding of \emph{homogeneous} \ac{ILRS} and~\ac{ISRS} codes, respectively, i.e. interleaved codes where the component codes have the same code dimension.
All decoding schemes proposed in this paper can be generalized to \emph{heterogeneous} interleaved codes, where the component codes may have a different dimensions $k_1,\dots,k_\intOrder$, in a straight-forward manner like e.g. in~\cite{wachter2013decoding,bartz2017algebraic}.
The resulting decoding regions are then $t<\frac{\intOrder}{\intOrder+1}(n-\bar{k}+1)$ for list decoding and $t\leq\frac{\intOrder}{\intOrder+1}(n-\bar{k})$ for probabilistic unique decoding where $\bar{k}\defeq\frac{1}{\intOrder}\sum_{l=1}^{\intOrder}k_l$.

\subsection{Outlook \& Future Work}

For future work it would be interesting to see how the results generalize for codes and decoder over (particular) rings.

It would also be interesting to generalize further interpolation based decoding schemes for rank-metric codes, such as e.g. the interpolation-based decoder for nonlinear rank-metric codes~\cite{li2019interpolation}, the decoder in~\cite{kadir2021interpolation}, the decoder for additive generalized twisted Gabidulin codes~\cite{kadir2020decoding} and the decoder for several optimal rank-metric codes from~\cite{kadir2022encoding} to the interleaved sum-rank-metric code setting.

Another interesting direction of future work could be to consider decoding of interleaved variants of the codes from~\cite{martinez2020general}, which can be constructed using smaller field sizes.

\section*{Acknowledgements}

H.~Bartz acknowledges the financial support by the Federal Ministry of Education and Research of Germany in the programme of ``Souverän. Digital. Vernetzt.'' Joint project 6G-RIC, project identification number: 16KISK022.

S.~Puchinger was with the Department of Applied Mathematics and Computer Science, Technical University of Denmark (DTU), Lyngby, Denmark and the Department of Electrical and Computer Engineering, Technical University of Munich, Munich, Germany. 
Within this period he was supported by the European Union’s Horizon 2020 research and innovation program under the Marie Sklodowska-Curie grant agreement no. 713683 and by the European Research Council (ERC) under the European Union’s Horizon 2020 research and innovation programme (grant agreement no. 801434).

\bibliographystyle{AIMS} 
\providecommand{\href}[2]{#2}
\providecommand{\arxiv}[1]{\href{http://arxiv.org/abs/#1}{arXiv:#1}}
\providecommand{\url}[1]{\texttt{#1}}
\providecommand{\urlprefix}{URL }

\newpage

\appendix

\section{Proofs and statements from Section~\ref{sec:ILRS}}\label{app:proofs_ILRS}

\subsection{Proof of Lemma~\ref{lem:properties_LODecMat_ILRS}}

\begin{proof}

\begin{itemize}
  \item Ad \ref{itm:transformed_LO_matrix}): For the proof of first statement we use the fact, that the $\Fqm$-row space of $\lambda_{n-t-1}(\vecbeta)_\a$ forms an \ac{LRS} code of length $n$ and dimension $n-t-1$.
  Since any codeword $\c_i$ is in the row space of $\lambda_k(\vecbeta)_\a$, we have (by \eqref{eq:rho_i+j_property})
  \begin{align*}
    \opexp{\a}{\c_i}{j} \in \Rowspace{\lambda_{k+j}(\vecbeta)_\a}
  \end{align*}
  for all $j$. In particular,
  \begin{align*}
  \Rowspace{\lambda_{n-t-k}(\c_i)_\a} \subseteq \Rowspace{\lambda_{n-t-1}(\vecbeta)_\a},
  \end{align*}
  so by elementary row operations, we have
  \begin{align*}
  \LODecMat 
  =
  \begin{pmatrix}
  \lambda_{n-t-1}(\vecbeta)_{\a} \\
  \lambda_{n-t-k}\!\left(\c_1\right)_{\a} + \lambda_{n-t-k}\!\left(\e_1\right)_{\a} \\
  \vdots \\
  \lambda_{n-t-k}\!\left(\c_\intOrder\right)_{\a} + \lambda_{n-t-k}\!\left(\e_\intOrder\right)_{\a} \\
  \end{pmatrix}
  \overset{\text{row op.}}{\sim}
  \begin{pmatrix}
  \lambda_{n-t-1}(\vecbeta)_{\a} \\
  \lambda_{n-t-k}\!\left(\e_1\right)_{\a} \\
  \vdots \\
  \lambda_{n-t-k}\!\left(\e_\intOrder\right)_{\a} \\
  \end{pmatrix}
  =
  \begin{pmatrix}
  \lambda_{n-t-1}(\vecbeta)_{\a} \\
  \lambda_{n-t-k}(\E)_{\a}
  \end{pmatrix}.
  \end{align*}

  \item Ad \ref{itm:LO_lemma_non-zero_colums_of_EU}):
  Since the $\Fq$-rank partition of $\E$ is $\t$, there are invertible matrices $\W^{(i)} \in \Fq^{n_i \times n_i}$ such that the rightmost $n_i-t_i$ columns of
  \begin{align*}
  \E^{(i)} \W^{(i)} \in \Fqm^{\intOrder \times n_i}
  \end{align*}
  are zero. This implies that also the rightmost $n_i-t_i$ columns of
  \begin{align*}
  \lambda_{n-t-k}\left(\E^{(i)}\right)_{a_i} \W^{(i)} \in \Fqm^{(n-t-k)\intOrder \times n_i}
  \end{align*}
  are zero. Since the $\Fqm$-rank of $\lambda_{n-t-k}(\E)_{\a}$ is $t$ and since
  \begin{align*}
  \lambda_{n-t-k}(\E)_{\a} \cdot \diag\!\left(\W^{(1)},\dots, \W^{(\ell)}\right)
  \end{align*}
  has exactly $t$ non-zero columns, these non-zero columns are $\Fqm$-linearly independent (i.e., the $t_i$ leftmost columns in each block).
  This implies \ref{itm:LO_lemma_non-zero_colums_of_EU}).

  \item Ad \ref{itm:LO_lemma_rank_L}):  
  Since, by \ref{itm:LO_lemma_non-zero_colums_of_EU}), $\lambda_{n-t-k}(\E)_{\a} \cdot \diag\!\left(\W^{(1)},\dots, \W^{(\ell)}\right)$ has exactly $t$ non-zero columns, the $\Fqm$-rank of $\LODecMat$, which, by \ref{itm:transformed_LO_matrix}), equals the rank of
  \begin{align*}
  \begin{pmatrix}
  \lambda_{n-t-1}(\vecbeta)_{\a} \\
  \lambda_{n-t-k}(\E)_{\a}
  \end{pmatrix} 
  \cdot
  \diag\!\left(\W^{(1)},\dots, \W^{(\ell)}\right),
  \end{align*}
  is given by $t$ plus the rank of the matrix $\B \in \Fqm^{(n-t-1) \times n-t}$ consisting of the columns of $\lambda_{n-t-1}(\vecbeta)_{\a} \cdot \diag\!\left(\W^{(1)},\dots, \W^{(\ell)}\right)$ in which 
  $\lambda_{n-t-k}(\E)_{\a} \cdot \diag\!\left(\W^{(1)},\dots, \W^{(\ell)}\right)$ is non-zero (i.e., the rightmost $n_i-t_i$ columns in block $i$).
  This can be easily seen by permuting the columns, such that the matrix is in block-triangular form 
  \begin{align*}
  \LODecMat \underset{\text{$\Fqm$-lin. row op.}}{\overset{\text{$\Fq$-lin. col. op.}}{\sim}}
  \begin{pmatrix}
  \lambda_{n-t-1}(\vecbeta)_{\a} \\
  \lambda_{n-t-k}(\E)_{\a}
  \end{pmatrix} 
  \cdot
  \diag\!\left(\W^{(1)},\dots, \W^{(\ell)}\right)
  \overset{\text{col. permu.}}{\sim}
  \begin{pmatrix}
  \star & \B \\
  \tilde{\E} & \0
  \end{pmatrix},
  \end{align*}
  where $\tilde{\E} \in \Fqm^{(n-t-k)\intOrder \times t}$ are the non-zero columns of $$\lambda_{n-t-k}(\E)_{\a} \cdot \diag\!\left(\W^{(1)},\dots, \W^{(\ell)}\right)$$ (note that $\rk_{q^m}(\tilde{\E}) = t$).
  Since
  \begin{align*}
  \lambda_{n-t-1}(\vecbeta)_{\a} \cdot \diag\!\left(\W^{(1)},\dots, \W^{(\ell)}\right)
  \end{align*}
  is a generator matrix of an $[n,n-t-1]$ LRS code (which is \ac{MDS} in the Hamming metric), $\B$ has rank $n-t-1$. Hence, the overall rank of $\LODecMat$ is $t+(n-t-1)=n-1$.

\end{itemize}
\end{proof}

\subsection{Proof of Lemma~\ref{lem:LO_correctness_lemma}}

\begin{proof}
\hfill
\begin{itemize}
  \item Ad \ref{itm:LO_lemma_rank_of_kernel_element}):
  Due to \ref{itm:LO_lemma_non-zero_colums_of_EU}) in Lemma~\ref{lem:properties_LODecMat_ILRS}, any vector in the right kernel of
  $\lambda_{n-t-k}(\E)_{\a} \cdot \diag\!\left(\W^{(1)},\dots, \W^{(\ell)}\right)$
  must be zero in the first $t_i$ positions of the $i$-th block, for every $i$. In particular, the leftmost $t_i$ positions of
  \begin{align*}
  \h^{(i)} \left({\W^{(1)}}^{-1}\right)^\top \in \Fqm^{n_i}
  \end{align*}
  are zero. This implies that
  \begin{align*}
  \rk_{q}\!\left(\h^{(i)}\right) \leq n_i - t_i.
  \end{align*}
  On the other hand, $\h$ is in the right kernel of the matrix $\lambda_{n-t-1}(\vecbeta)_{\a}$, which is a generator matrix of an $[n,n-t-1]$ LRS code. This code is MSRD, and hence its dual code has parameters $[n,t+1,n-t]$ (i.e., it is also MSRD). Since $\h$ is non-zero, its sum-rank weight must therefore be at least $n-t$. This can only be the case for
  \begin{align*}
  \rk_{q}\!\left(\h^{(i)}\right) = n_i - t_i.
  \end{align*}

  \item Ad \ref{itm:LO_lemma_Ti_matrices}): Expand $\h^{(i)}$ into an $m \times n_i$ matrix over $\Fq$, which has $\Fq$-rank $n_i-t_i$ by \ref{itm:LO_lemma_rank_of_kernel_element}). Then, we can perform elementary column operations on this matrix to bring it into reduced column echelon form, where the $n_i-t_i$ non-zero columns are the rightmost ones. The matrix $\T^{(i)}$ is then chosen to be the matrix that, by multiplication from the right, performs the used sequence of elementary column operations. Note that the $n_i-t_i$ non-zero entries of $\h^{(i)}\T^{(i)}$ are linearly independent over $\Fq$.
  
  \item Ad \ref{itm:LO_lemma_EDi_zero}): Consider the matrix
  \begin{align*}
  \A = (\A^{(1)} \mid \A^{(2)} \mid \dots \mid \A^{(\shots)})
  \in \Fqm^{\intOrder(n-k-t) \times (n-t)}
  \end{align*}
  and vector 
  \begin{align*}
  \b = (\b^{(1)} \mid \b^{(2)} \mid \dots \mid \b^{(\shots)})
  \in \Fqm^{n-t} \ ,
  \end{align*}
  where
  \begin{align*}
  \A^{(i)} &:= \left[\lambda_{n-t-k}(\E^{(i)})_{a_i} \D^{(i)}\right]_{\{t_i+1,\dots,n_i\}} \ , \\
  \b^{(i)} &:= \left[\h^{(i)} \T^{(i)}\right]_{\{t_i+1,\dots,n_i\}} \ .
  \end{align*}
  Since $\h \cdot \diag\!\left(\T^{(1)},\dots, \T^{(\ell)}\right)$ is in the right kernel of $$\lambda_{n-t-k}(\E)_{\a} \cdot \diag\!\left(\D^{(1)},\dots, \D^{(\ell)}\right)$$ and the $t_i$ leftmost positions of $\h^{(i)} \T^{(i)}$ are zero, the vector $\b$ is in the right kernel of $\A$.
  We prove that $\A$ is the zero matrix.
  
  Let $\that := \SumRankWeight(\A)$ and $\ttilde := \rk_{q^m}(\A)$.
  Since $$\SumRankWeight(\lambda_{n-t-k}(\E)_{\a}) = \rk_{q^m}(\lambda_{n-t-k}(\E)_{\a}) = t$$ and the columns of $\A$ are $\Fq$-linear combinations of the columns of $\lambda_{n-t-k}(\E)_{\a}$, we must have $\that = \ttilde$.
  Hence, there are invertible matrices $\V^{(i)} \in \Fq^{(n_i-t_i) \times (n_i-t_i)}$ such that
  \begin{align*}
  \A \cdot \diag\!\left( \V^{(1)}, \dots, \V^{(\ell)}\right)
  \end{align*}
  has exactly $\ttilde$ non-zero columns, say $\mathcal{A} \subseteq \{1,\dots,n-t\}$, which are $\Fqm$-linearly independent. Hence, the vector
  \begin{align*}
  \b \cdot \diag\!\left( \left({\V^{(1)}}^{-1}\right)^\top, \dots, \left({\V^{(\ell)}}^{-1}\right)^\top\right)
  \end{align*}
  is zero in all positions in $\mathcal{A}$. Since, by construction, the entries of $\b^{(i)}$ are linearly independent, we must have $\mathcal{A} = \emptyset$, $\ttilde=0$, and hence $\A = \0$. This proves \ref{itm:LO_lemma_EDi_zero}).
  
  \item Ad \ref{itm:LO_lemma_retrieve_error}): Consider the transformed and punctured received word $\Rtilde=(\Rtilde^{(1)} \mid \Rtilde^{(2)} \mid \dots \mid \Rtilde^{(\shots)})$
  defined by
  \begin{align*}
  \Rtilde^{(i)} := \left[ \R^{(i)} \D^{(i)} \right]_{\{t_i+1,\dots,n_i\}} \overset{\text{\ref{itm:LO_lemma_EDi_zero})}}{=} \left[ \C^{(i)} \D^{(i)} \right]_{\{t_i+1,\dots,n_i\}}.
  \end{align*}
  Hence, the $j$-th row of $\Rtilde^{(i)}$ can be written as
  \begin{align*}
  \begin{pmatrix}
  \tilde{r}^{(i)}_{j,t_i+1} & \cdots & \tilde{r}^{(i)}_{j,n_i}
  \end{pmatrix}
  &= \left[\c^{(i)}_j\D^{(i)}\right]_{\{t_i+1,\dots,n_i\}} \\
  &= \left[\begin{pmatrix}
  f_j\!\left(\beta^{(i)}_1\right)_{a_i} & \cdots & f_j\!\left(\beta^{(i)}_{n_i}\right)_{a_i}
  \end{pmatrix}
  \D^{(i)}\right]_{\{t_i+1,\dots,n_i\}} \\
  &= \left[\begin{pmatrix}
  f_j\!\left(\tilde{\beta}_1^{(i)}\right)_{a_i} & \cdots & f_j\!\left(\tilde{\beta}_{n_i}^{(i)}\right)_{a_i}
  \end{pmatrix}\right]_{\{t_i+1,\dots,n_i\}},
  \end{align*}
  where in the last equality we used the $\Fq$-linearity of the evaluation map $f_j(\cdot)_{a_i}$ for a fixed $a_i$.
  Hence, we can recover $f_j$ by interpolation as stated in \ref{itm:LO_lemma_retrieve_error}). Note that the $\tilde{\beta}_\mu^{(i)}$ are linearly independent by definition and $\sum_{i=1}^{\ell} (n_i-t_i) = n-t \leq k$, so the interpolation is well-defined.\footnote{In fact, we only need $k$ out of $n-t$ interpolation points, which leads to a more efficient algorithm.}  \hfill 
\end{itemize}
\end{proof}

\subsection{Proof of Lemma~\ref{lem:LO_technical_lemma_for_failure_prob}}

\begin{proof}
First, we show that if there exists a vector $\b$ satisfying~\eqref{eq:LO_lemma_existence_h}, then~\eqref{eq:LO_lemma_Fqm_rank<t} holds.
Let $\hvec$ be as in \eqref{eq:LO_lemma_existence_h} and denote by $\m_i$ the $i$-th row of $\M$.
Then, we can write $\lambda_{n-t-k}^{\aut^{-1}}(\hvec)_{\aut^{-1}(\a)} \M^\top = \0$ equivalently as
\begin{equation}\label{eq:M_mat_row_wise}
  \opexp{\aut^{-1}(\a)}{\hvec}{-j} \m_i^\top = 0,  \qquad \forall \, i=1,\dots,s, \quad \forall \, j=0,\dots,n-t-k-1.
\end{equation}
Applying $\aut^{j}$ to the $j$-th equation in~\eqref{eq:M_mat_row_wise} and using the property from~\cite[Lemma~1]{hormann2024syndrome}, we get
\begin{equation}\label{eq:M_mat_row_wise_row_op}
  \hvec \cdot \opexp{\a}{\m_i}{j}^\top = 0  \quad \forall \, i=1,\dots,s, \quad \forall \, j=0,\dots,n-t-k-1
\end{equation}
or equivalently in vector-matrix form as
\begin{equation*}
  \hvec \cdot \lambda_{n-t-k}(\M)_{\a}^\top = \0.
\end{equation*}
Hence, $\hvec \neq \0$ is in the right kernel of $\lambda_{n-t-k}(\M)_{\a}$, which implies that the $\Fqm$-rank of $\lambda_{n-t-k}(\M)_{\a}$ is smaller than $t$.

Now we show that if~\eqref{eq:LO_lemma_Fqm_rank<t} holds, there exists a vector $\b$ satisfying~\eqref{eq:LO_lemma_existence_h}.
Let $\rk_{q^m}(\lambda_{n-t-k}(\M)_{\a})<t$.
Then, there is a non-zero vector $\hvec \in \Fqm^t$ such that $\hvec^\top$ is in the right kernel of $\lambda_{n-t-k}(\M)_{\a}$. Clearly, all $\Fqm$-multiples of $\hvec^\top$ are also in this right kernel. Thus, we have
\begin{align}
  \big(\myalpha\hvec\big) \lambda_{n-t-k}(\M)_{\a}^\top &= \0, \quad \forall \, \myalpha \in \Fqm \notag \\
  \Rightarrow \quad \myalpha \hvec \opexp{\a}{\m_i}{j}^\top &= 0,
  \quad \forall \, \myalpha \in \Fqm, \quad \forall \, i=1,\dots,s, \quad \forall \, j=0,\dots,n-t-k-1. \label{eq:multiples_b_tilde_row_wise}
\end{align}
Applying $\aut^{-j}$ to the $j$-th equation in~\eqref{eq:multiples_b_tilde_row_wise} and using the property from~\cite[Lemma~1]{hormann2024syndrome}, we get
\begin{equation}\label{eq:multiples_b_tilde_row_wise_row_op}
  \opexp{\aut^{-1}\!(\a)}{\myalpha\hvec}{-j} \m_i^\top = 0, \quad \begin{array}{l} \forall \, \myalpha \in \Fqm, \\ \forall \, i=1,\dots,s, \\ \forall \, j=0,\dots,n-t-k-1,\end{array}
\end{equation}
or equivalently in matrix form
\begin{equation*}
  \lambda_{n-t-k}^{\aut^{-1}}\Big(\myalpha'\hvec\Big)_{\!\aut^{-1}(\a)} \M^\top = \0, \quad \forall \, \myalpha' \in \Fqm.
\end{equation*}

Thus, $\hvec \neq \0$ satisfies the equality in \eqref{eq:LO_lemma_existence_h} and it is left to show that $\hvec$ has sum-rank weight $\SumRankWeight(\hvec)>n-t-k$.

Suppose, towards a contradiction, that $\SumRankWeight(\hvec) = r \leq n-t-k$ and rank partition $\r = (r_1,\dots,r_\ell)$ with $r_i \geq 0$ and $\sum_{i=1}^{\ell} r_i=r$.
Then, there is a block diagonal matrix $\T = \diag(\T_1,\dots,\T_\ell) \in \Fq^{t \times t}$ such that every $\T_i \in \Fq^{t_i \times t_i}$ is invertible and such that in every block $i=1,\dots,\ell$, exactly $r_i$ entries of $\hvec\T$ are non-zero.
It is easy to see that these $r_i$ non-zero entries are $\Fq$-linearly independent (within a block).
We denote the indices of the non-zero positions of $\hvec\T$ in the $i$-th block by $\Tset_i \subseteq \{1,\dots,n\}$ and $\Tset := \cup_{i=1}^\ell \Tset_i$. Note that $\Tset$ is the Hamming support of the entire vector $\hvec\T$.

By Proposition~\ref{prop:rankGenOpVandermonde}, the $r$ columns of $\lambda_{n-t-k}^{\aut^{-1}}(\hvec\T)_{\aut^{-1}(\a)}$ indexed by $\Tset$ have full $\Fqm$-rank, i.e.,
\begin{equation*}
\rk_{q^m} \left[\lambda_{n-t-k}^{\aut^{-1}}(\hvec\T)_{\aut^{-1}(\a)}\right]_{\Tset} = r.
\end{equation*}
Due to
\begin{equation*}
\lambda_{n-t-k}^{\aut^{-1}}(\hvec)_{\aut^{-1}(\a)} \M^\top 
= \lambda_{n-t-k}^{\aut^{-1}}(\hvec \T)_{\aut^{-1}(\a)} \left(\M \left(\T^{-1}\right)^\top\right)^\top= \0
\end{equation*}
and due to the fact that the columns of $\lambda_{n-t-k}^{\aut^{-1}}(\hvec \T)_{\aut^{-1}(\a)}$ indexed by the complement of $\Tset$ are zero, we must have that the columns of $\M (\T^{-1})^\top$ indexed by $\Tset$ are zero. 
Hence, we have
\begin{align*}
\SumRankWeight(\M) =\SumRankWeight\left(\M \left(\T^{-1}\right)^\top\right)^\top \leq t-r < t,
\end{align*}
where the first equality is true since $\left(\T^{-1}\right)^\top$ is a block-diagonal matrix with invertible $(t_i \times t_i)$ matrices over $\Fq$ on the diagonal, and the last inequality holds since $\hvec \neq \0$, so $r>0$. This is a contradiction to the assumption $\SumRankWeight(\M) = t$.

Overall, $\hvec$ fulfills both properties in \eqref{eq:LO_lemma_existence_h}, which concludes the proof.
\end{proof} 

\subsection{Proof of Lemma~\ref{lem:LO_probability_full_Fqm_rank}}
\begin{proof}
Let $\t$ be the rank partition of $\E$.
Then there are invertible matrices $\T^{(i)} \in \Fq^{n_i \times n_i}$ such that $\E^{(i)}\T^{(i)}$ has exactly $t_i$ non-zero entries.
Denote by $\M \in \Fqm^{\intOrder \times t}$ the non-zero columns of $\E\cdot\diag(\T^{(1)}, \dots, \T^{(\ell)})$. By construction, we have
\begin{equation}
\rk_{q^m} \left(\lambda_{n-t-k}(\E)_{\a}\right) = \rk_{q^m}\left(\lambda_{n-t-k}(\M)_{\a} \right). \label{eq:LO_relation_M_E}
\end{equation}
It is readily seen that drawing $\E$ uniformly at random with a fixed rank partition $\t$, i.e.,
\begin{equation*}
\E \sample \Eset^{(\t)} := \left\{\E \in \Fqm^{\intOrder \times n} \, : \, \text{$\E$ has rank partition $\t$} \right\}
\end{equation*}
results in $\M$ being drawn uniformly at random from the set
\begin{align*}
\M \sample \Mset^{(\t)} := \left\{\M \in \Fqm^{\intOrder \times t} \, : \, \SumRankWeight(\M) = t\right\}.
\end{align*}
By \eqref{eq:LO_relation_M_E} and Lemma~\ref{lem:LO_technical_lemma_for_failure_prob}, with $\E$ and $\M$ drawn as above, we have
\begin{align*}
\Pr\!\left(\rk_{q^m}(\lambda_{n-t-k}(\E)_{\a})<t\right) &= \Pr\!\left(\rk_{q^m}\left(\lambda_{n-t-k}(\M)_{\a} \right) < t \right) \\
&= \Pr\!\left(\text{\eqref{eq:LO_lemma_existence_h} is satisfied for $\M$} \right).
\end{align*}
We upper-bound the latter probability.
First, let $\hvec \in \Fqm^t$ be fixed with $\SumRankWeight(\hvec)>n-t-k$.
We count the number of matrices $\M$ with
\begin{align}
\lambda_{n-t-k}^{\aut^{-1}}(\hvec)_{\aut^{-1}(\a)} \M^\top = \0. \label{eq:LO_M_condition}
\end{align} 
Note that $\lambda_{n-t-k}^{\aut^{-1}}(\hvec)_{\aut^{-1}(\a)} \in \Fqm^{(n-t-k) \times t}$, and, by Proposition~\ref{prop:rankGenOpVandermonde} and $\SumRankWeight(\hvec)>n-t-k$, we have
\begin{equation*}
\rk_{q^m}\!\left( \lambda_{n-t-k}^{\aut^{-1}}(\hvec)_{\aut^{-1}(\a)} \right) = n-t-k.
\end{equation*}
Hence, the right kernel of $\lambda_{n-t-k}^{\aut^{-1}}(\hvec)_{\aut^{-1}(\a)}$ has cardinality $(q^m)^{t-(n-t-k)} = q^{m(2t-n+k)}$, and there are at most $q^{m\intOrder(2t-n+k)}$ many matrices $\M$ satisfying \eqref{eq:LO_M_condition}.
On the other hand, we have
\begin{align*}
\left| \left\{\M \in \Fqm^{\intOrder \times t} \, : \, \SumRankWeight(\M) = t\right\} \right| &= \prod_{i=1}^{\ell} \left|\left\{ \M^{(i)} \in \Fqm^{\intOrder \times t_i} \, : \, \rk_q\!\left(\M^{(i)}\right) = t_i \right\}\right| \\
&\overset{(\ast)}{\geq} \prod_{i=1}^{\ell} q^{\intOrder m t_i}\gammaq^{-1} = q^{\intOrder m t}\gammaq^{-\ell},
\end{align*}
where in $(\ast)$ we use \cite[Lemma~3.13]{overbeck2007public}.
In summary, the probability that \eqref{eq:LO_M_condition} is satisfied for a specific $\hvec$ is upper-bounded by
\begin{align}
\Pr\!\left(\text{$\M$ satisfies \eqref{eq:LO_M_condition} for a specific $\hvec$} \right) &\leq \frac{q^{m\intOrder(2t-n+k)}}{\left| \left\{\M \in \Fqm^{\intOrder \times t} \, : \, \SumRankWeight(\M) = t\right\} \right|} \notag\\
&\leq \gammaq^{\shots}q^{m\intOrder(2t-n+k)} q^{-\intOrder m t} \notag \\
&= \gammaq^{\shots} q^{sm(t-n+k)}. \label{eq:LO_Pr_condition_M_fixed_h}
\end{align}

We union-bound this probability over the choices of $\hvec$ with $\SumRankWeight >n-t-k$. Note that \eqref{eq:LO_M_condition} is the same condition for two $\hvec$ vectors for which the row space of $\lambda_{n-t-k}(\hvec)_{\a}$ is the same. Since this row space is trivially the same for two vectors $\hvec$ and $\hvec'$ with $\hvec = \myalpha \hvec'$ for $\myalpha \in \Fqm^\ast$, we multiply \eqref{eq:LO_Pr_condition_M_fixed_h} by the following number:\footnote{For the existing failure probability bound in the special case $\ell=1$, Overbeck~\cite{overbeck2007public} uses $\quadbinomqm{t}{1} \approx q^{m(t-1)}$, which is in fact a relatively tight \emph{lower} bound (see \eqref{eq:bounds_gaussian_binomial}). Hence, the result in~\cite{overbeck2007public} rather gives an estimate than a strict upper bound. To obtain an expression that is a \emph{strict} upper bound, we use the right-hand side of \eqref{eq:bounds_gaussian_binomial}.}
\begin{align}
\frac{\left| \left\{ \hvec \in \Fqm^t \, : \, \SumRankWeight >n-t-k \right\} \right|}{q^m-1} \leq \frac{q^{mt}-1}{q^m-1} = \quadbinomqm{t}{1} \leq \gammaq q^{m(t-1)} \label{eq:counting_h_vectors}
\end{align}
Overall, we have
\begin{align*}
\Pr\!\left(\text{\eqref{eq:LO_lemma_existence_h} is satisfied for $\M$} \right) &\leq \gammaq^{\ell+1} q^{sm(t-n+k)}q^{m(t-1)} \\
&= \gammaq^{\ell+1} q^{m( t(s+1)-s(n-k)-1 )} \\
&= \gammaq^{\ell+1} q^{-m\left((s+1)(\tmax-t)+1\right)}.
\end{align*}
Note that this expression is independent of the rank partition $\t$, so it is also an upper bound for the probability $\Pr\!\left(\rk_{q^m}(\lambda_{n-t-k}(\E)_{\a})<t\right)$ with $\E$ drawn as in the lemma statement (i.e., from the set of all errors of sum-rank weight $t$, and not from the subset $\Eset^{(\t)}$ with a specific rank partition $\t$).
\end{proof}

\subsection{Proof of Proposition~\ref{prop:RFproperties}}
\begin{proof}
 \begin{itemize}
  \item Ad~\ref{itm:cardinalityGroebStar}): 
    By Lemma~\ref{lem:existence_ILRS} there exists at least one nonzero polynomial $Q\in\Qspace$ which implies that $\cardGroebStar\geq1$.
    Now suppose that $\LP{Q}=0$ for some $Q\in\Qspace\setminus\{0\}$, i.e. we have that $\deg(Q_0)>\max\{\deg(Q_j)+k-1\}$. 
    Since $t<\frac{\intOrder}{\intOrder+1}(n-k+1)$ and $Q\in\Qspace$ we have that $Q(x,f_1,\dots,f_\intOrder)=0$ (see Theorem~\ref{thm:decRadiusILRS}) which is possible only if $\deg(Q_0)\leq\max\{\deg(Q_j)+k-1\}$.
    Therefore we must have that $\deg_\w(Q)\geq\degConstraint$ and thus $Q\notin\Qspace$.
    Hence, there are no polynomials in $\Qspace$ (and so in $\GroebStar$) with leading position $0$ and therefore we must have that $\cardGroebStar\leq\intOrder$ since by assumption the leading positions of $\GroebStar$ are distinct and $\LP{Q}\in\{0,\dots,\intOrder\}$ for any $Q\in\MultSkewPolyringZeroDer$.
  \item Ad~\ref{itm:rankRFmat}):
    Let $\set{J}=\LP{\GroebStar}$ and suppose w.l.o.g. that $\deg_\vec{w}(Q^{(r)})=\degConstraint-1$ for all $r=1,\dots,\cardGroebStar$.
    In case some $Q^{(r)}$ have weighted degree less than $\degConstraint-1$ we can increase the degree to $\degConstraint-1$ by taking the left product with an appropriate $a\in\SkewPolyringZeroDer$ without changing $\LP{Q^{(r)}}$ and the number of solutions of $Q^{(r)}\left(x,f_1(x),\dots,f_\intOrder(x)\right)=0$ for all $r\in\intervallincl{1}{\cardGroebStar}$.
    Let $\Mat{Q}_{S}^{(j)}\in\Fqm^{k \times k}$ be the submatrix of $\RFmat$ consisting of the $k$ columns corresponding to the unknowns $f_{j,0},\aut^{-1}(f_{j,1}),\dots,\aut^{-(k-1)}(f_{j,k-1})$ and $k$ rows containing inverse automorphisms of the (leading) coefficient $q^{(r)}_{j,\degConstraint-k}$ for each $j\in \set{J}$ and some $r\in\intervallincl{1}{\cardGroebStar}$.
    Then each $\Mat{Q}_{S}^{(j)}$ is a $k\times k$ upper triangular matrix with elements
    \begin{equation*}
      \aut^{-(\degConstraint-k)}\left(q_{j,\degConstraint-k}^{(r)}\right),\aut^{-(\degConstraint-k+1)}\left(q_{j,\degConstraint-k}^{(r)}\right),\dots,\aut^{-(\degConstraint-1)}\left(q_{j,\degConstraint-k}^{(r)}\right)
    \end{equation*}
    on the diagonal since by definition the leading positions of $\GroebStar$ are distinct for all $Q^{(r)}\in\GroebStar$.
    Let $j_{r}$ for $r=1,\dots,\cardGroebStar$ be the indices of the leading terms of the polynomials in $\GroebStar$ and let $\cardGroebStar=|\GroebStar|$. Then we can set up an upper block triangular truncated root-finding subsystem of the form
    %
    \begin{equation}\label{eq:rootFindingSubSystem}
    \underbrace{
    \begin{pmatrix}
     \Mat{Q}_{S}^{(j_{1})} & \dots & \dots &
     \\
     & \Mat{Q}_{S}^{(j_{2})} & & \vdots
     \\
     & & \ddots & \vdots
     \\
     & & & \Mat{Q}_{S}^{(j_{\cardGroebStar})}
    \end{pmatrix}
    }_{\hat{\Mat{Q}}}
    \cdot
    \begin{pmatrix}
     f_{j_{1},0}    \\
     \vdots   \\
     \aut^{-(k-1)}\left(f_{j_{1}, k-1}\right) 
     \\[8pt] \hline
     \vdots 
     \\\hline
     f_{j_{\cardGroebStar}, 0}    \\
     \vdots   \\
     \aut^{-(k-1)}\left(f_{j_{\cardGroebStar}, k-1}\right)  \\
     \end{pmatrix}
     = -\hat{\vec{q}}_{0}
    \end{equation}
    where $\hat{\vec{q}}_{0}$ obtained by considering the corresponding entries of $\vec{q}_{0}$.
    We have $\rk_{q^m}\left(\Mat{Q}_{S}^{(j)}\right)=k$ for all $j\in {\set{J}}$ and conclude that $\rk_{q^m}(\RFmat)\geq\rk_{q^m}(\Mat{\hat{Q}})=\sum_{j\in \set{J}}k$ where the first inequality follows because we considered only a submatrix of $\RFmat$.
  \item Ad~\ref{itm:numSolRFsystem}): 
    By~\ref{itm:cardinalityGroebStar}) and~\ref{itm:rankRFmat}) the rank of the root-finding matrix $\RFmat$ satisfies $\rk_{q^m}(\RFmat)\geq \cardGroebStar k$.
    Hence, the dimension of the solution space of the $\Fqm$-linear root-finding system in~\eqref{eq:rootFindingEquationILRS} is at most $\intOrder k-\cardGroebStar k=k(\intOrder-\cardGroebStar)$.
  \item Ad~\ref{itm:unqiueSolRFsystem}): 
    The $\Fqm$-linear root-finding system in~\eqref{eq:rootFindingEquationILRS} has a unique solution if and only if the rank of the root-finding matrix $\RFmat$ in~\eqref{eq:rootFindingMatrixILRS} is full, i.e. if $\rk_{q^m}(\RFmat)=\intOrder k$.
    By~\ref{itm:rankRFmat}) this is satisfied if $\cardGroebStar=\intOrder$. 
    Now assume that the root-finding system set up with all polynomials $\widetilde{Q}^{(r)}\in\Qspace\setminus\{0\}$ has a unique solution whereas the root-finding system set up with $\GroebStar$ has no unique solution.
    Then there exists at least one $\widetilde{Q}^{(r)}\in\Qspace$ which cannot be represented as $\SkewPolyringZeroDer$-linear combination of the polynomials $Q^{(1)},\dots,Q^{(\cardGroebStar)}$ in $\GroebStar$, which contradicts the assumption that the left $\SkewPolyringZeroDer$-linear span of the polynomials in $\GroebStar$ contains $\Qspace$.
    Therefore, we conclude that the root-finding system in~\eqref{eq:rootFindingEquationILRS} has a unique solution if and only if $\cardGroebStar=\intOrder$.
 \end{itemize}
\end{proof}

\newpage
\section{Proofs and statements from Section~\ref{sec:fast_interpolation}}\label{app:proofs_fast_interpolation}

\subsection{Preliminaries on Skew Polynomial Matrices}

For a matrix $\MABout \in \SkewPolyringZeroDer^{a \times b}$ and a vector $\s \in \ZZ^a$, we define the $\s$-shifted column degree of $\MABout$ to be the tuple
\begin{equation*}
\cdeg_\s(\MABout) = (d_1,\dots,d_b) \in \left(\ZZ \cup \{-\infty\}\right)^b
\end{equation*}
where\ $d_j$ is the maximal shifted degree in the $j$-th column, i.e.,
$$d_j := \textstyle\max_{i=1,\dots,a}\{\deg \MABoutentry_{ij} + s_i \}.$$
We write $\cdeg(\MABout) := \cdeg_\0(\MABout)$, where $\0 := (0,\dots,0)$.
Analogously, for $\s \in \ZZ^b$, we define the ($\s$-shifted) row degree of $\MABout$ to be
\begin{equation*}
\rdeg_\s (\MABout) := \cdeg_\s\!\left(\MABout{}^\top\right) \quad \text{ and } \quad \rdeg \MABout := \cdeg\!\left(\MABout{}^\top\right).
\end{equation*}
The degree of the matrix, i.e.~the maximal degree among its entries, is denoted:
\begin{equation*}
\deg (\MABout) := \max_{i,j}\{\deg \MABoutentry_{ij}\}.
\end{equation*}
If $\v \in \SkewPolyringZeroDer^{1 \times a} \setminus \{\0\}$ is a row vector and $\s = (s_1,\dots,s_a) \in \ZZ^a$ a shift, we define the $\s$-pivot index of $\v$ to be the largest index $i$ with $1 \leq i \leq a$ such that $\deg v_i + s_i = \rdeg_\s(\v)$, and analogously for column vectors.
Note, that the $\s$-pivot index of a vector $\v\in\SkewPolyringZeroDer^{1 \times a}$ coincides with the leading position $\textrm{LP}_{\prec_{\s}}(V)$, where $V\in\MultSkewPolyringZeroDerAny{a}$ is the multivariate polynomial corresponding to $\v$.
If $a \geq b$ (or $a \leq b$, respectively), then we say that $\MABout$ is in column (row) $\s$-ordered weak Popov form if the $\s$-pivot indices of its columns (rows) are strictly increasing in the column (row) index.

Given a matrix $\MABin \in \SkewPolyringZeroDer^{a \times b}$ and an ``order'' $d \in \ZZ_{\geq 0}$, a left approximant basis is a matrix $\MABout \in \SkewPolyringZeroDer^{a \times a}$ such that $\MABout \MABin \equiv 0 \modr x^d$, and such that $\MABout$ is in a certain normal form while satisfying that any vector $\vec b \in \SkewPolyringZeroDer^{1 \times a}$ such that $\vec b \MABin \equiv 0 \modr x^d$ is in the left $\SkewPolyringZeroDer$-row space of $\MABout$.
An analogous definition is given for right approximant bases.

\begin{definition}[Left/Right Approximant Bases~\cite{bartz2020fast}]\label{def:minimal_approximant_basis}
Let $\MABin \in \SkewPolyringZeroDer^{a \times b}$ and $d \in \ZZ_{\geq 0}$.
\begin{itemize}
\item For $\s \in \ZZ^b$, a right {\MABnameFullStandard} is a full-rank matrix $\MABout \in \SkewPolyringZeroDer^{b \times b}$ s.t.\
    \begin{enumerate}
        \item $\MABout$ is in $\s$-ordered column weak Popov form.
        \item The columns of $\MABout$ are a basis of all right approximants of $\MABin$ of order $d$.
    \end{enumerate}
\item For $\s \in \ZZ^a$, a left {\MABnameFullStandard} is a full-rank matrix $\MABout \in \SkewPolyringZeroDer^{a \times a}$ s.t.\
    \begin{enumerate}
        \item $\MABout$ is in $\s$-ordered row weak Popov form.
        \item The rows of $\MABout$ are a basis of all right approximants of $\MABin$ of order $d$.
    \end{enumerate}
\end{itemize}
We denote by $\RMABnameShortStandard$ (right case) and $\LMABnameShortStandard$ (left case) the sets of all such bases, respectively.
If the input is not relevant, we simply write (left or right) approximant basis.
\end{definition}

By fixing the basis $\{1,y_1,\dots,y_\intOrder\}$ each multivariate skew polynomial $Q\in\MultSkewPolyringZeroDer$ of the form
\begin{equation*}
    Q(x,y_1,\dots,y_\intOrder)=Q_0(x)+Q_1(x)y_1+\dots+Q_\intOrder(x)y_\intOrder
\end{equation*}
may be uniquely represented by a vector\footnote{We index the vector $\Q$ starting from zero to be compliant with the conventional notation used in the literature for interpolation-based decoding.} $\Q=(Q_0,Q_1,\dots,Q_{\intOrder})\in\SkewPolyringZeroDer^{\intOrder+1}$ such that
\begin{equation*}
    Q(x,y_1,\dots,y_\intOrder)=\Q 
    \begin{pmatrix}
    1 
    \\ 
    y_1 
    \\ 
    \vdots 
    \\ 
    y_\intOrder
    \end{pmatrix}.
\end{equation*}
Note, that in this case we have that $\deg_{\vec{w}}(Q)=\rdeg_{\vec{w}}(\Q)$.

Given a set $\set{B} \defeq \{(b_i,a_i):i=1,\dots,n\}$, the minimal polynomial $\minpolyOpNoX{\set{B}}$ vanishing on the elements $b_1,\dots,b_n$ from $\Fqm$ with respect to the corresponding evaluation parameters $a_1,\dots,a_n$, i.e.
\begin{equation}
  \opev{\minpolyOpNoX{\set{B}}}{b_i}{a_i}=0,\quad\forall i=1,\dots,n,
\end{equation}
is defined as
\begin{equation}
  \minpolyOp{\set{B}}{}=\lclm\left(x-\frac{\aut(b_i)a_i}{b_i}\right)_{\mystack{1\leq i\leq n}{b_i\neq 0}},
\end{equation}
where $\lclm(\cdot)$ denotes the \ac{lclm} of the polynomials in the bracket.
We have $\deg(\minpolyOp{\set{B}}{})\leq n$ with equality if and only if the $b_i$ belonging to the same evaluation parameter $a_i$ are $\Fq$-linearly independent and the distinct $a_i$ are from different conjugacy classes of $\Fqm$.

\begin{example}[Minimal Skew Polynomial]
  Consider the elements $b_1,b_2,b_3,b_4$ from $\mathbb{F}_{3^5}$ and let $a_1=a_2=1$ and $a_3=a_4=\pe$.
  Since $1$ and $\pe$ are representatives from all $q-1=2$ nontrivial conjugacy classes of $\mathbb{F}_{3^5}$, we have that $\deg(\minpolyOp{\set{B}}{})=4$ where $\set{B}=\{(b_i,a_i):i=1,\dots,4\}$ if and only if the two elements $b_1$ and $b_2$ as well as the two elements $b_3$ and $b_4$ are $\F_3$-linearly independent.  
\end{example}

\subsection{Fast Interpolation via Minimal Approximant Bases}\label{subsec:fast_interpolation}
We now generalize the results on the fast (operator) interpolation algorithm based on minimal approximant bases~\cite[Algorithm~6]{bartz2020fast} to the generalized operator evaluation.
Let 
\begin{equation}\label{eq:def_int_point_matrix}
 \U=\left(\shot{\U}{1},\shot{\U}{2},\dots,\shot{\U}{\shots}\right)\in\prod_{i=1}^{\shots}\Fqm^{n_i\times (\intOrder+1)}
\end{equation}
be a tuple containing the matrices 
\begin{equation}
  \mat{U}^{(i)}=
  \begin{pmatrix}
   u_{1,1}^{(i)} & u_{1,2}^{(i)} & \dots & u_{1,\intOrder+1}^{(i)}
   \\ 
   u_{2,1}^{(i)} & u_{2,2}^{(i)} & \dots & u_{2,\intOrder+1}^{(i)}
   \\ 
   \vdots & \vdots & \ddots & \vdots
   \\ 
   u_{n_i,1}^{(i)} & u_{n_i,2}^{(i)} & \dots & u_{n_i,\intOrder+1}^{(i)}
  \end{pmatrix}
  \in\Fqm^{n_i\times (\intOrder+1)}
\end{equation}
where $\rk_q(\mat{U}^{(i)})=n_i$ for all $i=1,\dots,\shots$.
Then for all $j=1,\dots,n_i$ and $i=1,\dots,\shots$, the $j$-th row of each matrix $\shot{\U}{i}$ corresponds to the interpolation point associated with the generalized operator evaluation map $\mathscr{E}_j^{(i)}$.
Similar to~\cite[Problem~13]{bartz2020fast}, we now define the generalized operator vector interpolation problem in Problem~\ref{prob:general_interpolation_problem}.

\begin{problem}[Generalized Operator Vector Interpolation]\label{prob:general_interpolation_problem}
Given $\intOrder,n,D \in \ZZ_{> 0}$, $\vec{w} \in \ZZ_{\geq 0}^{\intOrder+1}$, $\a\in\Fqm^{\shots}$ and $\mat{U}\in \prod_{i=1}^{\shots}\Fqm^{n_i \times (\intOrder+1)}$ as defined in~\eqref{eq:def_int_point_matrix}, where the rows of each $\shot{\U}{i}$ are $\Fq$-linearly independent.
Consider the $\Fqm$-vector space $\Qspace$ (left scalar multiplication) of vectors $\Q = \left(Q_0,Q_1,\dots,Q_\intOrder\right)\in\SkewPolyringZeroDer^{\intOrder+1}$ that satisfy the following two conditions:
\begin{align}
\sum_{l=1}^{\intOrder+1} \opev{Q_{l-1}}{U_{j,l}^{(i)}}{a_i} &= 0, &&\forall \, j=1,\dots,n_i,\,i=1,\dots,\shots \label{eq:interpolation_problem_eval} \\
\rdeg_{\vec{w}}(\Q) &< \degConstraint. \label{eq:interpolation_problem_deg}
\end{align}
Find left $\SkewPolyringZeroDer$-linearly independent $\Q^{(1)},\dots,\Q^{(\IntParam)} \in \Qspace \setminus \{\0\}$ whose left $\SkewPolyringZeroDer$-span contains $\Qspace$. 
\end{problem}

Note, that the conditions~\eqref{eq:interpolation_problem_eval} and~\eqref{eq:interpolation_problem_deg} are equivalent to 
\begin{align*}
    \mathscr{E}_j^{(i)}(Q)&=0,\qquad \forall j=1,\dots,n_i,\,i=1,\dots,\shots
    \\ 
    \deg_{\vec{w}}(Q)&<\degConstraint,
\end{align*}
respectively, where $Q\in\MultSkewPolyringZeroDer$ is the multivariate skew polynomial corresponding to $\Q$.
Hence, the interpolation problem in the interpolation-based decoding procedures for \ac{ILRS} codes (Problem~\ref{prob:skewIntProblemGenOpILRS}) is an instance of the generalized operator vector interpolation problem in Problem~\ref{prob:general_interpolation_problem}.

We now show how to speed up the interpolation step for Problem~\ref{prob:general_interpolation_problem} (and thus also Problem~\ref{prob:skewIntProblemGenOpILRS}) by computing a so-called \emph{left approximant bases} of a matrix $\A$ that is constructed from interpolation and minimal polynomials depending on the interpolation points~\cite{bartz2020fast}.
To construct such a matrix $\A$, we first need to transform the interpolation points as described in Lemma~\ref{lem:intProblemSubspace_Zi_matrices}. 
Since we apply $\Fq$-linear elementary row operations on $\shot{\U}{i}$, the interpolation conditions do not change due to the $\Fq$-linearity of the generalized operator evaluation. 

\begin{lemma}\label{lem:intProblemSubspace_Zi_matrices}
Consider an instance of Problem~\ref{prob:general_interpolation_problem} with $\U=(\shot{\U}{1},\shot{\U}{2},\dots,\shot{\U}{\shots})$.
Using $\Fq$-linear elementary row operations, we can transform each $\U^{(i)}$ into a matrix of the form
\begin{align}
\widetilde{\mat{U}}^{(i)} = \left(
\def\arraystretch{1.4}
\begin{array}{ccccc}
\multicolumn{1}{c|}{\0_{\nu_1^{(i)}\times \eta_1^{(i)}}} & & \widetilde{\mat{U}}^{(i,1)} & & \\\hline
\multicolumn{2}{c|}{\0_{\nu_2^{(i)}\times \eta_2^{(i)}}} & & \widetilde{\mat{U}}^{(i,2)} &   \\\hline
\multicolumn{3}{c|}{\0_{\nu_3^{(i)}\times \eta_3^{(i)}}} & \multicolumn{2}{|c}{\widetilde{\mat{U}}^{(i,3)}} \\\hline
\multicolumn{5}{c}{\vdots} \\\hline
\multicolumn{4}{c|}{\0_{\nu_{\varrho^{(i)}}^{(i)}\times \eta_{\varrho^{(i)}}^{(i)}}} & \multicolumn{1}{|c}{\widetilde{\mat{U}}^{(i,\varrho^{(i)})}} 
\end{array}
\right), \label{eq:intProblemSubspace_Zi_matrices} 
\end{align}
where $1 \leq \varrho^{(i)} \leq \intOrder+1$ and we have $\widetilde{\mat{U}}^{(i,r)} \in \Fqm^{\nu_r^{(i)}\times (\intOrder+1-\eta_r^{(i)}) }$ for all $i=1,\dots,\shots$ and $r=1,\dots,\varrho^{(i)}$, with
\begin{itemize}
\item $0\leq \eta_1^{(i)} < \eta_2^{(i)} < \cdots < \eta_{\varrho^{(i)}}^{(i)} < \intOrder+1$,
\item $1 \leq \nu_i^{(i)} \leq n$ such that $\sum_{r=1}^{\varrho^{(i)}} \nu_r^{(i)} = n_i$, and
\item the entries of the first column of $\widetilde{\mat{U}}^{(i,r)}$ are linearly independent over $\Fq$ for each $i=1,\dots,\shots$ and $r=1,\dots,\varrho^{(i)}$.
\end{itemize}
Each matrix $\widetilde{\mat{U}}^{(i)}$ can be obtained with $O\left(\intOrder m n_i^{\omega-1}\right)$ operations in $\Fq$.
\end{lemma}
We define
\begin{itemize}
 \item $\set{A}_i=\left\{\eta_1^{(i)},\eta_2^{(i)},\dots,\eta_{\varrho^{(i)}}^{(i)}\right\}$
 \item $\set{A}=\bigcup_{i=1}^{\shots}\set{A}_i=\{\eta_1,\eta_2,\dots,\eta_\varrho\}$ (where $\varrho=|\set{A}|$)
 \item $\set{J}_r\defeq\left\{i:\eta_r\in\set{A}_i\right\}$ (set of shot indices $i$ that have the current pivot position $\eta_r+1$)
 \item For all $i\in\set{J}_r$ we define an $r_i$ s.t. $\eta_{r_i}^{(i)}=\eta_r$. Then we have that all matrices $\widetilde{\mat{U}}^{(i,r_i)}$ have the same pivot position $\eta_r+1$ for all $i\in\set{J}_r$ and $r=1,\dots,\varrho$.
\end{itemize}

\begin{lemma}\label{lem:interpolation_problem_kernel_basis}
Let $\widetilde{\mat{U}}^{(1)},\dots,\widetilde{\mat{U}}^{(\varrho^{(i)})}$ be defined as in Lemma~\ref{lem:intProblemSubspace_Zi_matrices} for all $i=1,\dots,\shots$.
Then, $\Q = (Q_0,\dots,Q_\intOrder) \in \SkewPolyringZeroDer^{\intOrder+1}$ satisfies Condition~\eqref{eq:interpolation_problem_eval} in Problem~\ref{prob:general_interpolation_problem} if and only if there exists a vector $\vec{\chi} \in \SkewPolyringZeroDer^{\varrho}$ such that
\begin{align}
\begin{pmatrix}
\Q & \vec{\chi}
\end{pmatrix}
\cdot
\MABin
= \0, \label{eq:intProblemSubspace_approximant_bases_reformulation}
\end{align}
where $\MABin \in \SkewPolyringZeroDer^{(\intOrder+1+\varrho) \times \varrho}$ is a matrix whose $r$-th column, for $r=1,\dots,\varrho$, is of the form
\begin{align*}
\left(
\begin{array}{c}
\0_{\eta_r \times 1} \\
\hline
1 \\
R^{(r)}_{\eta_r+2} \\
\vdots \\
R^{(r)}_{\intOrder+1} \\
\hline
\0_{(r-1) \times 1} \\
\hline
G^{(r)} \\
\hline
\0_{(\varrho-r) \times 1} \\
\end{array}
\right)
\end{align*}
where, for all $r=1,\dots,\varrho$ we have
\begin{align*}
G^{(r)} &:= \minpolyOpNoX{\set{B}_r} 
\quad\text{with}\quad 
\set{B}_r = \left\{\left(\widetilde{U}_{\kappa,1}^{(i,r_i)}, a_i\right):i\in\set{J}_r, \kappa=1,\dots,\nu_{r_i}^{(i)}\right\} \qquad \text{and}
\\
R^{(r)}_j &:= \IPop{\set{B}_{r,j}}
\quad\text{with}\quad 
\set{B}_{r,j}=\left\{\left(\widetilde{U}_{\kappa,1}^{(i,r_i)}, \widetilde{U}_{\kappa,j-\eta_r}^{(i,r_i)}, a_i\right):i\in\set{J}_r, \kappa=1,\dots,\nu_{r_i}^{(i)}\right\}
\end{align*}
for all $j=\eta_r+2,\dots,\intOrder+1$.
\end{lemma}

\begin{proof}
A vector $\Q = (Q_0,\dots,Q_{\intOrder}) \in \SkewPolyringZeroDer^{\intOrder+1}$ satisfies Condition~\eqref{eq:interpolation_problem_eval} in Problem~\ref{prob:general_interpolation_problem} on interpolation points in $\U=(\shot{\U}{1},\shot{\U}{2},\dots,\shot{\U}{\shots})$ if and only if each sub-block $(Q_{\eta_r},\dots,Q_{\intOrder})$ satisfies \eqref{eq:interpolation_problem_eval} on the rows of $\widetilde{\mat{U}}^{(i,r_i)}$.
Using $G^{(r)}$ and $R^{(r)}_j$ as above, we can rewrite this condition, restricted to $\widetilde{\mat{U}}^{(i,r)}$, as
\begin{align}
&\sum_{j=1}^{\intOrder+1} Q_{j-1}\!\left(U_{\kappa,j}^{(i)}\right)_{a_i} = 0 \quad \forall \kappa=1,\dots,\nReceiveShot{i},i=1,\dots,\shots \notag \\
\Leftrightarrow&\sum_{j=\eta_r+1}^{\intOrder+1} Q_{j-1}\!\left(\widetilde{U}_{\kappa,j-\eta_r}^{(i,r_i)}\right)_{a_i} = 0 \quad \forall \kappa=1,\dots,\nu_{r_i}^{(i)},i\in\set{J}_r \label{eq:i-th_interpolation_condition} \\
&\Leftrightarrow \: Q_{\eta_r}\!\left(\widetilde{U}_{\kappa,1}^{(i,r_i)}\right)_{a_i} + \sum_{j=\eta_r+2}^{\intOrder+1} Q_{j-1}\!\left(R_j^{(r)}\!\left(\widetilde{U}_{\kappa,1}^{(i,r_i)}\right)_{a_i}\right)_{a_i} = 0 \quad \forall \kappa,i \notag \\
&\Leftrightarrow \: \left(Q_{\eta_r} + \sum_{j=\eta_r+2}^{\shots+1} Q_{j-1} R_j^{(r)} \right)\!\left(\widetilde{U}_{\kappa,1}^{(i,r_i)}\right)_{a_i} = 0 \quad \forall \kappa,i \notag \\
&\Leftrightarrow \: Q_{\eta_r} + \sum_{j=\eta_r+2}^{\shots+1} Q_{j-1} R_j^{(r)} \overset{(a)}{\equiv} 0 \quad \modr \underbrace{\minpolyOp{\set{B}_r}{}}_{= \, G^{(r)}} \notag \\
&\Leftrightarrow \: \exists\, \chi_r \in \SkewPolyringZeroDer \, : \, Q_{\eta_r}\, + \sum_{j=\eta_r+2}^{\shots+1} Q_{j-1} R_j^{(r)} + \chi_r G^{(r)} = 0 \notag \\
&\Leftrightarrow \: \exists\, \chi_r \in \SkewPolyringZeroDer \, : \, \begin{pmatrix}
Q_{\eta_r} & \cdots & Q_{\intOrder} & \chi_r
\end{pmatrix} \cdot
\begin{pmatrix}
1 \\
R^{(r)}_{\eta_r+2} \\
\vdots \\
R^{(r)}_{\intOrder+1} \\
G^{(r)}
\end{pmatrix} = 0 \label{eq:left_component_kernels}
\end{align}
where $\set{B}_r = \left\{\big(\widetilde{U}_{\kappa,1}^{(i,r_i)}, a_i\right):i\in\set{J}_r, \kappa=1,\dots,\nu_{r_i}^{(i)}\big\}$.
This is equivalent to \eqref{eq:intProblemSubspace_approximant_bases_reformulation} since the $\chi_r$'s are independent of each other, but the $Q_j$ are the same for each $r$.
\end{proof}

Note, that the generalized operator evaluation of a skew polynomial modulo the minimal polynomial in step $(a)$ is considered in \cite[Lemma~1]{bartz2021fastSkewKNH}.

\begin{algorithm}[ht]
    \caption{\algoname{Fast Generalized Operator Interpolation Algorithm}}
    \label{alg:fast_interpolation}
    \SetKwInOut{Input}{Input}\SetKwInOut{Output}{Output}
    \Input{Instance of Problem~\ref{prob:general_interpolation_problem}: 
    $\intOrder,\shots,n,D \in \ZZ_{> 0}$, shift vector $\vec{w} \in \ZZ_{\geq 0}^{\intOrder+1}$, and $\mat{U}\in \prod_{i=1}^{\shots}\Fqm^{n_i \times (\intOrder+1)}$ as defined in~\eqref{eq:def_int_point_matrix}, where the rows of each $\shot{\U}{i}$ are $\Fq$-linearly independent.
    }
    \Output{If it exists, a solution of Problem~\ref{prob:general_interpolation_problem}. Otherwise, ``no solution''.}

    \BlankLine

    \For{$i=1,\dots,\shots$}{
        \If{elements in first column of $\shot{\U}{i}$ are $\Fq$-lin.\ ind.}{
            $\widetilde{\U}^{(i,1)} \gets \shot{\U}{i}$, $\varrho^{(i)} \gets 1$, $\nu_1^{(i)} \gets n_i$, $\eta_1^{(i)} \gets 0$
            } \Else{
            $\widetilde{\U}^{(i,r)} \in \Fqm^{\nu_r^{(i)} \times (\intOrder+1-\eta_r^{(i)})}$ for $r=1,\dots,\varrho^{(i)}$ $\gets$ compute as in Lemma~\ref{lem:intProblemSubspace_Zi_matrices} \label{line:intProblemSubspace_compute_Zi} \\
        }
    }
    Define 
        $\set{A}_i=\left\{\eta_1^{(i)},\eta_2^{(i)},\dots,\eta_{\varrho^{(i)}}^{(i)}\right\}$,
        $\set{A}=\bigcup_{i=1}^{\shots}\set{A}_i=\{\eta_1,\eta_2,\dots,\eta_\varrho\}$ and 
        $\set{J}_r\defeq\left\{i:\eta_r\in\set{A}_i\right\}$
    \\ 
    Define an $r_i$ s.t. $\eta_{r_i}^{(i)}=\eta_r$ For all $i\in\set{J}_r$.
    \\ 
    \For{$r=1,\dots,\varrho$}{
            $G^{(r)} \defeq \minpolyOpNoX{\set{B}_r}$ where $\set{B}_r = \left\{\left(\widetilde{U}_{\kappa,1}^{(i,r_i)}, a_i\right):i\in\set{J}_r, \kappa=1,\dots,\nu_{r_i}^{(i)}\right\}$ \\
            \For{$j=\eta_i+2,\dots,\intOrder+1$}{
                $R^{(r)}_j \defeq \IPop{\set{B}_{r,j}}$ with $\set{B}_{r,j}=\left\{\left(\widetilde{U}_{\kappa,1}^{(i,r_i)}, \widetilde{U}_{\kappa,j-\eta_r}^{(i,r_i)}, a_i\right):i\in\set{J}_r, \kappa=1,\dots,\nu_{r_i}^{(i)}\right\}$
            }
        }
    $\MABin \gets$ set up matrix from the $G^{(r)}$ and $R^{(r)}_j$ as in Lemma~\ref{lem:interpolation_problem_kernel_basis} \\
    $w_\mathrm{min} \gets \min_{l=1,\dots,\intOrder+1} \{w_l\}$ \\
    $d \gets D-w_\mathrm{min}+n$ \\
    $\s \gets (w_1,\dots,w_{\intOrder+1},w_\mathrm{min},\dots,w_\mathrm{min}) \in \ZZ_{\geq 0}^{\intOrder+1+\varrho}$ \\
    $\B \gets$ left \MABnameFullStandard \hfill \label{line:interpolation_minimal_approx_basis}  \myAlgoComment{solved by~\cite[Algorithm~4]{bartz2020fast}}
    $\{i_1,\dots,i_{\cardGroebStar}\} \gets$ indices of rows of $\B$ with $\s$-shifted row degree $<D$ \\
    \If{$\cardGroebStar>0$}{
        \For{$j=1,\dots,\cardGroebStar$}{
            $\Q^{(j)} \gets \left(B_{i_j,1}, \dots,B_{i_j,\intOrder+1}\right)$
        }
        \Return{$\Q^{(1)},\dots,\Q^{(\cardGroebStar)}$}
    } \Else{
        \Return{``no solution''}
    }
\end{algorithm}

\begin{theorem}[Correctness of Algorithm~\ref{alg:fast_interpolation}]\label{thm:fast_interpolation_correctness_complexity}
Algorithm~\ref{alg:fast_interpolation} is correct.
For the complexity, assume $D \in \Theta(n)$.
If the first column of the input matrices $\shot{\U}{1},\dots,\shot{\U}{\shots}$ consists of $\Fq$-linearly independent elements, it can be implemented with complexity
\begin{align*}
\softoh{\intOrder^\omega \OMul{n}}
\end{align*}
operations in $\Fqm$.
Otherwise, it costs 
\begin{align*}
\softoh{\intOrder^\omega \OMul{n}}
\end{align*}
operations in $\Fqm$ plus $O\left(\intOrder m n^{\omega-1}\right)$ operations in $\Fq$.
\end{theorem}

\begin{proof}
 The correctness of the algorithm follows from Lemma~\ref{lem:interpolation_problem_kernel_basis} and \cite[Lemma~21]{bartz2020fast}.
 The annihilator polynomials $G^{(r)}$ and interpolation polynomials $R^{(r)}_j$ can be computed in $\softO(\OMul{\nu_i})$ operations in $\Fqm$.
 Computing all the polynomials $G^{(r)}$ and $R^{(r)}_j$ with $r=1,\dots,\varrho$ and $j=r+1,\dots,\intOrder+1$ hence costs at most
 \begin{align*}
  \softoh{ \intOrder \sum_{i=1}^{\shots} \sum_{r=1}^{\varrho^{(i)}} \OMul{\nu_r^{(i)}}} 
  \subseteq 
  \softoh{ \intOrder \sum_{i=1}^{\shots} \OMul{n_i}}
  \subseteq 
  \softoh{\intOrder\OMul{n}}
 \end{align*}
 operations in $\Fqm$, since $\sum_{r=1}^{\varrho^{(i)}} \nu_r^{(i)} = n_i$ and $\OMul{\cdot}$ is a convex function.

Checking whether the first column of $\U^{(i)}$ has $\Fq$-rank $n_i$ can be done by computing the minimal polynomial of the entries $u_{1,1}^{(i)}, \dots, u_{n_i,1}{(i)}$ and checking if the degree equals $n_i$.
This check can be done in $\softoh{\OMul{n_i}}$ operations in $\Fqm$.
Overall, this requires $\softoh{\OMul{n}}$ operations in $\Fqm$.
Only if the entries are linearly independent, we need to compute the matrices $\U^{(i,r)}$ in Line~\ref{line:intProblemSubspace_compute_Zi}.
This costs $\oh{\intOrder m \sum_{i=1}^{\shots}n_i^{\omega-1}}\subseteq\oh{\intOrder m n^{\omega-1}}$ operations in $\Fq$.

By definition of $G^{(r)}$ and $R^{(r)}_j$, we have $\deg \MABin \leq n$. Due to $d \leq D+n$, Line~\ref{line:interpolation_minimal_approx_basis} costs $\softoh{\intOrder^\omega \OMul{n}}$ operations in $\Fqm$.
\end{proof}

\end{document}